\documentclass[12pt,letterpaper]{article}
\usepackage[T1]{fontenc}
\usepackage[utf8]{inputenc}
\usepackage{microtype}
\usepackage{setspace}
\usepackage{sectsty}
\allsectionsfont{\normalfont\bfseries}
\usepackage{fancyhdr}
\usepackage[hang,flushmargin]{footmisc}
\usepackage[font=small,labelfont=bf]{caption}
\usepackage[english]{babel}
\usepackage{csquotes}
\usepackage[margin=1.25in]{geometry}
\usepackage{amsmath}
\usepackage{amsfonts}
\usepackage{amssymb}
\usepackage{bbm}
\usepackage{amsthm}
\usepackage{graphicx}
\usepackage{placeins}
\usepackage{subcaption}
\usepackage[mathscr]{euscript}
\usepackage{rotating}
\usepackage{pdflscape}
\usepackage{booktabs}
\usepackage[authoryear]{natbib}
\usepackage{enumerate}
\usepackage{threeparttable}
\usepackage{verbatim}
\usepackage{url}
\usepackage{array}
\usepackage{adjustbox}
\usepackage[titletoc,title]{appendix}
\usepackage[hidelinks,breaklinks]{hyperref}
\hypersetup{colorlinks=true,allcolors=blue}
\usepackage{framed}
\usepackage[dvipsnames]{xcolor}

\newtheorem{prop}{PROPOSITION}

\theoremstyle{plain}
\newtheorem{assumption}{ASSUMPTION}
\newtheorem{define}{DEFINITION}
\newtheorem{example}{EXAMPLE}
\newcommand{\assumptiondyn}{Assumption~1-Dyn}

\usepackage{mathpazo} 

\begin{document}

\thispagestyle{plain}

\begin{center}
{\Large\bfseries Identification and Estimation of Demand Models\\
with Endogenous Product Entry and Exit\par}
\vspace{0.65em}
Victor Aguirregabiria\textsuperscript{\(\dagger\)}
\hspace{1.8em}
Alessandro Iaria\textsuperscript{\(\ddagger\)}
\hspace{1.8em}
Senay Sokullu\textsuperscript{\(\S\)}

\vspace{0.35em}
July 27, 2026
\end{center}

\vspace{-0.4em}

\begin{abstract}
\noindent Firms introduce products in markets where they anticipate stronger demand, using information unobserved by researchers. This creates endogenous selection in demand estimation. In differentiated-product oligopolies, multidimensional demand unobservables and strategic entry can render the ordinary propensity score insufficient for selection correction. Existing approaches either restrict firms' information at entry or jointly estimate demand, pricing, and entry under strong supply-side assumptions. We derive a new mixture representation of the selection-bias function using latent propensity scores: entry probabilities conditional on observables and a latent market state generating dependence across entry decisions. This representation yields a convenient two-step semiparametric estimator that corrects for selection and price endogeneity while accommodating richer information at entry. The approach makes weaker supply-side assumptions and is simpler to implement because it avoids repeatedly solving the full model. Applied to the US airline industry, the method yields more elastic demand and less market power than estimates ignoring endogenous entry.

\vspace{0.25em}
\noindent \textbf{Keywords:} Demand estimation; Product entry; Selection bias; Airline markets.

\vspace{0.25em}
\noindent \textbf{JEL codes:} C14, C34, C35, C57, D22, L13, L93.
\end{abstract}

\vfill

\begingroup
\footnotesize
\singlespacing
\noindent\textit{Acknowledgments.} We are grateful for helpful comments from Roy Allen, Gaurab Aryal, Giovanni Compiani, Andreea Enache, Christos Genakos, Ying Fan, Michele Fioretti, Philip Haile, Nail Kashaev, Mathieu Marcoux, Mateusz Mysliwski, Anders Munk-Nielsen, David Pacini, Bertel Schjerning, Philipp Schmidt-Dengler, Jesse Shapiro, Yutec Sun, Yuanyuan Wan, Ao Wang, and Christine Zulehner; as well as from seminar participants at the Universities of Bolzano, Bonn, Copenhagen, CREST, Duke, ENSAI-Rennes, Georgia, Glasgow, Harvard, Helsinki GSE, Cambridge (Judge), Mannheim, NHH-Bergen, Penn State, Rochester, Sciences Po, TSE, and Vienna; and from participants at the Advances in Demand Analysis Workshop (2024), BSE Summer Forum on Structural Microeconometrics (2024), CEPR IO Meeting (2024), Cowles Conference on Models \& Measurement (2024), IAAE (2023), MaCCI Summer Institute (2022), Midwest Econometrics Group Meeting (2023), and Oxford Econometrics Workshop (2025).

\vspace{0.25em}
\noindent\textsuperscript{\(\dagger\)}University of Toronto and CEPR, \href{mailto:victor.aguirregabiria@utoronto.ca}{victor.aguirregabiria@utoronto.ca}. \quad
\textsuperscript{\(\ddagger\)}University of Bristol and CEPR, \href{mailto:alessandro.iaria@bristol.ac.uk}{alessandro.iaria@bristol.ac.uk}. \quad
\textsuperscript{\(\S\)}University of Bristol, \href{mailto:senay.sokullu@bristol.ac.uk}{senay.sokullu@bristol.ac.uk}.
\endgroup

\newpage

\section{Introduction}

Demand systems for differentiated products are commonly estimated using data across markets and periods in which product availability varies. Because firms tend to introduce products where they anticipate stronger demand, partly on the basis of information unobserved by researchers, this variation reflects endogenous product selection and can substantially bias demand estimates. The problem arises in many industries, including airlines \citep{berry2006airline,berry2010tracing,aguirregabiria2012dynamic}, supermarket chains \citep{smith2004supermarket}, radio stations \citep{sweeting2013dynamic}, personal computers \citep{eizenberg2014upstream}, and ice cream \citep{draganska_2009}.

The selection problem in this model differs from conventional cases because demand unobservables are multi-dimensional and enter firms' expected profits non-additively. The resulting entry rule need not be monotone or admit the single-index representation under which the ordinary propensity score---the probability of product entry given exogenous observables---is sufficient to control for selection bias \citep{angrist_1997}. Multiple equilibria in the entry and pricing games create an additional source of non-monotonicity when different equilibria are selected across markets. Consequently, standard identification results and two-step estimation methods based on the ordinary propensity score do not generally apply \citep[e.g.,][]{ahn_powell_1993,das2003,aradillas2007pairwise,newey_2009}.

Existing approaches to this selection problem either rule out endogenous selection by assumption or accommodate it through joint estimation of demand, pricing, and entry under strong supply-side assumptions. In this paper, we establish more general conditions for the sequential identification and estimation of demand parameters when product entry is endogenous. Our approach leverages cross-product dependence in firms' market-entry decisions to identify entry probabilities that condition not only on observable characteristics but also on a latent market state capturing unobserved interdependencies among firms' choices. We refer to these probabilities as latent propensity scores. Unlike the ordinary propensity score, which averages over this latent state, latent propensity scores retain the information needed to characterize the selection-bias function in the demand equation.

Our identification result proceeds in two steps. First, we use the joint distribution of entry decisions to recover the latent propensity scores that summarize how this latent state shifts firms' entry probabilities. This step exploits a key feature of the model: conditional on observable characteristics and the latent market state, firms' entry decisions are independent, so the dependence across products observed in the data is generated by this latent state. Second, we show that the selection-bias function in the demand equation can be represented in terms of these latent propensity scores and the conditional means of the demand shocks given this latent state. We develop two complementary approaches that make this representation useful for identification: one based on a finite-mixture representation and one based on continuous normal heterogeneity with index restrictions.

Building on our constructive proof of identification, we propose a transparent and computationally simple two-step estimator that jointly corrects for endogenous product selection and price endogeneity in demand estimation. In the first step, we estimate the entry-probability components needed to construct the selection-bias function by semiparametric Maximum Likelihood Estimation (MLE). In the second step, we estimate demand by a Generalized Method of Moments (GMM) that controls for selection bias. This yields consistent estimates without specifying and jointly estimating a full equilibrium model of demand, pricing, and entry.

We illustrate the method using data from the airline industry. The results demonstrate the importance of accounting for endogenous product entry and highlight the limitations of conventional selection-correction approaches. Specifically, demand estimates that do not correct for endogenous entry substantially underestimate the magnitude of own- and cross-price elasticities and overstate market power. We also detect selection bias---though weaker than for demand---in the estimation of marginal-cost functions derived from Bertrand--Nash pricing equations. Moreover, our reduced-form estimation of entry probabilities---capturing rich correlations in firms’ entry decisions---provides economically meaningful insights. In particular, specifications that ignore correlated unobservables in market-entry decisions substantially understate the deterrent effect of competitors' presence on entry---a bias that would overstate market contestability in the evaluation of mergers.

Our approach contributes to two related strands of literature on differentiated-product demand with endogenous product entry. One strand estimates structural oligopoly models sequentially, beginning with demand  \citep{draganska_2009,aguirregabiria2012dynamic,fan_2013,sweeting2013dynamic,eizenberg2014upstream,schaumans_verboven_2015,fan_yang_2020,bontemps2023price,caoui_steck_2026,liu_luo_2025demand}. To avoid selection bias in this first step, these studies impose informational restrictions that rule out selection on demand unobservables, at the cost of potential misspecification. A second strand allows for such selection by jointly estimating demand, marginal costs, and entry costs using nested fixed-point algorithms \citep{ciliberto2021market,li2022repositioning}. While powerful, this approach makes demand identification depend on strong functional-form and distributional assumptions and is computationally demanding. We show that endogenous selection in demand estimation can instead be accommodated within a sequential approach: cross-firm dependence in entry decisions identifies the selection-bias controls without requiring the joint estimation of the full model or solving the entry game inside the demand estimator.

Our method contributes to the literature on semiparametric estimation of sample selection models \citep{das2003,newey_2009,powell_2001,aradillas2007pairwise}. We extend two-step propensity-score selection-function approaches to settings where the unobservables in the selection equation violate the standard monotonicity condition. When selection decisions arise within a system of simultaneous equations and non-monotonic unobservables generate dependence across them, we show that it is still possible to identify a selection function correcting for selection bias. Our approach applies in several contexts with this feature, such as labor market models with two-sided matching \citep{choo_siow_2006,galichon_salanie_2022}, joint household decisions \citep{browning_2014}, or peer effects models with endogenous network formation \citep{graham_2017,depaula_2018}.

The remainder of the paper is organized as follows. Sections \ref{sec:model}--\ref{sec:identification} introduce the model, characterize the selection problem, and present our identification and estimation methods. Section \ref{sec:application} applies our methods to the US airline industry, and Section \ref{sec:conclusions} concludes. Additional details and results appear in the Online Appendix.

\section{Model \label{sec:model}}
We study endogenous product entry in the canonical Berry--Levinsohn--Pakes (BLP) framework \citep{berry1995automobile}. For notational simplicity, we focus on single-product firms. In Online Appendix~\ref{sec_multiproduct}, however, we discuss how our arguments can be adapted to multi-product firms. There are $J$ firms indexed by $j\in\mathcal{J}=\{1,\ldots,J\}$ and $T$ markets indexed by $t\in\{1,\ldots,T\}$. A market may be a geographic location, a period, or a combination of both. Consumers in market $t$ can buy only the products available in that market. Market-entry decisions, prices, and quantities arise from the equilibrium of a two-stage game. In the first stage, each firm chooses whether to enter the market to maximize its expected profit. In the second stage, active firms set prices in a Bertrand--Nash equilibrium, with quantities determined by demand. The game is played separately across markets.\footnote{Treating markets as separate games is standard in empirical industrial organization. Important exceptions are structural entry models that allow potential entrants to internalize network externalities across markets \citep[e.g.,][]{bontemps2023price,jia2008happens,aguirregabiria2012dynamic}. These models do not study the endogenous sample-selection problem considered here.} The baseline model is static; however, Online Appendix~\ref{sec_dynamic_entry_exit} discusses how the same logic can be extended to dynamic games of firms' product entry and exit. To help keep track of the notation, Online Appendix Tables~\ref{tab:notation_main} and~\ref{tab:notation_appendix} summarize the main symbols used in the paper.

\subsection{Demand\label{sec:demand_model}}

The indirect utility of consumer $h$ in market $t$ from buying product $j$ is:
\begin{equation}
    U_{hjt}
    \equiv
    \delta_{jt}
    +
    \left( p_{jt}, \boldsymbol{x}_{jt}^{D \prime} \right)
    \boldsymbol{\Omega}_{\boldsymbol{\sigma}}
    \boldsymbol{\upsilon}_{ht}
    +
    \varepsilon_{hjt},
\end{equation}
where $p_{jt}$ is price and $\delta_{jt}$ is mean utility. For the baseline static model, let $\boldsymbol{x}_{t}$ collect all exogenous market- and product-level variables observed by the researcher, including the characteristics of every potential product regardless of entry, and let $\boldsymbol{x}_{jt}^{D}\in\mathbb{R}^{K}$ denote the subvector entering product $j$'s demand. The vector $\boldsymbol{\upsilon}_{ht}\in\mathbb{R}^{K+1}$ has a known mean-zero distribution $F_{\upsilon}$, and $\boldsymbol{\Omega}_{\boldsymbol{\sigma}}$ is a $(K+1)\times(K+1)$ matrix that is a known, continuously differentiable function of the parameter vector $\boldsymbol{\sigma}$. Thus, $\boldsymbol{\Omega}_{\boldsymbol{\sigma}}\boldsymbol{\upsilon}_{ht}$ is the vector of consumer-specific random-coefficient deviations. The shock $\varepsilon_{hjt}$ is independent of $\boldsymbol{\upsilon}_{ht}$ and is i.i.d. over $(h,j,t)$ with a type-I extreme-value distribution.

Following the standard linear specification, the mean utility of product $j$ is:
\begin{equation}
    \delta_{jt}
    \equiv
    \alpha p_{jt}
    +
    \boldsymbol{x}_{jt}^{D \prime} \boldsymbol{\beta}
    +
    \xi_{jt},
\label{eq:Delta}
\end{equation}
where $\alpha\in\mathbb{R}$ and $\boldsymbol{\beta}\in\mathbb{R}^{K}$ are parameters, and $\xi_{jt}$ captures product characteristics unobserved by the researcher. Throughout the paper, we normalize $\mathbb{E}(\xi_{jt})=0$. The outside option is indexed by $j=0$, with $U_{h0t}=\varepsilon_{h0t}$.

Let $a_{jt}\in\{0,1\}$ indicate whether product $j$ is offered in market $t$, and let $\boldsymbol{a}_{t}\equiv(a_{jt}:j\in\mathcal{J})$ collect the offer indicators. The outside option $j=0$ is always offered. Every consumer chooses the product that maximizes their utility. Define $\boldsymbol{\delta}_{t}\equiv(\delta_{jt}:j\in\mathcal{J})$. The market share $s_{jt}$ of product $j$ in market $t$, equal to the proportion of consumers choosing that product, is
\begin{equation}
    s_{jt}
    =
    d_{jt}(\boldsymbol{\delta}_{t},\boldsymbol{a}_{t},\boldsymbol{\sigma})
    \equiv
    \int
    \frac{
        a_{jt}
        e^{
            \delta_{jt}
            +
            (p_{jt},\boldsymbol{x}_{jt}^{D \prime})
            \boldsymbol{\Omega}_{\boldsymbol{\sigma}}
            \boldsymbol{\upsilon}
        }
    }{
        1+
        \sum_{i=1}^{J}
        a_{it}
        e^{
            \delta_{it}
            +
            (p_{it},(\boldsymbol{x}_{it}^{D})^{\prime})
            \boldsymbol{\Omega}_{\boldsymbol{\sigma}}
            \boldsymbol{\upsilon}
        }
    }
    \,
    dF_{\upsilon}(\boldsymbol{\upsilon}),
\label{eq_market_shares}
\end{equation}
where, to simplify notation, the market subscript on $d_{jt}$ is intended to capture the dependence of $d_{jt}$ on prices and observed demand characteristics.

For our analysis, we restrict the demand system to the products that are offered. Define $\mathcal{J}_{t}^{\boldsymbol{a}}\equiv\{j\in\mathcal{J}:a_{jt}=1\}$, $\boldsymbol{s}_{t}^{\boldsymbol{a}}\equiv(s_{jt}:j\in\mathcal{J}_{t}^{\boldsymbol{a}})$, and $\boldsymbol{\delta}_{t}^{\boldsymbol{a}}\equiv(\delta_{jt}:j\in\mathcal{J}_{t}^{\boldsymbol{a}})$. We write this subsystem as
\begin{equation}
    \boldsymbol{s}_{t}^{\boldsymbol{a}}
    =
    \boldsymbol{d}_{t}^{\boldsymbol{a}}
    (\boldsymbol{\delta}_{t}^{\boldsymbol{a}},\boldsymbol{\sigma}).
\label{subsystem}
\end{equation}
Proposition~\ref{proposition_1} establishes that, for any configuration of $\boldsymbol{a}_{t}$, this subsystem is invertible with respect to $\boldsymbol{\delta}_{t}^{\boldsymbol{a}}$ \citep{berry1994estimating}.

\bigskip

\begin{prop} \label{proposition_1}
    Suppose that the outside option $j=0$ is always offered. Fix a market $t$,
    its prices and observed demand characteristics, and a configuration
    $\boldsymbol{a}_{t}\in\{0,1\}^{J}$. Define the set of feasible interior
    market shares for the offered products as
    \[
        \mathcal{S}^{\boldsymbol{a}}
        \equiv
        \left\{
            \boldsymbol{s}^{\boldsymbol{a}}
            \in(0,1)^{|\mathcal{J}_{t}^{\boldsymbol{a}}|}:
            \sum_{j\in\mathcal{J}_{t}^{\boldsymbol{a}}}s_j<1
        \right\}.
    \]
    Then, for every fixed value of $\boldsymbol{\sigma}$, the mapping
    $\boldsymbol{d}_{t}^{\boldsymbol{a}}(\cdot,\boldsymbol{\sigma})$
    is a bijection from
    $\mathbb{R}^{|\mathcal{J}_{t}^{\boldsymbol{a}}|}$ onto
    $\mathcal{S}^{\boldsymbol{a}}$. Hence, for every
    $\boldsymbol{s}_{t}^{\boldsymbol{a}}\in\mathcal{S}^{\boldsymbol{a}}$,
    there exists a unique $\boldsymbol{\delta}_{t}^{\boldsymbol{a}}$ such that
    equation~\eqref{subsystem} holds, and the inverse
    $\left(\boldsymbol{d}_{t}^{\boldsymbol{a}}\right)^{-1}
    (\cdot,\boldsymbol{\sigma})$
    is well defined on $\mathcal{S}^{\boldsymbol{a}}$.
    \qquad $\blacksquare$
\end{prop}

\noindent\textit{Proof of Proposition \ref{proposition_1}:} See Online Appendix~\ref{app:proof_proposition_1}.

\bigskip

\noindent For every product-market observation with $a_{jt}=1$, let
$d_{jt}^{-1}(\cdot,\boldsymbol{\sigma})
\equiv
\left[(\boldsymbol{d}_{t}^{\boldsymbol{a}})^{-1}
(\cdot,\boldsymbol{\sigma})\right]_{j}$
denote the $j$-th component of the inverse demand system. Then
\begin{equation}
    d_{jt}^{-1}\left(
    \boldsymbol{s}_{t}^{\boldsymbol{a}},
    \boldsymbol{\sigma}
    \right)
    =
    \alpha p_{jt}
    +
    \boldsymbol{x}_{jt}^{D \prime}\boldsymbol{\beta}
    +
    \xi_{jt}.
\label{regression equation}
\end{equation}
The following example illustrates Proposition \ref{proposition_1} in the case of the nested logit model.

\bigskip

\begin{example}[Nested logit model]\label{example_1}
Let the products be partitioned into nests $\{\mathcal J_g\}_{g=1}^G$, with $g(j)$ denoting $j$'s nest; the outside option forms a singleton nest. Given $\boldsymbol a_t$ with $a_{jt}=1$, define $s_{0t}\equiv1-\sum_{i\in\mathcal J_t^{\boldsymbol a}}s_{it}$ and $s_{g,t}^{\boldsymbol a}\equiv\sum_{i\in\mathcal J_g:a_{it}=1}s_{it}$. In the standard nested logit model \citep{berry1994estimating}, with nesting parameter $\sigma\in[0,1)$, the share inversion is
\[
\begin{aligned}
d_{jt}^{-1}(\boldsymbol s_t^{\boldsymbol a},\sigma)
&=\ln\!\left(\frac{s_{jt}}{s_{0t}}\right)
-\sigma\ln\!\left(\frac{s_{jt}}{s_{g(j),t}^{\boldsymbol a}}\right)\\
    &=\delta_{jt}
    =\alpha p_{jt}+\boldsymbol{x}_{jt}^{D \prime}\boldsymbol{\beta}+\xi_{jt}.
\end{aligned}
\]
The shares above are positive, so the inverse is well defined whenever $a_{jt}=1$. \qquad $\blacksquare$
\end{example}

\noindent Equation~\eqref{regression equation} for product $j$ is available for estimation precisely when $a_{jt}=1$. For values of $\boldsymbol{x}_{t}$ such that
$\Pr(a_{jt}=1\mid\boldsymbol{x}_{t})>0$, the selection-bias function is
$\mathbb{E}(\xi_{jt}\mid\boldsymbol{x}_{t},a_{jt}=1)$. Subtracting this
conditional mean from $\xi_{jt}$ yields a residual with conditional mean zero
given $(\boldsymbol{x}_{t},a_{jt}=1)$.

Because firms make interdependent entry decisions in a game with
common unobservables, rivals' decisions
$\boldsymbol{a}_{-jt}\equiv(a_{it}:i\in\mathcal{J},\,i\neq j)$ can be
informative about $\xi_{jt}$. Hence, the configuration-specific selection-bias
function
$\mathbb{E}(\xi_{jt}\mid\boldsymbol{x}_{t},a_{jt}=1,\boldsymbol{a}_{-jt})$
may vary with $\boldsymbol{a}_{-jt}$. Rival entry does not, however, change the
structural form of product $j$'s inverted demand equation. For every
configuration $\boldsymbol{a}_{t}$ with $a_{jt}=1$, applying the inverse for
that offered set yields equation~\eqref{regression equation} with the same
parameters and demand shock $\xi_{jt}$ as in equation~\eqref{eq:Delta}, even
though the inverse mapping, equilibrium prices, and market shares generally
depend on the configuration. This differs from censored simultaneous-equation
settings in which censoring changes the subsystem determining the observed
outcome. By iterated expectations, the selection-bias function
$\mathbb{E}(\xi_{jt}\mid\boldsymbol{x}_{t},a_{jt}=1)$ is the average of the
configuration-specific functions over the distribution of
$\boldsymbol{a}_{-jt}$ conditional on
$(\boldsymbol{x}_{t},a_{jt}=1)$.

Controlling for
$\mathbb{E}(\xi_{jt}\mid\boldsymbol{x}_{t},a_{jt}=1)$ is sufficient to
remove selection bias, but one could instead control for the finer
selection-bias function
$\mathbb{E}(\xi_{jt}\mid\boldsymbol{x}_{t},a_{jt}=1,\boldsymbol{a}_{-jt})$.
The conditioning set used for the selection-bias function determines which
instruments are orthogonal to the resulting demand residual by construction.
Under the coarser correction, the residual is orthogonal to every function of
$\boldsymbol{x}_{t}$ over the selected sample $a_{jt}=1$, but need not be
orthogonal to functions of $\boldsymbol{a}_{-jt}$. The maintained assumptions
therefore justify instruments constructed from the characteristics
$\boldsymbol{x}_{t}$ observed for all potential products, but do not generally
justify instruments that depend on realized rival entry. For example,
$\sum_{i\neq j}\boldsymbol{x}_{it}^{D}$ is admissible on exogeneity grounds,
whereas $\sum_{i\neq j}\boldsymbol{x}_{it}^{D}a_{it}$ is not generally admissible.
Under the finer correction, the demand residual has conditional mean zero given
$(\boldsymbol{x}_{t},a_{jt}=1,\boldsymbol{a}_{-jt})$ and is therefore
orthogonal to functions of both $\boldsymbol{x}_{t}$ and
$\boldsymbol{a}_{-jt}$.

Although the finer selection-bias function broadens the class of admissible
instruments, conditioning on $\boldsymbol{a}_{-jt}$, which can take up to
$2^{J-1}$ configurations, creates additional dimensionality and support
challenges for identification and estimation. We therefore focus on the
coarser selection-bias function
$\mathbb{E}(\xi_{jt}\mid\boldsymbol{x}_{t},a_{jt}=1)$.

As discussed in Online Appendix~\ref{sec_multiproduct},
Proposition~\ref{proposition_1} continues to hold for multi-product firms
because it depends only on the demand system, not on product ownership.
Multi-product ownership changes the representation of the entry game, which
must be formulated over firm-level product portfolios, but the product-level
selection-bias function remains
$\mathbb{E}(\xi_{jt}\mid\boldsymbol{x}_{t},a_{jt}=1)$.

\subsection{Price competition}

Let $\boldsymbol{x}_{jt}^{\mathrm{mc}}$ and $\boldsymbol{x}_{jt}^{\mathrm{fc}}$ be product $j$'s variable- and fixed-cost subvectors of $\boldsymbol{x}_{t}$, respectively, and let $\Pi_{jt}$ denote its profit when active in market $t$:
\begin{equation}
    \Pi_{jt}
    \; = \;
    p_{jt} \; q_{jt}
    \; - \;
    c(q_{jt}, \, \boldsymbol{x}_{jt}^{\mathrm{mc}}, \, \omega_{jt})
    \; - \; fc_{jt},
\end{equation}
where $q_{jt}\equiv H_t s_{jt}$ is the quantity sold and $H_t$ is market size,
$c(q_{jt},\boldsymbol{x}_{jt}^{\mathrm{mc}},\omega_{jt})$ is the variable cost function,
$fc_{jt}$ is the fixed entry cost, and $\omega_{jt}$ is a variable-cost unobservable. Throughout the paper, we normalize $\mathbb{E}(\omega_{jt})=0$.

Conditional on the entry configuration $\boldsymbol{a}_{t}$, and under the
maintained assumption of single-product firms, every interior pure-strategy
Bertrand--Nash equilibrium satisfies the first-order conditions
\begin{equation}
    p_{jt}
    =
    mc_{jt}
    -
    s_{jt}
    \left[
        \dfrac{\partial s_{jt}}
        {\partial p_{jt}}
    \right]^{-1}
    \quad\text{for every }j\in\mathcal{J}_{t}^{\boldsymbol{a}},
\end{equation}
where
$mc_{jt}\equiv
\partial c(q_{jt},\boldsymbol{x}_{jt}^{\mathrm{mc}},\omega_{jt})/\partial q_{jt}$
is marginal cost. The own-price derivative is computed after substituting
equation~\eqref{eq:Delta} into the market-share function
\eqref{eq_market_shares}; it therefore accounts for the effect of price through
both mean utility and the random-coefficient term, while holding rivals' prices
fixed. All terms are evaluated at the equilibrium price vector.

Let $\boldsymbol{\xi}_{t}\equiv(\xi_{jt}:j\in\mathcal{J})$ and
$\boldsymbol{\omega}_{t}\equiv(\omega_{jt}:j\in\mathcal{J})$. For any rivals'
entry profile $\boldsymbol{a}_{-j}\in\{0,1\}^{J-1}$, let
\[
VP_j(
    \boldsymbol{a}_{-j},
    \boldsymbol{x}_{t},
    \boldsymbol{\xi}_{t},
    \boldsymbol{\omega}_{t}
)
\]
denote firm $j$'s \textit{indirect variable profit}: the variable profit
$p_{jt}q_{jt}-c(q_{jt},\boldsymbol{x}_{jt}^{\mathrm{mc}},\omega_{jt})$
evaluated at the Bertrand--Nash equilibrium conditional on
$(a_{jt}=1,\boldsymbol{a}_{-j},\boldsymbol{x}_{t},
\boldsymbol{\xi}_{t},\boldsymbol{\omega}_{t})$.\footnote{If the pricing game
admits multiple equilibria, we take as given an equilibrium-selection rule,
possibly indexed by an additional market-level unobservable. We otherwise
leave this rule unrestricted. The indirect variable profit is evaluated at the
equilibrium selected by this rule, and we suppress the selection index from
the notation.}

\subsection{Market-entry game}

We model product entry as the outcome of a static game among the $J$ potential
entrants. The model nests complete-information games such as those in
\cite{ciliberto2009market} and \cite{ciliberto2021market}, as well as
incomplete-information games with common-knowledge unobservables, as in
\cite{grieco2014discrete} and \cite{aguirregabiria2019identification}. It
allows firms' information about demand and variable-cost unobservables at entry
to range from knowing only their distributions to observing their realizations,
with imperfect common signals as intermediate cases. This flexibility is
central because firms' information determines the scope for
endogenous selection on demand and variable-cost unobservables. Online
Appendix~\ref{sec_dynamic_entry_exit} discusses how the framework can be
adapted to a dynamic game of product entry and exit.

\bigskip

\begin{assumption}
\label{assumption_1}
At the time of entry in market $t$, firm $j$'s information set is
$(\boldsymbol{x}_{t},\boldsymbol{\kappa}_{t},
\boldsymbol{\eta}_{jt})$.
\begin{itemize}
    \item [a.] The vector $\boldsymbol{x}_{t}$ is observable to the researcher
    and common knowledge among the firms.
    
    \item [b.] The latent market state $\boldsymbol{\kappa}_{t}$, unobserved by
    the researcher, collects all information about demand and variable-cost
    unobservables $(\boldsymbol{\xi}_{t},\boldsymbol{\omega}_{t})$ and fixed
    entry costs that is common knowledge among firms. It may include the entire
    vector $(\boldsymbol{\xi}_{t},\boldsymbol{\omega}_{t})$, in which case
    firms know all demand and variable-cost unobservables when making their
    entry decisions.
    
    \item [c.] The vector $\boldsymbol{\eta}_{jt}$ represents firm $j$'s
    private information about its fixed entry cost. These vectors are
    independently and identically distributed across firms with common
    distribution $F_{\boldsymbol{\eta}}$ and are independent of
    $(\boldsymbol{\xi}_{t},\boldsymbol{\omega}_{t},
    \boldsymbol{\kappa}_{t},\boldsymbol{x}_{t})$. If $F_{\boldsymbol{\eta}}$ is degenerate,
    the entry game has complete information.
    
    \item [d.] All variables unobserved by the researcher---
    $(\boldsymbol{\xi}_{t},\boldsymbol{\omega}_{t},
    \boldsymbol{\kappa}_{t},
    \{\boldsymbol{\eta}_{jt}:j\in\mathcal{J}\})$---are jointly independent of
    the exogenous observables in $\boldsymbol{x}_{t}$.
    \qquad $\blacksquare$
\end{itemize}
\end{assumption}

\noindent Assumption~\ref{assumption_1} separates firms' information at entry
into the observed common state $\boldsymbol{x}_{t}$, the latent market state
$\boldsymbol{\kappa}_{t}$, and firm-specific private information
$\boldsymbol{\eta}_{jt}$. The content of $\boldsymbol{\kappa}_{t}$ determines
what firms know about demand and variable-cost unobservables and hence the scope
for entry to select products on those unobservables. Private information
$\boldsymbol{\eta}_{jt}$ permits incomplete-information entry; when
$F_{\boldsymbol{\eta}}$ is degenerate, the game has complete information. Any demand and
variable-cost unobservables not observed at entry become common knowledge before
active firms set prices in the Bertrand--Nash equilibrium described above.
Finally, part [d] imposes the standard exogeneity of $\boldsymbol{x}_{t}$ with respect to
all variables unobserved by the researcher. This exogeneity condition underpins
identification in both demand and entry models.

We normalize the payoff from remaining inactive to zero. To simplify notation
in the following expected-profit expressions, we use
$\boldsymbol{\xi}_{t}$ locally as shorthand for the combined vector of demand
and variable-cost unobservables
$(\boldsymbol{\xi}_{t},\boldsymbol{\omega}_{t})$; when the distinction matters,
we write the two components separately. For a hypothetical profile of rivals'
entry decisions $\boldsymbol{a}_{-j}\in\{0,1\}^{J-1}$, let
$\pi_{j}(\boldsymbol{a}_{-j},\boldsymbol{x}_{t},
\boldsymbol{\kappa}_{t},\boldsymbol{\eta}_{jt})$ denote firm $j$'s expected
profit from entering, given its information. Under
Assumption~\ref{assumption_1},
\begin{equation}
    \pi_{j}(\boldsymbol{a}_{-j},\boldsymbol{x}_{t},
    \boldsymbol{\kappa}_{t},\boldsymbol{\eta}_{jt})
    =
    \int
    VP_{j}(\boldsymbol{a}_{-j},\boldsymbol{x}_{t},
    \boldsymbol{\xi}_{t})
    \,dF_{j,\xi}
    \left(\boldsymbol{\xi}_{t}\mid\boldsymbol{\kappa}_{t}\right)
    -
    fc(\boldsymbol{x}_{jt}^{\mathrm{fc}},\boldsymbol{\kappa}_{t},
    \boldsymbol{\eta}_{jt}).
\label{eq_expected_profit}
\end{equation}
Here $F_{j,\xi}(\cdot\mid\boldsymbol{\kappa}_{t})$ denotes firm $j$'s
conditional beliefs about the combined vector of demand and variable-cost
unobservables. Parts [c]--[d] of Assumption~\ref{assumption_1} allow
$\boldsymbol{x}_{t}$ and $\boldsymbol{\eta}_{jt}$ to be omitted from the
conditioning set. The fixed entry cost is known to firm $j$ at entry and
therefore lies outside the integral; write
$fc_{jt}\equiv
fc(\boldsymbol{x}_{jt}^{\mathrm{fc}},\boldsymbol{\kappa}_{t},
\boldsymbol{\eta}_{jt})$. If $\boldsymbol{\kappa}_{t}$ reveals all demand and
variable-cost unobservables at entry, this conditional distribution is
degenerate and the integral reduces to
$VP_{j}(\boldsymbol{a}_{-j},\boldsymbol{x}_{t},
\boldsymbol{\xi}_{t})$.

As noted in Assumption \ref{assumption_1}[c], the entry game is one of incomplete information unless each $\boldsymbol{\eta}_{jt}$ has a degenerate distribution. We describe its equilibria as Bayesian Nash Equilibria (BNE), a solution concept that reduces to a complete-information Nash equilibrium when each $\boldsymbol{\eta}_{jt}$ is degenerate. Given $(\boldsymbol{x}_{t},\boldsymbol{\kappa}_{t})$, a BNE of this game can be represented as a $J$-tuple of entry probabilities, one for each firm, $(P_{jt}:j\in\mathcal{J})$. To describe this BNE, we first define a firm's expected profit function that accounts for its uncertainty about other firms' entry decisions.
\begin{equation}
    \pi_{j}^{P}(\boldsymbol{x}_{t},\boldsymbol{\kappa}_{t},\boldsymbol{\eta}_{jt})
    =\sum_{\boldsymbol{a}_{-j}\in\{0,1\}^{J-1}}\left(
    {\displaystyle\prod\limits_{i\neq j}}
    \left[  P_{it}\right]  ^{a_{i}}
    \left[  1-P_{it}\right]  ^{1-a_{i}}\right)
    \pi_{j}(\boldsymbol{a}_{-j},\boldsymbol{x}_{t},
    \boldsymbol{\kappa}_{t},\boldsymbol{\eta}_{jt}).
\end{equation}
Firm $j$'s best response is to enter the market if and only if this expected profit is nonnegative. We can then define a BNE in this game as follows.\footnote{If the entry game admits multiple BNE, we take an equilibrium-selection rule as given. We interpret $\boldsymbol{\kappa}_{t}$ as including its index, which we suppress from the notation.}

\bigskip

\begin{define}
\label{definition_bne}
\textbf{Bayesian Nash Equilibrium.} Under Assumption \ref{assumption_1} and given $(\boldsymbol{x}_{t},\boldsymbol{\kappa}_{t})$, a BNE can be represented as a $J$-tuple of probabilities $\{ P_{jt} \equiv P_{j}(\boldsymbol{x}_{t},\boldsymbol{\kappa}_{t}): j \in \mathcal{J} \}$ that solves the following system of $J$ best-response equations in the space of probabilities:
\begin{equation}
    P_{jt} \; = \; 
    \displaystyle \int
    \mathbb{1} \left\{
        \pi_{j}^{P}(\boldsymbol{x}_{t},\boldsymbol{\kappa}_{t},\boldsymbol{\eta}_{jt})
        \geq 0
    \right\}
    \;
    dF_{\boldsymbol{\eta}} \left( \boldsymbol{\eta}_{jt} \right).
    \qquad \blacksquare
\end{equation}

\end{define}

\subsection{Special cases in the literature}

Several existing models of demand with endogenous product entry are nested within our framework. To highlight this generality, Table~\ref{tab:entry_info_structure} summarizes selected contributions, focusing on the features relevant to this paper: the presence and nature of endogenous selection on unobservables, particularly whether it operates through demand-side unobservables. Most empirical applications of endogenous product entry assume that firms do not know demand or variable-cost unobservables at entry. These models therefore rule out---by assumption---selection on the demand and variable-cost unobservables entering variable profits. Two important exceptions are the models in \cite{ciliberto2021market} and \cite{li2022repositioning}.

\begin{table}[!htbp]
    \centering
    \caption{Models of endogenous product entry with differentiated-product demand}
    \label{tab:entry_info_structure}
    \footnotesize
    \resizebox{\textwidth}{!}{%
    \begin{tabular}{p{2.3cm} | p{2.8cm} | p{2.8cm} | p{3.0cm} | p{2.8cm} | p{1.9cm} }
    \hline \hline
    \multicolumn{1}{c|}{Paper}
    & {Industry -- \; \; \; \; \; Product Entry}
    & {Selection on demand or variable-cost unobservables}
    & {What firms know about $\boldsymbol{\xi}_{t}$ at entry}
    & {Unobservables in entry cost}  & {Entry game} \\
    \hline
    \qquad \qquad \qquad \qquad
    \cite{aguirregabiria2012dynamic}
    & \qquad \qquad \qquad \qquad US airlines -- \; \; \; Route (city-pair)
    & \qquad \qquad \qquad \qquad NO, once 
    airline \& route FEs are accounted for.
    & \qquad \qquad \qquad \qquad 
    $\boldsymbol{\xi}_{t}$ is unknown. \qquad Firms know airline \& route FEs but not demand shocks $\boldsymbol{\xi}_{t}$.
    & \qquad \qquad \qquad \qquad Only $\boldsymbol{\eta}_{jt}$. Private information shocks assumed independent of $\boldsymbol{\xi}_{t}$.
    & \qquad \qquad \qquad \qquad 
    Incomplete information dynamic game
    \\
    \hline
    \qquad \qquad \qquad \qquad
    \cite{sweeting2013dynamic}
    & \qquad \qquad \qquad \qquad US radio -- \; \; \; \; \; \; Station genre 
    & \qquad \qquad \qquad \qquad NO, once 
    observable lagged variables are accounted for.
    & \qquad \qquad \qquad \qquad
    $\boldsymbol{\xi}_{t}$ is unknown. Follows AR(1). Firms know lagged but not current $\boldsymbol{\xi}_{t}$.
    & \qquad \qquad \qquad \qquad
    Only $\boldsymbol{\eta}_{jt}$. Private information shocks assumed independent of $\boldsymbol{\xi}_{t}$.
    & \qquad \qquad \qquad \qquad
    Incomplete information dynamic game
    \\
    \hline
    \qquad \qquad \qquad \qquad
    \cite{eizenberg2014upstream}
    & \qquad \qquad \qquad \qquad US home PC  -- \; \; \; \; \; \; PC models 
    & \qquad \qquad \qquad \qquad NO. Selection on entry-cost unobservables, which are assumed independent of $\boldsymbol{\xi}_{t}$.
    & \qquad \qquad \qquad \qquad
    $\boldsymbol{\xi}_{t}$ is unknown. Firms only observe the realizations of $\boldsymbol{\xi}_{t}$ after committing to product choices.
    & \qquad \qquad \qquad \qquad
    Includes common-knowledge $\boldsymbol{\kappa}_{t}$, assumed independent of $\boldsymbol{\xi}_{t}$.
    & \qquad \qquad \qquad \qquad
    Complete information static game
    \\
    \hline
    \qquad \qquad \qquad \qquad
    \cite{fan_yang_2020}
    & \qquad \qquad \qquad \qquad US smartphones -- \; \; \; \; \; \; Phone models
    & \qquad \qquad \qquad \qquad NO. Selection on entry-cost unobservables, which are assumed independent of $\boldsymbol{\xi}_{t}$.
    & \qquad \qquad \qquad \qquad
    $\boldsymbol{\xi}_{t}$ is unknown. \qquad Firms know brand FEs but not demand shocks $\boldsymbol{\xi}_{t}$.
    & \qquad \qquad \qquad \qquad
    Includes common-knowledge $\boldsymbol{\kappa}_{t}$, assumed independent of $\boldsymbol{\xi}_{t}$.
    & \qquad \qquad \qquad \qquad
    Complete information static game
    \\
    \hline
    \qquad \qquad \qquad \qquad
    \cite{ciliberto2021market}
    & \qquad \qquad \qquad \qquad US airlines -- \; \; \; \; \; \; Route (city-pair) 
    & \qquad \qquad \qquad \qquad YES
    & \qquad \qquad \qquad \qquad $\boldsymbol{\xi}_{t}$ is known. \qquad Demand unobservables are known to firms when making product choices. 
    & \qquad \qquad \qquad \qquad 
    Includes common-knowledge $\boldsymbol{\kappa}_{t}$ that can be correlated with $\boldsymbol{\xi}_{t}$.
    & \qquad \qquad \qquad \qquad 
    Complete information static game
    \\
    \hline
    \qquad \qquad \qquad \qquad
    \cite{li2022repositioning}
    & \qquad \qquad \qquad \qquad US airlines -- \; \; \; \; \; \; Route (city-pair)
    & \qquad \qquad \qquad \qquad YES
    & \qquad \qquad \qquad \qquad $\boldsymbol{\xi}_{t}$ is known. \qquad Demand unobservables are known to firms when making product choices. 
    & \qquad \qquad \qquad \qquad 
    Includes common-knowledge $\boldsymbol{\kappa}_{t}$ that can be correlated with $\boldsymbol{\xi}_{t}$.
    & \qquad \qquad \qquad \qquad 
    Complete information static game
    \\
    \hline
    \qquad \qquad \qquad \qquad
    \cite{bontemps2023price}
    & \qquad \qquad \qquad \qquad US airlines -- \; \; \; \; \; \; Airline's network of non-stop routes
    & \qquad \qquad \qquad \qquad NO. Selection on entry-cost unobservables, which are assumed independent of $\boldsymbol{\xi}_{t}$.
    & \qquad \qquad \qquad \qquad
    $\boldsymbol{\xi}_{t}$ is unknown. \qquad Firms don't know demand unobservables when making network choices.
    & \qquad \qquad \qquad \qquad
    Includes common-knowledge $\boldsymbol{\kappa}_{t}$, assumed independent of $\boldsymbol{\xi}_{t}$.
    & \qquad \qquad \qquad \qquad
    Complete information static game
    \\
    & & & & & \\
    \hline \hline
\end{tabular}}
\end{table}

\section{Structure of the selection problem \label{sec:selection_problem}}

\subsection{Selection-bias function}

Recall from equation \eqref{regression equation} that, for any observation with $a_{jt}=1$, the demand inversion yields $d_{jt}^{-1}(\boldsymbol{s}_{t}^{\boldsymbol{a}}, \boldsymbol{\sigma}) = \alpha \, p_{jt} + \boldsymbol{x}_{jt}^{D \prime} \boldsymbol{\beta} + \xi_{jt}$. Whether product $j$ is offered is determined by firm $j$'s equilibrium entry decision:
\begin{equation}
    a_{jt} \; = \;
    \mathbb{1}\left\{
        \pi_{j}^{P}(\boldsymbol{x}_{t},\boldsymbol{\kappa}_{t},\boldsymbol{\eta}_{jt}) \geq 0
    \right\}.
\end{equation}
Since the sample of observations for which the demand equation holds is selected, we decompose the unobservable $\xi_{jt}$ into its conditional mean and a residual: $\xi_{jt} = \lambda_{j}(\boldsymbol{x}_{t}) + \widetilde{\xi}_{jt}$, where $\lambda_{j}(\boldsymbol{x}_{t}) \equiv \mathbb{E}(\xi_{jt} \mid \boldsymbol{x}_{t}, a_{jt}=1)$ is the selection-bias function and $\mathbb{E}(\widetilde{\xi}_{jt} \mid \boldsymbol{x}_{t}, a_{jt}=1)=0$ by construction. This gives the following regression equation for any observation with $a_{jt}=1$:
\begin{equation}
    d_{jt}^{-1}\left(\boldsymbol{s}_{t}^{\boldsymbol{a}},\boldsymbol{\sigma}\right)
    \; = \;
    \alpha \; p_{jt} +
    \boldsymbol{x}_{jt}^{D \prime} \; \boldsymbol{\beta} +
    \lambda_{j}(\boldsymbol{x}_{t}) +
    \widetilde{\xi}_{jt}.
\label{regression equation with selection term}
\end{equation}
Let $F_{\kappa}$ denote the marginal distribution of $\boldsymbol{\kappa}_{t}$. The structure of the selection-bias function plays a key role in the identification of $(\alpha, \, \boldsymbol{\beta}, \, \boldsymbol{\sigma})$. To characterize this structure, define the \textit{ordinary} propensity score---the probability of product $j$ being offered, conditional on observables---as:
\begin{equation}
    \overline{P}_{j}\left(\boldsymbol{x}_{t}\right)
    \;\equiv\;
    \Pr\left(a_{jt} =1\;|\;\boldsymbol{x}_{t}\right)
    \;=\; {\displaystyle\int}
    \mathbb{1}\left\{
        \pi_{j}^{P}(\boldsymbol{x}_{t},
        \boldsymbol{\kappa}_{t},\boldsymbol{\eta}_{jt}) \geq 0
    \right\}
    \;
    dF_{\kappa}(\boldsymbol{\kappa}_{t}) \,
    dF_{\boldsymbol{\eta}}(\boldsymbol{\eta}_{jt}).
\end{equation}
For values of $\boldsymbol{x}_{t}$ such that $\overline{P}_{j}(\boldsymbol{x}_{t}) > 0$, the selection-bias function can be written as:
\begin{equation}
    \lambda_{j}(\boldsymbol{x}_{t})
    \; = \;
    \frac{
        \mathbb{E}\left(
            \xi_{jt} \, a_{jt}
            \;\middle|\;
            \boldsymbol{x}_{t}
        \right)
    }{
        \overline{P}_{j}\left(\boldsymbol{x}_{t}\right)
    }.
\end{equation}
It is well known that estimating equation \eqref{regression equation with selection term} by instrumental variables---treating $\lambda_{j}(\boldsymbol{x}_{t}) + \widetilde{\xi}_{jt}$ as the composite error---is generally invalid. If $\lambda_{j}(\boldsymbol{x}_{t})$ is omitted, instruments constructed from $\boldsymbol{x}_{t}$ generally enter the composite error through the omitted selection-bias function and therefore violate the exclusion restriction \citep{wooldridge_pd_book_2010}. A natural alternative is a selection-function approach that explicitly controls for $\lambda_{j}(\boldsymbol{x}_{t})$. However, without additional structure, $\lambda_{j}(\boldsymbol{x}_{t})$ remains an unrestricted function of the full observed state $\boldsymbol{x}_{t}$. In particular, the direct effect of $\boldsymbol{x}_{jt}^{D}$ on consumer demand ($\boldsymbol{\beta}$) cannot be separated from its indirect effect through selection; absent further restrictions, the demand parameters are not identified.

In this setting, the standard strategy in the literature is to impose conditions under which the selection-bias function depends only on the ordinary propensity score $\overline{P}_{j}(\boldsymbol{x}_{t})$, that is, $\lambda_{j}(\boldsymbol{x}_{t})=g_{j}\!\left(\overline{P}_{j}(\boldsymbol{x}_{t})\right)$ for some function $g_{j}(\cdot)$. This restriction reduces the dimensionality of the nonparametric problem: $g_j$ remains unrestricted, but its argument is the scalar $\overline{P}_{j}(\boldsymbol{x}_{t})$ rather than the full vector $\boldsymbol{x}_{t}$. With sufficient variation in the demand regressors conditional on the ordinary propensity score, their direct effects can be distinguished from the selection effect captured by $g_j$. Estimation then follows a standard two-step procedure. In the first step, the ordinary propensity score $\overline{P}_{j}(\boldsymbol{x}_{t})$ is estimated using data on $(a_{jt},\boldsymbol{x}_{t})$. In the second step, the structural parameters are estimated using, for example, the series estimators in \cite{das2003} and \cite{newey_2009}, or the pairwise differencing approaches in \cite{powell_2001} and \cite{aradillas2012pairwise}. After controlling for $g_j\!\left(\overline{P}_{j}(\boldsymbol{x}_{t})\right)$, price endogeneity due to simultaneity can be handled using familiar BLP-type instruments constructed from the exogenous characteristics of other potential products.

\subsection{Failure of the ordinary propensity score}

The restriction that the selection-bias function depend solely on the ordinary propensity score need not hold in models of endogenous product entry and oligopoly competition, even under simple specifications. Intuitively, entry decisions in these environments depend on equilibrium profitability, which is jointly determined by the demand and variable-cost unobservables of all competing products. As a result, the selection rule need not admit the single-index structure under which the ordinary propensity score controls for selection bias.

Proposition~\ref{proposition_2} summarizes benchmark conditions under which the ordinary propensity score is sufficient. Under the maintained exogeneity conditions, the monotonicity condition used in local average treatment effects (LATE) is equivalent to a scalar-threshold representation of the entry indicator, as in Theorem~1 of \cite{vytlacil_2002}. Either of these equivalent conditions is sufficient for the ordinary propensity score to control for selection bias, following Propositions~2 and~3 in \cite{angrist_1997}.

\bigskip

\begin{prop}
\label{proposition_2}
    Under the independence of $(\boldsymbol{\xi}_{t}, \boldsymbol{\kappa}_{t}, \boldsymbol{\eta}_{jt})$ from $\boldsymbol{x}_{t}$ implied by Assumption~\ref{assumption_1}[d], consider the following conditions:
    \begin{itemize}
    \item [a.] Monotonicity: For any two observable states $\boldsymbol{x}$ and $\boldsymbol{x}^{\prime}$, either 
    $\mathbb{1}\{ \pi_{j}^{P}(\boldsymbol{x}, \boldsymbol{\kappa},\boldsymbol{\eta}_{j}) \ge 0 \} \geq \mathbb{1}\{ \pi_{j}^{P}(\boldsymbol{x}^{\prime}, \boldsymbol{\kappa},\boldsymbol{\eta}_{j}) \ge 0 \}$ for almost every $(\boldsymbol{\kappa},\boldsymbol{\eta}_{j})$, or the reverse inequality holds for almost every $(\boldsymbol{\kappa},\boldsymbol{\eta}_{j})$.
    \item [b.] Single-index representation: There are real-valued functions $h_{1j}$ and $h_{2j}$ such that, for every observable state $\boldsymbol{x}$,
    $\pi_{j}^{P}(\boldsymbol{x}, \boldsymbol{\kappa},\boldsymbol{\eta}_{j}) \geq 0 \Longleftrightarrow
    h_{1j}(\boldsymbol{x}) \; \geq \;
    h_{2j}(\boldsymbol{\kappa},\boldsymbol{\eta}_{j})$ for almost every $(\boldsymbol{\kappa},\boldsymbol{\eta}_{j})$.
    \item [c.] Conditional independence: the joint distribution of $(\xi_{jt}, a_{jt})$ conditional on $(\boldsymbol{x}_{t}, \overline{P}_{j}(\boldsymbol{x}_{t}))$ coincides with its distribution conditional on $\overline{P}_{j}(\boldsymbol{x}_{t})$ alone.
    \end{itemize}
    Under the conditions of Theorem~1 in \cite{vytlacil_2002}, with the separability extension stated there for uncountable support, conditions (a) and (b) are equivalent. Furthermore, either condition (a) or (b) implies condition (c). Condition (c) is precisely the requirement that the ordinary propensity score $\overline{P}_{j}(\boldsymbol{x}_{t})$ be sufficient for the joint distribution of $(\xi_{jt},a_{jt})$. For the selected demand equation and values with $\overline{P}_{j}(\boldsymbol{x}_{t})>0$, it implies the weaker conditional-mean restriction that $\lambda_j(\boldsymbol{x}_{t})$ depends on $\boldsymbol{x}_{t}$ only through $\overline{P}_{j}(\boldsymbol{x}_{t})$. \qquad $\blacksquare$
\end{prop}

\noindent The following example provides a simple oligopoly model with endogenous entry satisfying Assumption~\ref{assumption_1} in which both condition (c) and the weaker conditional-mean restriction for the selected demand equation fail.

\bigskip

\begin{example}[\textbf{\textit{Failure of ordinary propensity-score sufficiency}}]
\label{example_2}
Consider a two-product setting satisfying Assumption \ref{assumption_1}. Product $k$ is always offered, while product $j$ enters endogenously, and demand follows the standard logit model. We omit the market subscript and fix market size $H>0$. Constant marginal costs $mc_i(\boldsymbol{x}_i^{\mathrm{mc}})$, exogenous price-cost margins $PCM_i(\boldsymbol{x})$, and the fixed entry cost $fc_j(\boldsymbol{x}_j^{\mathrm{fc}})$ depend only on observables. Define $v_i(\boldsymbol{x})\equiv(\boldsymbol{x}_i^{D})^{\prime}\boldsymbol{\beta}+\alpha[mc_i(\boldsymbol{x}_i^{\mathrm{mc}})+PCM_i(\boldsymbol{x})]$ for $i\in\{j,k\}$. Demand shocks $\boldsymbol{\xi}=(\xi_j,\xi_k)$ are common knowledge at entry and there is no private information, so $\boldsymbol{\kappa}=\boldsymbol{\xi}$ and $\boldsymbol{\eta}_j$ is degenerate. Assume $0<fc_j(\boldsymbol{x}_j^{\mathrm{fc}})<PCM_j(\boldsymbol{x})H$ on the support considered below. Firm $j$'s entry profit and the equivalent entry rule are
\[
\scalebox{0.935}{$
\begin{aligned}
\pi_j^P
&=PCM_j(\boldsymbol{x})H
\frac{e^{v_j(\boldsymbol{x})+\xi_j}}
{1+e^{v_j(\boldsymbol{x})+\xi_j}+e^{v_k(\boldsymbol{x})+\xi_k}}
-fc_j(\boldsymbol{x}_j^{\mathrm{fc}}),\\
\pi_j^P\geq0
&\Longleftrightarrow
[PCM_j(\boldsymbol{x})H-fc_j(\boldsymbol{x}_j^{\mathrm{fc}})]e^{v_j(\boldsymbol{x})+\xi_j}
\geq fc_j(\boldsymbol{x}_j^{\mathrm{fc}})[1+e^{v_k(\boldsymbol{x})+\xi_k}],\\
a_j
&=\mathbb{1}\!\left\{\xi_j\geq
\ln[1+e^{v_k(\boldsymbol{x})+\xi_k}]-B_j(\boldsymbol{x})\right\},\quad
B_j(\boldsymbol{x})\equiv v_j(\boldsymbol{x})+
\ln\!\frac{PCM_j(\boldsymbol{x})H-fc_j(\boldsymbol{x}_j^{\mathrm{fc}})}{fc_j(\boldsymbol{x}_j^{\mathrm{fc}})}.
\end{aligned}
$}
\]

\noindent Suppose $\xi_j$ and $\xi_k$ are independent and jointly independent of $\boldsymbol{x}$. Let $\xi_j$ have mean zero, $\mathbb{E}|\xi_j|<\infty$, and a continuously differentiable CDF $F$ with density $f(z)>0$ for every $z\in\mathbb{R}$. Let $\xi_k\in\{u_L,u_H\}$ with equal probability, where $u_H=-u_L>0$. Assume also that $(v_k(\boldsymbol{x}),B_j(\boldsymbol{x}))$ has a density that is strictly positive on a nonempty open set. For $m\in\{L,H\}$, define
\[
\begin{aligned}
t_m(v,b)&\equiv\ln(1+e^{v+u_m})-b,
&r_m(v)&\equiv\frac{e^{v+u_m}}{1+e^{v+u_m}},\\
\mathsf P(v,b)&\equiv1-\tfrac12[F(t_L(v,b))+F(t_H(v,b))],
&\mathsf N(v,b)&\equiv\tfrac12\sum_{m\in\{L,H\}}\int_{t_m(v,b)}^\infty zf(z)\,dz.
\end{aligned}
\]
The threshold rule implies $\overline P_j(\boldsymbol{x})=\mathsf P(v_k(\boldsymbol{x}),B_j(\boldsymbol{x}))$ and $\mathbb{E}(\xi_ja_j\mid\boldsymbol{x})=\mathsf N(v_k(\boldsymbol{x}),B_j(\boldsymbol{x}))$. Since $\partial t_m/\partial v=r_m$ and $\partial t_m/\partial b=-1$, direct differentiation gives
\[
\det\!\left(\frac{\partial(\mathsf P,\mathsf N)}{\partial(v,b)}\right)
=\tfrac14 f(t_L)f(t_H)(r_H-r_L)(t_H-t_L)>0,
\]
because $u_H>u_L$ implies $r_H>r_L$ and $t_H>t_L$.

The inverse function theorem and the local-density assumption imply that $(\mathsf P,\mathsf N)$ has a two-dimensional absolutely continuous component. It therefore cannot satisfy $\mathsf N=n(\mathsf P)$ almost surely for any measurable scalar function $n$, as condition (c) in Proposition \ref{proposition_2} would require. Thus condition (c) fails. Since $0<\overline P_j(\boldsymbol{x})<1$, the ratio $\lambda_j(\boldsymbol{x})=\mathbb{E}(\xi_ja_j\mid\boldsymbol{x})/\overline P_j(\boldsymbol{x})$ is not a function of the ordinary propensity score either. Finally, because either condition (a) or condition (b) would imply condition (c), both fail. \qquad $\blacksquare$
\end{example}

\noindent Example \ref{example_2} shows that, even in simple environments, the ordinary propensity score need not be sufficient in the sense of condition (c): with $\boldsymbol{\kappa}_{t}=\boldsymbol{\xi}_{t}$, the rival's demand shock, $\xi_{kt}$, affects firm $j$'s entry threshold through oligopoly competition. As a result, selection depends on the full latent market state $\boldsymbol{\kappa}_{t}$, whereas the ordinary propensity score averages over $\boldsymbol{\kappa}_{t}$ and loses relevant information.

\subsection{Latent propensity scores and the selection-bias function}

We now derive a representation of the selection-bias function $\lambda_{j}(\boldsymbol{x}_{t})$ using the equilibrium entry probabilities from Definition~\ref{definition_bne}, which satisfy
\[
    P_{j}(\boldsymbol{x}_{t}, \boldsymbol{\kappa}_{t})
    \;=\;
    \Pr\left( a_{jt} = 1 \;\middle|\; \boldsymbol{x}_{t}, \boldsymbol{\kappa}_{t} \right).
\]
We refer to $P_{j}(\boldsymbol{x}_{t}, \boldsymbol{\kappa}_{t})$ as the \textit{latent propensity score}, to highlight its dependence on the latent market state $\boldsymbol{\kappa}_{t}$. By Assumption~\ref{assumption_1}[d], the ordinary propensity score is the average of the latent propensity scores over the marginal distribution of $\boldsymbol{\kappa}_{t}$:
\begin{equation}
    \overline{P}_j\left( \boldsymbol{x}_{t} \right)
    \; = \;
    \displaystyle \int
    P_{j}\left(  \boldsymbol{x}_{t}, \boldsymbol{\kappa} \right)
    \, dF_{\kappa}\left(\boldsymbol{\kappa}\right).
\end{equation}
Proposition \ref{proposition_3} shows that, under only the information and exogeneity conditions in Assumption \ref{assumption_1}, the selection-bias function admits a mixture representation in terms of these latent propensity scores. Unlike the ordinary propensity-score approach, this representation does not require any monotonicity or single-index assumption.

\bigskip

\begin{prop} \label{proposition_3}
    Under Assumption \ref{assumption_1}, and for values of $\boldsymbol{x}_{t}$ such that $\overline{P}_j\left( \boldsymbol{x}_{t} \right) > 0$, the selection-bias function 
    $\lambda_j(\boldsymbol{x}_{t})$ admits the following mixture representation:
    \begin{equation}
        \lambda_{j}(\boldsymbol{x}_{t}) 
        \; = \;
        \displaystyle
        \int 
        \left[ 
            \frac{P_{j}\left(  \boldsymbol{x}_{t}, \boldsymbol{\kappa} \right)}{\overline{P}_j\left( \boldsymbol{x}_{t} \right)}
        \right]
        \, \mu_{j}(\boldsymbol{\kappa}) \,
        dF_{\kappa}(\boldsymbol{\kappa}),
    \label{eq_prop_3_mixture}
    \end{equation}
    with $\mu_{j}(\boldsymbol{\kappa}) \equiv \mathbb{E}( \xi_{jt} \mid \boldsymbol{\kappa}_{t} = \boldsymbol{\kappa})$. $\qquad \blacksquare$
\end{prop}

\noindent \textit{Proof:} See Online Appendix~\ref{app:proof_proposition_3}. 

\bigskip

\noindent Proposition~\ref{proposition_3} provides the first building block of our identification strategy by expressing the selection-bias function as a mixture involving latent propensity scores. The next section develops two routes from this representation to identification of the selection-bias functions and demand parameters.

\section{Identification and estimation \label{sec:identification}}

The researcher observes $T$ local markets, each with the same $J$ potential entrants. In every market $t$, the researcher observes the exogenous variables $\boldsymbol{x}_{t}$ and firms' entry decisions $\boldsymbol{a}_{t}\in\{0,1\}^{J}$; for every active firm $j$, the researcher also observes $p_{jt}$ and $s_{jt}$.

Let the vector of demand parameters be $\boldsymbol{\theta} \equiv (\alpha, \boldsymbol{\beta}^{\prime}, \boldsymbol{\sigma}^{\prime})^{\prime}$. We adopt a sequential identification strategy for $\boldsymbol{\theta}$. In the first step, we use the joint distribution of firms' entry decisions conditional on $\boldsymbol{x}_{t}$ to identify the probability objects required by the two identification routes described below. In the second step, we use the selected-sample demand equations to identify the demand parameters $\boldsymbol{\theta}$, together with the remaining components of the selection-bias functions.

The econometric model consists of three sets of objects: (i) the selected-sample demand equations in~\eqref{regression equation with selection term}; (ii) the selection-bias representation in~\eqref{eq_prop_3_mixture}; and (iii) the equilibrium-implied joint distribution of firms’ entry decisions conditional on $\boldsymbol{x}_{t}$. Under Assumption~\ref{assumption_1}, for every $\boldsymbol{a}\in\{0,1\}^{J}$, its conditional probability mass function has the mixture representation
\begin{equation}
    \Pr \left(\boldsymbol{a}_{t}=\boldsymbol{a}
    \mid \boldsymbol{x}_{t} \right)
    \; = \;
    \displaystyle
    \int 
    \Bigg(
        \displaystyle \prod_{j=1}^{J}
        P_{j}(\boldsymbol{x}_{t}, \boldsymbol{\kappa}) ^{a_{j}} \;
        \left[1-P_{j}(\boldsymbol{x}_{t}, \boldsymbol{\kappa}) \right]^{1-a_{j}}
    \Bigg) 
    \, dF_{\kappa}(\boldsymbol{\kappa}).
\end{equation}
Conditional on $(\boldsymbol{x}_{t}, \boldsymbol{\kappa}_{t})$, firms’ entry decisions depend only on their private shocks, which are assumed to be independent across firms. As a result, entry decisions are conditionally independent given $(\boldsymbol{x}_{t}, \boldsymbol{\kappa}_t)$, and by Assumption~\ref{assumption_1}[d] the conditional mixing distribution is the marginal distribution $F_{\kappa}$ and does not depend on $\boldsymbol{x}_{t}$. All residual dependence across firms’ entry decisions---beyond what is explained by observables---is therefore driven by the unobservables in $\boldsymbol{\kappa}_{t}$. This structure provides a key source of identifying variation. The joint distribution of $\boldsymbol{a}_{t}$ conditional on $\boldsymbol{x}_{t}$, and in particular cross-sectional dependence in entry decisions, therefore contains information about how the latent market state shifts firms' entry probabilities. The additional assumptions introduced below determine how this information is mapped into the selection-bias function.

We consider two complementary routes to an identified representation of the selection-bias function, each based on assumptions that may suit different economic applications. The first route---developed in Online Appendix~\ref{app: cont kappa}---assumes that the latent market state coincides with the vector of demand shocks, $\boldsymbol{\kappa}_{t}=\boldsymbol{\xi}_{t}$, and that $\boldsymbol{\xi}_{t}$ is jointly normal. Under additional index, exclusion, and smoothness restrictions, a multivariate version of Stein's lemma then delivers the closed form
\begin{equation}
    \lambda_{j}(\boldsymbol{x}_{t})
    \; = \;
    \sum_{k=1}^{J} \varrho_{jk} \,
    \frac{\partial \log \overline{P}_{j}(\boldsymbol{x}_{t})}{\partial x_{k1t}},
\label{eq:cont_kappa_lambda}
\end{equation}
where $x_{k1t}$ is a continuous regressor entering product $k$'s latent entry index with coefficient $\tau_{k1}$, $\Sigma_{jk}\equiv\operatorname{Cov}(\xi_{jt},\xi_{kt})$, and $\varrho_{jk}\equiv\Sigma_{jk}/\tau_{k1}$ (for details, see Proposition~\ref{proposition_normal} in Online Appendix~\ref{app: cont kappa}). The derivative regressors entering the selection-bias function can be constructed from a first-step estimate of $\overline{P}_{j}(\boldsymbol{x}_{t})$, and their coefficients are estimated jointly with the demand parameters in the second step. The second route, developed in the remainder of this section, assumes instead an exact finite-mixture representation of the joint entry distribution and the selection-bias functions.

\subsection{Two-step finite-mixture approach\label{sec:identification_finite_mixture}}

The support $\mathcal{K}$ of the latent market state $\boldsymbol{\kappa}_{t}$ need not be finite or discrete. For example, in our differentiated-products setting, $\boldsymbol{\kappa}_{t}$ may collect the product-specific demand and variable-cost unobservables, $\boldsymbol{\kappa}_{t} = (\xi_{jt}, \omega_{jt} : j \in \mathcal{J}) \in \mathbb{R}^{2J}$.

Equation \eqref{eq_prop_3_mixture} shows that the selection-bias function depends on the latent market state only through three objects: the latent propensity score $P_j(\boldsymbol{x}_{t}, \boldsymbol{\kappa})$, the marginal distribution $F_{\kappa}$, and the conditional mean function $\mu_j(\boldsymbol{\kappa})$. However, without additional structure, $\lambda_j(\boldsymbol{x}_{t})$ remains an unrestricted nuisance function of the observed state $\boldsymbol{x}_{t}$. We therefore need further dimensionality-reducing restrictions under which the selection-bias functions can be represented as linear combinations of functions recoverable from the joint distribution of entry decisions, with the remaining coefficients identified jointly with the demand parameters.

For an integer $L \geq 1$, let $\{ \mathcal{K}_1, \mathcal{K}_2, \dots, \mathcal{K}_L \}$ denote a partition of $\mathcal{K}$ into positive-probability cells. Each element of the partition can be interpreted as a \textit{latent market type}. For each latent market type $\ell = 1, 2, \dots, L$, define its probability
\begin{equation}
    \widetilde{f}_{\ell} \; \equiv \;
    \Pr \left( 
        \boldsymbol{\kappa}_{t} \in \mathcal{K}_{\ell}
    \right)
    \; = \;
    F_{\kappa}(\mathcal{K}_{\ell}).
\label{eq:def_type_probability}
\end{equation}
Define the type-specific propensity score
\begin{equation}
    \widetilde{P}_{j,\ell}(\boldsymbol{x}_{t}) \; \equiv \;
    \displaystyle
    \frac{1}{\widetilde{f}_{\ell}}
    \int_{\mathcal{K}_{\ell}}
    P_j(\boldsymbol{x}_{t}, \boldsymbol{\kappa}) \,
    dF_{\kappa}(\boldsymbol{\kappa}),
\end{equation}
as well as the type-specific average of the function
$\mu_j(\boldsymbol{\kappa})$:
\begin{equation}
    \widetilde{\mu}_{j,\ell}  \; \equiv \; \mathbb{E}( \xi_{jt} \mid \boldsymbol{\kappa}_{t} \in \mathcal{K}_\ell ) \; = \;
    \displaystyle
    \frac{1}{\widetilde{f}_{\ell}}
    \int_{\mathcal{K}_{\ell}}
    \mu_j(\boldsymbol{\kappa}) \,
    dF_{\kappa}(\boldsymbol{\kappa}).
\label{eq:def_type_mean}
\end{equation}
For every $\boldsymbol{a}=(a_j:j\in\mathcal{J})\in\{0,1\}^{J}$, these objects define the $L$-type finite-mixture conditional probability mass function
\begin{equation}
    \widetilde{\Pr}^{L}(\boldsymbol{a}_{t}=\boldsymbol{a}\mid\boldsymbol{x}_{t})
    \; \equiv \;
    \sum_{\ell=1}^{L}
    \widetilde{f}_{\ell} \,
    \Bigg(
        \prod_{j=1}^J
        \widetilde{P}_{j,\ell}(\boldsymbol{x}_{t})^{a_j}
        \left[1-\widetilde{P}_{j,\ell}(\boldsymbol{x}_{t})\right]^{1-a_j}
    \Bigg).
\label{eq:finite_mixture_entry_distribution}
\end{equation}
For every such partition, the law of total probability gives
\begin{equation}
    \overline{P}_{j}(\boldsymbol{x}_{t})
    \; = \;
    \sum_{\ell=1}^{L}
    \widetilde{P}_{j,\ell}(\boldsymbol{x}_{t}) \, \widetilde{f}_{\ell}.
\end{equation}
The same finite-mixture objects define the $L$-type selection-bias function
\begin{equation}
    \widetilde{\lambda}^{L}_{j}(\boldsymbol{x}_{t})
    \; \equiv \;
    \displaystyle
    \sum_{\ell=1}^{L}
    \left[
        \frac{\widetilde{P}_{j,\ell}(\boldsymbol{x}_{t})}
        {\overline{P}_{j}(\boldsymbol{x}_{t})}
    \right]
    \widetilde{\mu}_{j,\ell} \,
    \widetilde{f}_{\ell}.
\label{eq:finite_mixture_lambda}
\end{equation}
Expression \eqref{eq:finite_mixture_lambda} is the finite-mixture analog of the mixture representation in Proposition \ref{proposition_3}. For a given partition, the finite-mixture selection-bias function is summarized by $\{ \widetilde{f}_{\ell} \}_{\ell=1}^{L}$, $\{ \widetilde{P}_{j,\ell}(\boldsymbol{x}_{t})\}_{\ell=1}^{L}$, and
$\{\widetilde{\mu}_{j,\ell}\}_{\ell=1}^{L}$.

For a generic partition, the finite-mixture distribution in
\eqref{eq:finite_mixture_entry_distribution} and the finite-mixture
selection-bias functions in \eqref{eq:finite_mixture_lambda} need not
reproduce the true joint distribution of entry decisions and the selection-bias
functions. The next assumption imposes exact reproduction of both objects
together with an observable rank restriction.

\bigskip

\begin{assumption}[Exact finite-mixture representation and rank]
\label{assumption_2}
There exist an integer $L^*<\infty$ and a partition
$\{\mathcal{K}_1,\dots,\mathcal{K}_{L^*}\}$ of $\mathcal{K}$ into
positive-probability cells whose implied objects, defined in
\eqref{eq:def_type_probability}--\eqref{eq:def_type_mean}, satisfy, almost surely in
$\boldsymbol{x}_{t}$,
\begin{align*}
\Pr(\boldsymbol{a}_{t}=\boldsymbol{a}\mid\boldsymbol{x}_{t})
&=\widetilde{\Pr}^{L^*}(\boldsymbol{a}_{t}=\boldsymbol{a}\mid\boldsymbol{x}_{t})
&&\text{for every $\boldsymbol{a}\in\{0,1\}^J$},\\
\lambda_j(\boldsymbol{x}_{t})
&=\widetilde{\lambda}^{L^*}_j(\boldsymbol{x}_{t})
&&\text{for every $j$ with $\overline{P}_{j}(\boldsymbol{x}_{t})>0$.}
\end{align*}
Moreover, for some partition of $\mathcal{J}$ into two nonempty groups, the
matrix of their joint entry-profile probabilities conditional on
$\boldsymbol{x}_{t}$, with rows and columns indexed by the two groups' entry profiles, has
rank $L^*$ for almost every $\boldsymbol{x}_{t}$ in a set with positive
probability. Finally, the type probabilities are pairwise distinct.
\qquad $\blacksquare$
\end{assumption}

\noindent The exact representation and the rank condition identify
$L^*$ from the joint distribution of entry decisions. For the grouping in
Assumption~\ref{assumption_2}, the $L$-type conditional probability matrix
in \eqref{eq:finite_mixture_entry_distribution} is a sum of $L$ rank-one matrices and
therefore has rank at most $L$. The rank $L^*$ imposed by
Assumption~\ref{assumption_2} thus rules out every representation with fewer
than $L^*$ latent market types, while the exact representation in the
assumption supplies one with exactly $L^*$ types. Hence, $L^*$ is the
minimal number of latent market types that can reproduce the joint
distribution of entry decisions and is identified from that distribution.
Without the rank equality, the observable rank would provide only a lower
bound. This argument follows Proposition~4 in
\cite{kasahara2014non}, which establishes the rank equality generically when
$L^*$ does not exceed the number of entry profiles in either group.

The latent market state need not have discrete support: Assumption~\ref{assumption_2} requires only that a finite partition reproduce the joint distribution of entry decisions and the selection-bias functions, not that the resulting representation recover $F_{\kappa}$ itself or the latent propensity scores $P_j(\boldsymbol{x}_{t},\boldsymbol{\kappa})$ pointwise. It also
excludes latent heterogeneity driving selection that leaves no trace in
cross-firm dependence in entry decisions. If the heterogeneity relevant to
each firm's entry decision is firm-specific, independent across firms, and
affects only that firm's latent propensity score, entry decisions are
independent conditional on $\boldsymbol{x}_{t}$ and their joint distribution
admits a one-type representation. The entry data then cannot reveal this
heterogeneity even when it generates nonzero selection bias, so the
exact-representation requirement fails.

\subsection{First-step identification\label{sec:first_step_identification}}

The rank condition in Assumption~\ref{assumption_2} identifies $L^*$ but not
the type probabilities $\{\widetilde{f}_{\ell}\}_{\ell=1}^{L^*}$ and type-specific
propensity scores
$\{\widetilde{P}_{j,\ell}(\boldsymbol{x}_{t})\}_{\ell=1}^{L^*}$.
Conditional on $L^*$, Corollary~5 of \cite{allman2009identifiability}
establishes generic identification of these objects when
$J\geq2\lceil\log_2 L^*\rceil+1$. To describe the underlying rank condition,
for $L^*>1$, partition the products into three nonempty groups and, for each group and
almost every $\boldsymbol{x}_{t}$, form the matrix whose $\ell$-th row
contains the probabilities of that group's entry profiles for latent market
type $\ell$. Given the positive type probabilities in
Assumption~\ref{assumption_2}, Corollary~2 of
\cite{allman2009identifiability} identifies the type probabilities and the
probabilities of the three groups' entry profiles, up to a simultaneous
permutation of the latent market types, if the sum of the row Kruskal ranks of
these matrices is at least $2L^*+2$. Marginalizing these probabilities then
identifies each $\widetilde{P}_{j,\ell}(\boldsymbol{x}_{t})$.

The preceding identification result is pointwise in
$\boldsymbol{x}_{t}$. Assumption~\ref{assumption_1}[d] makes the type
probabilities common across values of $\boldsymbol{x}_{t}$, and their
pairwise distinctness in Assumption~\ref{assumption_2} permits the latent
market types to be matched across those values. Without loss of generality,
we label them so that
$\widetilde{f}_{1}>\cdots>\widetilde{f}_{L^*}>0$.
This labeling fixes the remaining global permutation, which merely relabels
the type-specific objects and leaves the finite-mixture distribution in
\eqref{eq:finite_mixture_entry_distribution} and the finite-mixture
selection-bias functions in \eqref{eq:finite_mixture_lambda} unchanged.

\subsection{Second-step identification}

Under Assumption~\ref{assumption_2} and the preceding first-step identification conditions, the selection-bias function is linear in the unknown type-specific averages $\{\widetilde{\mu}_{j,\ell}\}_{\ell=1}^{L^*}$. Their definition and the normalization $\mathbb{E}(\xi_{jt})=0$ imply $\sum_{\ell=1}^{L^*}\widetilde{\mu}_{j,\ell}\widetilde{f}_{\ell}=0$, so only $L^*-1$ coefficients are free. Taking type $L^*$ as the reference, so that $\widetilde{\mu}_{j,L^*}=-\sum_{\ell=1}^{L^*-1}(\widetilde{f}_{\ell}/\widetilde{f}_{L^*})\widetilde{\mu}_{j,\ell}$, the selected-sample demand equation becomes
\begin{equation}
\begin{aligned}
d_{jt}^{-1}\left(\boldsymbol{s}_{t}^{\boldsymbol{a}},\boldsymbol{\sigma}\right)
&=
\alpha p_{jt}
+\boldsymbol{x}_{jt}^{D \prime}\boldsymbol{\beta}
+\sum_{\ell=1}^{L^*-1}\widetilde{\mu}_{j,\ell}r_{j\ell t}
+\widetilde{\xi}_{jt},\\
r_{j\ell t}
&\equiv
\frac{
\widetilde{P}_{j,\ell}(\boldsymbol{x}_{t})
-\widetilde{P}_{j,L^*}(\boldsymbol{x}_{t})
}{
\overline{P}_{j}(\boldsymbol{x}_{t})
}
\,\widetilde{f}_{\ell},
\qquad \ell=1,\ldots,L^*-1.
\end{aligned}
\label{eq:second_step_reparam}
\end{equation}
Let $\widetilde{\boldsymbol\mu}_{j}\equiv
(\widetilde{\mu}_{j,1},\dots,\widetilde{\mu}_{j,L^*-1})^{\prime}$,
$\boldsymbol{r}_{jt}\equiv(r_{j1t},\dots,r_{j,L^*-1,t})^{\prime}$, and stack
$\widetilde{\boldsymbol\mu}\equiv
(\widetilde{\boldsymbol\mu}_{1}^{\prime},\dots,
\widetilde{\boldsymbol\mu}_{J}^{\prime})^{\prime}$. Define
$\boldsymbol\vartheta\equiv
(\boldsymbol\theta^{\prime},\widetilde{\boldsymbol\mu}^{\prime})^{\prime}$,
and let $\boldsymbol\vartheta_{0}$ denote its true value. Let
$\boldsymbol{R}_{jt}$ be the $J(L^*-1)$-vector whose $j$-th
$(L^*-1)$-dimensional block is $\boldsymbol{r}_{jt}$ and whose remaining
blocks are zero, so that
$\boldsymbol{R}_{jt}^{\prime}\widetilde{\boldsymbol\mu}
=\boldsymbol{r}_{jt}^{\prime}\widetilde{\boldsymbol\mu}_{j}$. When
$L^*=1$, all these selection-function vectors are interpreted as empty.\footnote{If $L^*=1$, Assumption~\ref{assumption_2} and the normalization $\mathbb{E}(\xi_{jt})=0$ imply $\lambda_j(\boldsymbol{x}_{t})=0$.}
For each $j$, let $\boldsymbol{z}_{jt}\in\mathbb{R}^{K_z}$, with
$K_z\geq\dim(\boldsymbol{\sigma})+1$, contain excluded instruments
constructed from $\boldsymbol{x}_{t}$, such as characteristics of potential
entrants other than $j$, and define
$\boldsymbol{w}_{jt}\equiv
(\boldsymbol{z}_{jt}^{\prime},\boldsymbol{x}_{jt}^{D \prime},
\boldsymbol{R}_{jt}^{\prime})^{\prime}$.
Up to an irrelevant positive normalization, the pooled selected-sample
moment function is
\begin{equation*}
\boldsymbol{m}(\boldsymbol\vartheta)\equiv \sum_{j=1}^{J}\Pr(a_{jt}=1)\,\mathbb{E}\Big(\boldsymbol{w}_{jt}\Big(d_{jt}^{-1}(\boldsymbol{s}_{t}^{\boldsymbol{a}},\boldsymbol{\sigma})-\alpha p_{jt}-\boldsymbol{x}_{jt}^{D \prime}\boldsymbol{\beta}-\boldsymbol{R}_{jt}^{\prime}\widetilde{\boldsymbol\mu}\Big)\,\Big|\,a_{jt}=1\Big).
\end{equation*}
By equation~\eqref{eq:second_step_reparam}, the residual inside
$\boldsymbol{m}(\boldsymbol\vartheta)$, evaluated at
$\boldsymbol\vartheta_{0}$, equals $\widetilde{\xi}_{jt}$. Because every component of
$\boldsymbol{w}_{jt}$ is a function of $\boldsymbol{x}_{t}$, iterated
expectations and
$\mathbb{E}(\widetilde{\xi}_{jt}\mid\boldsymbol{x}_{t},a_{jt}=1)=0$
establish $\boldsymbol{m}(\boldsymbol\vartheta_{0})=\boldsymbol{0}$.
The standard rank condition for local identification in
\cite{rothenberg_1971} then yields the following result.

\bigskip

\begin{prop}
\label{proposition_4}
Suppose that $\Pr(a_{jt}=1)>0$ for every $j$, $\boldsymbol\vartheta_{0}$
lies in the interior of the parameter space, the expectations defining
$\boldsymbol{m}$ are finite, $\boldsymbol{m}$ is continuously differentiable
in a neighborhood of $\boldsymbol\vartheta_{0}$, and
\[
\boldsymbol{M}(\boldsymbol\vartheta_{0})
\equiv
\left.
\frac{\partial\boldsymbol{m}(\boldsymbol\vartheta)}
{\partial\boldsymbol\vartheta^{\prime}}
\right|_{\boldsymbol\vartheta=\boldsymbol\vartheta_{0}}
\]
has full column rank. Then $\boldsymbol\vartheta_{0}$ is locally identified.
\qquad $\blacksquare$
\end{prop}

\noindent Proposition~\ref{proposition_4} completes the sequential identification argument. The first step identifies $L^*$, the type probabilities, and the type-specific propensity scores, and hence the selection-function regressors $\boldsymbol{r}_{jt}$. Assumption~\ref{assumption_2} guarantees that these regressors span the selection-bias functions. Conditional on these objects, the pooled demand moments locally identify the common demand parameters $\boldsymbol{\theta}$ jointly with the product-specific selection-function coefficients $\{\widetilde{\boldsymbol\mu}_{j}\}_{j=1}^{J}$. Thus, identification of $\boldsymbol{\theta}$ requires neither a specification of entry payoffs nor a pricing equation.

Importantly, as discussed after Proposition~\ref{proposition_1} in Section~\ref{sec:demand_model}, our selection correction conditions on $(\boldsymbol{x}_{t},a_{jt}=1)$, but not on the rivals' entry decisions $\boldsymbol{a}_{-jt}$. Admissible instruments $\boldsymbol{z}_{jt}$ can therefore be functions of the characteristics $\boldsymbol{x}_{t}$ of all potential entrants, but cannot generally depend on the rivals' actual entry decisions. Using such instruments would require either additional restrictions or the finer selection correction $\mathbb{E}\left(\xi_{jt}\mid\boldsymbol{x}_{t},a_{jt}=1,\boldsymbol{a}_{-jt}\right)$, complicating both identification and estimation.

\subsubsection{Identification of marginal costs and fixed entry costs\label{sec:identification_costs}}

Once demand parameters are identified and a form of competition is specified, price-cost margins can be recovered from firms’ profit maximization conditions. Combining these with observed prices yields realized marginal costs $mc_{jt}$.

Suppose that marginal cost, possibly after a known transformation (the logarithm in our application), is modeled as a function of observable product characteristics $\boldsymbol{x}_{jt}^{\mathrm{mc}}$ plus the additive variable-cost unobservable $\omega_{jt}$. Estimating this marginal-cost equation raises the same selection-bias issues as in demand estimation. In particular, the equation involves the selection-bias function $\lambda^{\text{mc}}_{j}(\boldsymbol{x}_{t}) = \mathbb{E}( \omega_{jt} \mid \boldsymbol{x}_{t}, a_{jt} = 1 )$. By Assumption~\ref{assumption_1}[c]--[d], the private entry shock $\boldsymbol{\eta}_{jt}$ is independent of $\boldsymbol{\omega}_{t}$ conditional on $(\boldsymbol{x}_{t},\boldsymbol{\kappa}_{t})$, and $\boldsymbol{x}_{t}$ is independent of the unobservables, so this selection-bias function has the same structure as in Proposition~\ref{proposition_3}:
\begin{equation}
    \lambda^{\text{mc}}_{j}(\boldsymbol{x}_{t}) 
    \; = \;
    \displaystyle
    \int 
    \left[ 
        \frac{P_{j}\left(  \boldsymbol{x}_{t}, \boldsymbol{\kappa} \right)}{\overline{P}_j\left( \boldsymbol{x}_{t} \right)}
    \right]
    \, \mu^{\text{mc}}_{j}(\boldsymbol{\kappa}) \,
    dF_{\kappa}(\boldsymbol{\kappa}),
\end{equation}
with $\mu^{\text{mc}}_{j}(\boldsymbol{\kappa}) \equiv \mathbb{E}( \omega_{jt} \mid \boldsymbol{\kappa}_{t} = \boldsymbol{\kappa} )$ and $\widetilde{\mu}^{\text{mc}}_{j,\ell}\equiv\mathbb{E}(\omega_{jt}\mid\boldsymbol{\kappa}_{t}\in\mathcal{K}_{\ell})$. The normalization $\mathbb{E}(\omega_{jt})=0$ implies $\sum_{\ell=1}^{L^*}\widetilde{\mu}^{\text{mc}}_{j,\ell}\widetilde{f}_{\ell}=0$. Under an extension of Assumption~\ref{assumption_2} requiring the same partition also to reproduce the marginal-cost selection-bias functions, the same reference-type reparameterization and finite-mixture identification strategy can be applied to recover the marginal-cost parameters and the corresponding selection-bias functions.

To recover fixed costs, we further associate each latent market type with the type-specific conditional means $\widetilde{\mu}_{j,\ell}$ and $\widetilde{\mu}^{\text{mc}}_{j,\ell}$ of the demand and variable-cost unobservables. Under this additional type-level restriction and an identified variable-cost specification, the demand and marginal-cost parameters and these type-specific conditional means determine equilibrium variable profits for any given $\boldsymbol{x}_{t}$, latent market type $\ell = 1,\ldots,L^*$, and counterfactual entry configuration $\boldsymbol{a} \in \{0,1\}^{J}$. In particular, for a given $\boldsymbol{x}_{t}$ and latent type $\ell$, we obtain equilibrium price-cost margins, market shares, and variable profits by setting $\xi_{jt} = \widetilde{\mu}_{j,\ell}$ and $\omega_{jt} = \widetilde{\mu}^{\text{mc}}_{j,\ell}$ for all $j \in \mathcal{J}$.

Let $VP_{j \ell}(a_j=1,\boldsymbol{a}_{-j}, \boldsymbol{x}_{t})$ denote the equilibrium variable profit of firm $j$ evaluated at $\ell \in \{1,\ldots,L^*\}$ when firm $j$ enters and its rivals' entry configuration is $\boldsymbol{a}_{-j} \in \{0,1\}^{J-1}$. Using the type-specific propensity scores and conditional independence of rival entry decisions within each latent type, expected variable profits at the time of entry decisions can be written as:
\begin{equation}
    VP^{P}_{j \ell}(\boldsymbol{x}_{t}) 
    \; = \;
    \displaystyle 
    \sum_{\boldsymbol{a}_{-j} \in \{0,1\}^{J-1}}
    VP_{j \ell}(a_j=1, \boldsymbol{a}_{-j},  \boldsymbol{x}_{t}) \,
    \prod_{i \neq j} 
    \widetilde{P}_{i,\ell}(\boldsymbol{x}_{t})^{a_{i}} \,
    \left[ 1 - \widetilde{P}_{i,\ell}(\boldsymbol{x}_{t}) \right]^{1-a_{i}}.
\end{equation}
So far, the private information entering fixed costs has been allowed to be the general vector $\boldsymbol{\eta}_{jt}$. To recover fixed costs from the type-specific propensity scores, we now assume that this private information can be summarized by a scalar $\eta_{jt}$ entering additively in fixed costs, with mean zero and known strictly increasing CDF $F_{\eta}$. Under the exclusion restriction that, conditional on $\ell$, firm $j$'s fixed cost depends on $\boldsymbol{x}_{t}$ only through $\boldsymbol{x}_{jt}^{\mathrm{fc}}$, suppose fixed costs take the form $fc_{j \ell}(\boldsymbol{x}_{jt}^{\mathrm{fc}}) + \sigma_{\eta_j}\eta_{jt}$, where $\sigma_{\eta_j}>0$. Then, the type-specific propensity scores satisfy $\widetilde{P}_{j,\ell}(\boldsymbol{x}_{t}) = F_{\eta}\left( \left[ VP^{P}_{j \ell}(\boldsymbol{x}_{t}) - fc_{j \ell} (\boldsymbol{x}_{jt}^{\mathrm{fc}}) \right] / \sigma_{\eta_j} \right)$. For $0<\widetilde{P}_{j,\ell}(\boldsymbol{x}_{t})<1$, this relationship can be inverted to obtain:
\begin{equation}
    F_{\eta}^{-1} \left( \widetilde{P}_{j,\ell}(\boldsymbol{x}_{t}) \right) 
    \; = \;
    \displaystyle 
    \frac{1}{\sigma_{\eta_j}} 
    VP^{P}_{j \ell}(\boldsymbol{x}_{t}) 
    \; - \;
    \frac{1}{\sigma_{\eta_j}} fc_{j \ell} (\boldsymbol{x}_{jt}^{\mathrm{fc}}).
\end{equation}
Provided that functions of the components of $\boldsymbol{x}_{t}$ excluded from $\boldsymbol{x}_{jt}^{\mathrm{fc}}$ shift $VP^{P}_{j\ell}(\boldsymbol{x}_{t})$ sufficiently while not entering $fc_{j\ell}(\boldsymbol{x}_{jt}^{\mathrm{fc}})$, this equation identifies the scale parameter $\sigma_{\eta_j}$ and the fixed-cost function $fc_{j \ell}(\boldsymbol{x}_{jt}^{\mathrm{fc}})$.

\subsection{Estimation \label{sec:estimation}}

We implement the finite-mixture approach with a two-step estimator for each candidate number of latent market types $L$. The first step estimates the type probabilities and type-specific propensity scores by semiparametric sieve Maximum Likelihood Estimation (MLE); the second constructs the selection-function regressors and estimates the demand and selection-function parameters by Generalized Method of Moments (GMM). We then discuss selection among candidate values of $L$. The continuous-$\boldsymbol{\kappa}_{t}$ approach has the same two-step structure: its derivative regressors are constructed from a first-step estimate of the ordinary propensity score and enter the same second-step regression, as detailed in Online Appendix~\ref{app: cont kappa}.

\subsubsection{First step: Type probabilities and type-specific propensity scores}

For a candidate number of latent market types $L$, we approximate each type-specific propensity score using a logistic sieve \citep{hirano2003efficient,chen_2007}. Let $\boldsymbol{b}_{t}\equiv\left(b_{1}(\boldsymbol{x}_{t}),\ldots,b_{N_X}(\boldsymbol{x}_{t})\right)^{\prime}$ collect $N_X$ basis functions. For $j=1,\ldots,J$ and $\ell=1,\ldots,L$, we specify
\begin{equation}
    \widetilde{P}_{j,\ell}(\boldsymbol{x}_{t})
    =
    \Lambda\!\left(\boldsymbol{b}_{t}^{\prime}\boldsymbol{\gamma}_{j,\ell}\right),
\label{eq_sp_logit}
\end{equation}
where $\Lambda$ is the logistic CDF and $\boldsymbol{\gamma}_{j,\ell}\in\mathbb{R}^{N_X}$.

Let $\boldsymbol{\gamma}$ collect the $JLN_X$ sieve coefficients and $\widetilde{\boldsymbol{f}}\equiv(\widetilde{f}_1,\ldots,\widetilde{f}_L)^{\prime}$ the type probabilities, with $\widetilde{f}_\ell>0$ and $\sum_{\ell=1}^L\widetilde{f}_\ell=1$. The finite-mixture distribution in equation~\eqref{eq:finite_mixture_entry_distribution} implies the sample log-likelihood
\begin{equation}
    \ln\mathcal{L}_{\text{1st}}(\widetilde{\boldsymbol{f}},\boldsymbol{\gamma})
    =
    \sum_{t=1}^{T}
    \ln\!\left[
        \sum_{\ell=1}^{L}\widetilde{f}_{\ell}
        \prod_{j=1}^{J}
        \widetilde{P}_{j,\ell}(\boldsymbol{x}_{t})^{a_{jt}}
        \left\{1-\widetilde{P}_{j,\ell}(\boldsymbol{x}_{t})\right\}^{1-a_{jt}}
    \right].
\label{eq:finite_mixture_logit}    
\end{equation}
We maximize \eqref{eq:finite_mixture_logit} using the Expectation-Maximization (EM) algorithm \citep{pilla_lindsay_2001}.\footnote{Recent applications of EM to nonparametric discrete-choice mixtures include \cite{bunting_2022,bunting_diegert_2022,hu_xin_2022,williams_2020}.} For fixed $L$ and $N_X$, the first-step model is finite-dimensional, and under correct specification, interiority, local identification, and nonsingularity, its MLE has the usual $\sqrt{T}$ asymptotics. With continuous covariates and $N_X\to\infty$, the sieve estimates generally converge more slowly and require the usual approximation, rate, and smoothness conditions.

\subsubsection{Second step: Demand parameters and selection-bias functions}

For a candidate $L$, interpret equation~\eqref{eq:second_step_reparam} and its associated selection-function vectors with $L$ in place of $L^*$. Given the first-step estimates $(\widehat{\widetilde{\boldsymbol{f}}},\widehat{\boldsymbol{\gamma}})$, we construct the selection-function regressors
\begin{equation}\label{eq:r_hat}
    \widehat{r}_{j\ell t}
    =
    \frac{
        \widehat{\widetilde{P}}_{j,\ell}(\boldsymbol{x}_{t})
        -
        \widehat{\widetilde{P}}_{j,L}(\boldsymbol{x}_{t})
    }
    {\widehat{\overline{P}}_{j}(\boldsymbol{x}_{t})}
    \widehat{\widetilde{f}}_{\ell},
    \qquad \ell=1,\ldots,L-1,
\end{equation}
where $\widehat{\overline{P}}_{j}(\boldsymbol{x}_{t})\equiv\sum_{\ell=1}^{L}\widehat{\widetilde{P}}_{j,\ell}(\boldsymbol{x}_{t})\widehat{\widetilde{f}}_{\ell}$ is the estimated ordinary propensity score. We then apply GMM to equation~\eqref{eq:second_step_reparam} to jointly estimate the demand parameters $\boldsymbol{\theta}$ and the product-specific selection-function coefficients $\{\widetilde{\boldsymbol\mu}_{j}\}_{j=1}^{J}$.
With $L$ and $N_X$ fixed as above, standard smoothness, moment, and rank conditions yield a $\sqrt{T}$-asymptotically linear joint two-step estimator for fixed $J$, with markets as the independent sampling units \citep{newey_2009}. Inference must account for first-step estimation error and its covariance with the second-step moments. The linearized common-weight bootstrap in Online Appendix~\ref{app:bootstrap} does so without re-estimating the first-step EM algorithm in every replication.

\subsubsection{Selecting the number of latent types \label{sec:estimating_L}}

Although Assumption~\ref{assumption_2} identifies $L^*$ in the population, implementation in finite samples requires choosing $L$ from a set of candidate values. A first-step likelihood-based Bayesian information criterion (BIC) and the precision of the entry estimates indicate which candidates are adequately supported by the joint distribution of entry decisions. Because $L$ determines the selection functions entering the selected-sample demand equation and our main objective is to estimate the structural demand parameters, we choose among the supported values of $L$ using a second-step residual-based BIC-style criterion and Hausman-type comparisons of the demand parameters.

\medskip

\noindent \textit{\textbf{First-step BIC.}} For given $L$, the likelihood-based BIC for the entry model is
\begin{equation}\label{eq:BIC_1}
	\mathrm{BIC}_{\text{1st}}(L)
	=
	-2\ln\mathcal{L}_{\text{1st}}
	(\widehat{\widetilde{\boldsymbol{f}}},
	\widehat{\boldsymbol{\gamma}})
	+d_{\text{1st}}(L)\ln(T),
\end{equation}
where $d_{\text{1st}}(L)$ is the number of estimated first-step parameters. If all sieve coefficients are type-specific, $d_{\text{1st}}(L)=JLN_X+L-1$; coefficients restricted to be common across latent types are counted only once. Smaller values of $\mathrm{BIC}_{\text{1st}}(L)$ favor a candidate $L$. We also examine the estimated type probabilities and the precision of the first-step estimates: a near-empty type or a nearly singular information matrix indicates that $L$ is too large for the available entry variation.

\medskip

\noindent \textit{\textbf{Second-step BIC-style criterion.}} Among the values of $L$ receiving adequate first-step support, our main information criterion is based on the fit of the selected-sample demand equation. Let $N_{\mathrm{sel}}$ denote the number of demand observations used in the second step. We define
\begin{equation}\label{eq:BIC_2}
	\mathrm{BIC}_{\text{2nd}}(L)
	=
	N_{\mathrm{sel}}\ln\widehat{\sigma}^2_{\xi}(L)
	+(L-1)J\ln(N_{\mathrm{sel}}),
\end{equation}
where $\widehat{\sigma}^2_{\xi}(L)\equiv N_{\mathrm{sel}}^{-1}\sum_{(j,t):a_{jt}=1}\widehat{\widetilde{\xi}}_{jt}(L)^2$ is the selected-sample residual mean square from equation~\eqref{eq:second_step_reparam}. The penalty counts the $J(L-1)$ selection-function coefficients (the number of demand parameters is constant across $L$). Smaller values of $\mathrm{BIC}_{\text{2nd}}(L)$ favor a candidate $L$.

\medskip

\noindent \textit{\textbf{Hausman-type tests on demand parameters.}} Information criteria do not answer the essential question of whether a larger $L$ yields demand estimates that are different from those at smaller $L$ or from estimates that ignore selection. We therefore complement them with Hausman-type tests on the demand parameters.

Because the selection-function coefficients vary in dimension with $L$, comparisons are based on the demand-parameter vector $\boldsymbol{\theta}$, which is common to all specifications. Consider two second-step specifications $A$ and $B$ that differ in how selection is handled. Under the null that both recover the same $\boldsymbol{\theta}$, we use the Hausman-type Wald statistic
\begin{equation}\label{eq:hausman_general}
	\mathcal{H}(A,B)
	=
	\big(\widehat{\boldsymbol{\theta}}^{B}
	-\widehat{\boldsymbol{\theta}}^{A}\big)^{\prime}
	\widehat{\boldsymbol{V}}_{AB}^{-1}
	\big(\widehat{\boldsymbol{\theta}}^{B}
	-\widehat{\boldsymbol{\theta}}^{A}\big),
\end{equation}
where $\widehat{\boldsymbol{V}}_{AB}$ estimates the variance-covariance matrix of $\widehat{\boldsymbol{\theta}}^{B}-\widehat{\boldsymbol{\theta}}^{A}$ and is computed using the linearized common-weight bootstrap of Online Appendix~\ref{app:bootstrap}. Under joint asymptotic normality and nonsingularity of this matrix, $\mathcal{H}(A,B)\xrightarrow{d}\chi^2(\dim(\boldsymbol{\theta}))$. The same construction applies to any $q$-dimensional common subvector of $\boldsymbol{\theta}$, with limiting distribution $\chi^2(q)$.

\section{Empirical application \label{sec:application}}

\subsection{Data and descriptive statistics}

We apply our method to estimate demand in the US airline industry. The challenge of endogenous product entry in demand estimation in this industry has recently been explored by \cite{ciliberto2021market} and \cite{li2022repositioning}.

\medskip 

\noindent \textit{\textbf{Data sources.}} Our sample consists of the publicly available DB1B and T100 datasets from the US Department of Transportation, complemented with population data from the Census Bureau's American Community Survey (ACS). Specifically, we use quarterly data spanning 2012-Q1 to 2013-Q4 for routes between the airports at the 100 largest Metropolitan Statistical Areas (MSA) in the United States. These account for 142 airports, as several of these MSAs are served by more than one airport.

\medskip 

\noindent \textit{\textbf{Airlines.}} The airlines included in our analysis are American (AA), Delta (DL), United (UA), US Airways (US), Southwest (WN), a combined group of Low-Cost Carriers (LCC), and a combined group of the remaining carriers (Others).\footnote{Following \cite{ciliberto2021market}, the list of airlines included in the group LCC is: Alaska, JetBlue, Frontier, Allegiant, Spirit, Sun Country, and Virgin. The group Others collects other smaller carriers.} Given the large number of carriers in Others, we do not consider this combined group as a player in the entry game. In the notation of our model, $j$ always indexes an airline.

\medskip

\noindent \textit{\textbf{Markets in the demand model.}} In the demand model, a market $t$ is defined as a \textit{directional airport pair} in a given quarter. For example, New York La Guardia (LGA) $\to$ Chicago O'Hare (ORD) in 2012-Q1 and ORD $\to$ LGA in 2012-Q1 are two distinct demand markets. In each demand market $t$, consumers choose among the airlines offering non-stop flights on that directional route (up to seven airlines, plus the outside option). Thus, $s_{jt}$ denotes the market share of airline $j$ in demand market $t$.

\medskip

\noindent \textit{\textbf{Markets in the entry model.}} In the entry model, with some abuse of notation, a market $t$ is defined as a \textit{non-directional} airport pair in a given quarter, where, for example, ORD to LGA in 2012-Q1 is the same market as LGA to ORD in 2012-Q1. Each non-directional entry market thus corresponds to two directional demand markets. There are potentially 10,011 non-directional airport pairs among the 142 airports, i.e., $142\times141/2$. However, many of these airport pairs have not had an incumbent airline with non-stop flights for several decades. These are typically airport pairs that are geographically too close or in smaller metropolitan areas (MSAs). In our sample, we consider a non-directional airport pair if it was served with non-stop flights either in at least 50 quarters between 1994 and 2018, or in at least one quarter of our 2012--2013 sample period. This results in 2,652 non-directional airport pairs and 20,859 entry markets.\footnote{Given 2,652 non-directional airport pairs and eight quarters, there are at most $2,652 \times 8 = 21,216$ potential entry markets. We discard from the analysis the 357 in which the airport pair had no non-stop flights in that quarter and was not served in at least 50 quarters between 1994 and 2018, leaving 20,859 markets.}

\medskip

\noindent \textit{\textbf{Potential entrants.}} An airline is a \textit{potential entrant} in a market if it operates non-stop flights out of both airports in that quarter, while it is an \textit{entrant} if it operates non-stop flights between the two airports in that quarter. Of the 20,859 markets, we exclude 1,315 for which none of the modeled airlines is a potential entrant, leaving 19,544 for the estimation of the entry model.

\medskip

\noindent \textit{\textbf{Market size and distance.}} Following the empirical literature on the airline industry, we define market size as the geometric mean of the populations in the MSAs of the two airports and market distance as the geodesic distance between the two airports.

\medskip

\noindent \textit{\textbf{Observable variables $\boldsymbol{x}_{t}$ for entry model.}} The vector of exogenous observable variables at the airline-market level includes: market size, market distance, the combined hub size at the origin and destination airports of the airline and of each of its six competitors, airline-by-quarter indicators, and airport indicators. We define the hub size of an airline at an airport as the number of non-stop routes that the airline operates from that airport, and its combined hub size for a given airport pair as the sum of its hub sizes at the two endpoints. 

Table \ref{tab_distribution_entrants} presents the distribution of the number of entrants and averages of the market size and distance. Notably, in more than $35\%$ of these markets, there are no airlines providing non-stop flights. Among the markets served by non-stop flights, more than $90\%$ are monopolies or duopolies. Furthermore, there is a strong positive correlation between the number of incumbents, market size, and distance.

{\setlength{\tabcolsep}{3pt}
\begin{table}[ht]
\caption{Distribution of Markets by Number of Entrants
\label{tab_distribution_entrants}}
\centering
\scalebox{0.90}{%
\begin{tabular}{r|ccc}
    \hline \hline
    & Frequency & Avg.\ market size & Avg.\ market distance \\
    Number of airlines
    & \# markets (\%) & in millions of people
    & in miles \\
    \hline
    0 airlines & 6,878 (35.19\%) & 2.66 & 638 \\
    1 airline & 8,770 (44.87\%) & 3.24 & 914 \\
    2 airlines & 2,708 (13.86\%) & 4.24 & 952 \\
    3 airlines & 871 (4.46\%) & 5.25 & 1109 \\
    4 airlines & 235 (1.20\%) & 5.24 & 1153 \\
    5 airlines & 72 (0.37\%) & 8.16 & 1255 \\
    $\geq$ 6 airlines & 10 (0.05\%) & 6.84 & 320 \\
    \hline
    Total & 19,544 (100.00\%) & 3.31 &    835 \\
    \hline \hline
\end{tabular}
}
\end{table}
}

Table~\ref{tab_freq_entry} reports each airline's entry frequency and the average size and distance of the markets it enters. Entry frequencies range from $24.9\%$ for WN to $9.6\%$ for AA; the correlations of entry with market size and distance also differ substantially across airlines. WN enters markets that do not differ significantly in size from those it does not enter ($3.34$ versus $3.30$ million people), whereas AA enters much larger markets ($5.12$ versus $3.12$ million people; averages for non-entered markets are not reported in the table). Entry also differs by distance: DL and US typically enter shorter-haul markets, averaging $874$ and $882$ miles, whereas markets served by LCC average $1{,}140$ miles.

\begin{table}[ht]
\caption{Entry Frequency by Airline \label{tab_freq_entry}}
\centering
\scalebox{0.90}{%
\begin{tabular}{r|ccc}
    \hline \hline
    & Frequency & Avg.\ market size & Avg.\ market distance \\
    Airline
    & \# markets (\%) & in millions of people
    & in miles \\
    \hline
    WN & 4,863 (24.9\%) & 3.34 & 987 \\
    DL & 3,332 (17.0\%) & 3.93 & 874 \\
    UA & 3,292 (16.8\%) & 4.36 & 962 \\
    LCC & 2,744 (14.0\%) & 4.36 & 1140 \\
    US & 2,059 (10.5\%) & 3.94 & 882 \\
    AA & 1,869 (9.6\%) & 5.12 & 964 \\
    \hline \hline
\end{tabular}
}
\end{table}

\subsection{First step: Estimation of the entry model\label{sec:empirical_entry_estim}}

For the entry decisions, we consider the semiparametric finite-mixture model in equation \eqref{eq:finite_mixture_logit}. We present results for $\boldsymbol{b}_{t}$ specified as a second-degree polynomial in $\boldsymbol{x}_{t}$.\footnote{The nine continuous regressors in $\boldsymbol{x}_{t}$---market size, market distance, and the combined origin-and-destination hub sizes of the airline and of its six competitors---are each rescaled to $[-1,1]$, and $\boldsymbol{b}_{t}$ is the full Chebyshev tensor product of total degree two: the nine linear terms, their nine squares, and all 36 pairwise interactions, i.e., 54 polynomial terms per airline.} Table \ref{tab_app_entry_model} presents the estimates of a probit model (with $L=1$) and of three nested specifications of the mixture logit model (with $L=2,3,4$).\footnote{As will become clear below, for $L=1$, we estimate a probit rather than a logit to construct selection-correction terms in the spirit of Heckman and for the case of continuous $\boldsymbol{\kappa}_t$ described in Online Appendix~\ref{app: cont kappa}.} For each specification, Table~\ref{tab_app_entry_model} reports the estimated probabilities of the latent market types and, for each type, the average elasticity of an airline's entry probability with respect to: market size, market distance, the airline's own hub size, and its competitors' hub size, where the latter is an average of the elasticities with respect to each of the six competitors' combined hub sizes.

{
\let\savedendtable\endtable
\renewcommand{\endtable}{
\par\vspace{0.5em}
\centering
\begin{minipage}{0.95\linewidth}
\footnotesize
\textit{Notes:} The table reports estimates of a probit entry model without latent market types ($L=1$) and of finite-mixture logit entry models with $L=2$, $L=3$, and $L=4$ latent market types. In all specifications, the market observables $\boldsymbol{x}_{t}$ include market size, market distance, the combined hub size at the origin and destination airports of the airline and of each of its six competitors, airline $\times$ quarter indicators, and airport indicators. For each latent type $\ell$, the table reports the estimated type probability $\widetilde{f}_{\ell}$ and the average elasticity of an airline's entry probability with respect to each continuous regressor in $\boldsymbol{x}_{t}$, where the elasticity denoted by ``Competitor hub size'' represent the average across the airline's six competitors' combined hub sizes (six distinct continuous regressors in $\boldsymbol{x}_{t}$). Standard errors are in parentheses. All specifications are estimated by maximum likelihood using the EM algorithm. BIC$_{\text{1st}}$ is the first-step criterion defined in equation~\eqref{eq:BIC_1}.
\end{minipage}
\par
\savedendtable}
\begin{table}[ht]
\caption{Estimation of Market Entry Model
\label{tab_app_entry_model}}
\centering\resizebox{0.9\textwidth}{!}{
    \begin{tabular}{l|cccc}
    \hline \hline
     & {Probit} & {Mixture Logit} & {Mixture Logit} & {Mixture Logit} \\
     & {$L=1$} & {$L=2$} & {$L=3$} & {$L=4$} \\
    \hline
    $(\widetilde{f}_1, \ldots, \widetilde{f}_L)$ & & $0.55,\,0.45$ & $0.38,\,0.34,\,0.28$ & $0.29,\,0.27,\,0.23,\,0.21$ \\
     & & $(0.01,\,0.01)$ & $(0.01,\,0.01,\,0.01)$ & \\
    & & & & \\
    Elasticities: & & & & \\
    \quad Market size & $1.63$ & $1.95,\,0.44$ & $10.91,\,4.71,\,-5.73$ & $1.56,\,19.99,\,-7.67,\,8.85$ \\
     & $(0.35)$ & $(0.38,\,0.68)$ & $(8.98,\,1.36,\,3.67)$ & \\
    & & & & \\
    \quad Distance & $-2.10$ & $-1.16,\,-17.25$ & $-11.95,\,-25.42,\,-15.30$ & $-42.91,\,-38.83,\,-13.28,\,-1.61$ \\
     & $(0.24)$ & $(0.23,\,1.50)$ & $(12.76,\,2.03,\,4.21)$ & \\
    & & & & \\
    \quad Hub size & $4.63$ & $4.39,\,10.39$ & $12.93,\,15.29,\,20.21$ & $21.05,\,50.71,\,75.22,\,52.31$ \\
     & $(0.18)$ & $(0.20,\,0.80)$ & $(7.17,\,1.10,\,3.63)$ & \\
    & & & & \\
    \quad Competitor hub size & $-0.20$ & $-0.87,\,-1.36$ & $-3.35,\,-3.00,\,-5.69$ & $-2.14,\,-5.07,\,-4.55,\,-3.85$ \\
     & $(0.15)$ & $(0.15,\,0.31)$ & $(2.74,\,0.46,\,1.38)$ & \\
    \hline
    Airline$\times$Quarter FE & Y & Y & Y & Y \\
    Airport FE & Y & Y & Y & Y \\
    Observations & $19,544$ & $19,544$ & $19,544$ & $19,544$ \\
    Parameters & $482$ & $807$ & $1,132$ & $1,457$ \\
    Log-likelihood & $-13493.6$ & $-11047.8$ & $-10002.2$ & $-9268.8$ \\
    BIC$_{\text{1st}}$ & $31749.6$ & $30069.1$ & $31189.1$ & $32933.4$ \\
    \hline \hline
    \end{tabular}}
\end{table}

}

Most estimates in Table~\ref{tab_app_entry_model} have the expected signs: entry probabilities rise with market size and own hub size and fall with distance and competitors' hub size. The estimated type probabilities are $0.55$ and $0.45$ for $L=2$ and $0.38$, $0.34$, and $0.28$ for $L=3$; all are sizable and precisely estimated. Elasticities vary markedly across types---for $L=3$, the distance elasticity ranges from $-11.95$ to $-25.42$---indicating substantial heterogeneity in entry behavior across markets. Most notably, the competitors' hub-size elasticity is small and statistically insignificant without latent types ($-0.20$; s.e. $0.15$) but sizable and significant with $L=2$ ($-0.87$ and $-1.36$). Unobserved market characteristics that raise all airlines' profitability induce a positive correlation between competitors' presence and entry, confounding competition's deterrent effect if left uncontrolled.

Our choice of $L$ uses the first- and second-step BICs \eqref{eq:BIC_1} and \eqref{eq:BIC_2}, the precision of the entry-model estimates, and the robustness of the demand estimates. In Table~\ref{tab_app_entry_model}, moving from $L=1$ to $L=2$ substantially raises the log-likelihood and lowers BIC$_{\text{1st}}$ from $31{,}749.6$ to $30{,}069.1$,\footnote{This improvement is not due to the probit at $L=1$: comparing a logit at $L=1$ with a finite-mixture logit at $L=2$ gives the same result.} showing that latent market types capture strong correlation among airline entry decisions unexplained by $\boldsymbol{x}_{t}$. Although $L=3$ further raises the likelihood, its additional parameters increase BIC$_{\text{1st}}$ to $31{,}189.1$; hence $L=2$ ranks first and $L=3$ second, still below $L=1$. At $L=4$, BIC$_{\text{1st}}$ rises to $32{,}933.4$, the largest value, and the estimates become extremely imprecise; we therefore omit their standard errors.\footnote{For $L=2$, $3$, and $4$, our EM algorithm uses $32$, $48$, and $64$ starts, respectively, combining random starts with starts constructed from the $L-1$ solution, and retains the highest-likelihood solution. It then performs twelve basin-hopping rounds that perturb the best solution's posterior-type probabilities, re-run EM, and retain any improvement.} Thus, the first step favors $L=2$ or $L=3$ and rules out $L=4$; the next section chooses between $L=2$ and $L=3$ using BIC$_{\text{2nd}}$ and the robustness of the demand estimates.

\subsection{Second step: Estimation of demand parameters}\label{sec:empirical_demand_estim}
For the demand system, we follow \cite{ciliberto2021market} and estimate a nested logit model in which the seven airlines active in the directional market $t$ form a single nest with nesting parameter $\sigma$, and the outside option is a separate alternative.
\begin{equation}\label{eq:empirical_nested_logit}
    \ln\left(\frac{s_{jt}}{s_{0t}}\right)
    \; = \;
    \alpha\; p_{jt} +
    \boldsymbol{x}_{jt}^{D \prime}\;\boldsymbol{\beta} +
    \sigma \;
    \ln \left( \frac{s_{jt}}{1-s_{0t}} \right) +
    \lambda_{j}(\boldsymbol{x}_{t}) +
    \widetilde{\xi}_{jt}.
\end{equation}
We compute airline $j$'s market share $s_{jt}$ in demand market $t$ as the number of passengers traveling that route on airline $j$'s non-stop flights (multiplied by $10$, since the data are a $10\%$ survey of total traffic), divided by market size---the geometric mean of the origin and destination MSA populations. The outside-option share is $s_{0t}=1-\sum_j s_{jt}$, where the sum runs over the airlines in the demand model. Price $p_{jt}$ is the average fare on airline $j$'s non-stop flights in market $t$, in hundreds of dollars. The vector of exogenous demand characteristics $\boldsymbol{x}_{jt}^{D}$ includes market distance (in thousands of miles), its square, airline $j$'s airport presence, and airline $\times$ quarter and airport indicators. Airport presence is the share of airline $j$'s network operated from the route endpoints: the sum of its hub sizes at the origin and destination divided by its total hub size across all airports that quarter. For the selection-bias function $\lambda_j(\boldsymbol{x}_t)$, we consider the specifications below: two Heckman-type benchmarks and the two corrections proposed in Section~\ref{sec:identification}---the continuous-$\boldsymbol{\kappa}_t$ correction of Online Appendix~\ref{app: cont kappa} and the finite-mixture correction with latent market types.

\medskip

\noindent \textit{\textbf{Heckman approach: single-index and normal unobservables.}} The unobservables of the entry model satisfy a single-threshold-crossing condition: $a_{jt} = \mathbb{1}\{ \eta_{jt} \leq \boldsymbol{\kappa}_{t}^{\prime} \, \boldsymbol{\rho}_{j} +
\boldsymbol{b}_{t}^{\prime} \boldsymbol{\gamma}_{j}^{P} \}$, where $\boldsymbol{\eta}_{t}\equiv(\eta_{1t},\ldots,\eta_{Jt})^{\prime}$ collects the scalar private shocks; $\boldsymbol{\xi}_{t}$, $\boldsymbol{\kappa}_{t}$, and $\boldsymbol{\eta}_{t}$ are jointly normally distributed; $\boldsymbol{\rho}_{j}$ and $\boldsymbol{\gamma}_{j}^{P}$ are parameters; and $\boldsymbol{b}_{t}$ contains basis functions of $\boldsymbol{x}_{t}$. Let $v_{jt} \equiv \eta_{jt} - \boldsymbol{\kappa}_{t}^{\prime} \, \boldsymbol{\rho}_{j}$. The standard Heckman correction is
\begin{equation*}
    \lambda_{j}(\boldsymbol{x}_{t})
    \; = \;
    \mathbb{E}\left(
    \xi_{jt} \mid
    v_{jt} \leq \boldsymbol{b}_{t}^{\prime} \boldsymbol{\gamma}_{j}^{P}
    \right)
    \; = \;
    \varrho_{j}^{\mathrm{H}} \;
    \frac{\phi \left( \Phi^{-1}[ \overline{P}_j(\boldsymbol{x}_{t})]\right)}
    {\overline{P}_j(\boldsymbol{x}_{t})},
\end{equation*}
where $\phi$ and $\Phi$ are the standard normal density and CDF, and the airline-specific composite coefficient $\varrho_{j}^{\mathrm{H}} \equiv -\operatorname{Cov}(\xi_{jt},v_{jt})/\sqrt{\operatorname{Var}(v_{jt})}$. We construct the inverse Mills ratio from $\overline{P}_{j}(\boldsymbol{x}_{t}) \equiv \Phi(\boldsymbol{b}_{t}^{\prime}\boldsymbol{\gamma}_{j}^{P})$, the ordinary propensity score implied by the probit entry model with $L=1$ from Table~\ref{tab_app_entry_model}. With six airlines included in the entry model (all except Others), this correction introduces six additional control variables, one inverse Mills ratio for each modeled airline.

\medskip

\noindent \textit{\textbf{Semiparametric Heckman approach.}} We maintain the single-threshold-crossing structure of the entry model $a_{jt} = \mathbb{1}\{ \eta_{jt} \leq \boldsymbol{\kappa}_{t}^{\prime} \, \boldsymbol{\rho}_{j} + \boldsymbol{b}_{t}^{\prime} \boldsymbol{\gamma}_{j}^{P} \}$ but allow $\boldsymbol{\xi}_{t}$, $\boldsymbol{\kappa}_{t}$, and $\boldsymbol{\eta}_{t}$ to have unrestricted distributions with continuous support. The selection-bias term is then a deterministic function of $\boldsymbol{b}_{t}^{\prime} \boldsymbol{\gamma}_{j}^{P}$ or, equivalently, of the propensity score $\overline{P}_{j}(\boldsymbol{x}_{t}) \equiv F_{v,j}(\boldsymbol{b}_{t}^{\prime} \boldsymbol{\gamma}_{j}^{P})$, where $F_{v,j}$ is the (invertible) CDF of $v_{jt}$. Because the basis functions in $\boldsymbol{b}_{t}$ can approximate any smooth function, the probit propensity score with $L=1$ from Table~\ref{tab_app_entry_model} entails no loss of flexibility. Following \cite{newey_2009}, we approximate the selection-bias function with a cubic polynomial in Heckman's inverse Mills ratio, with airline-specific coefficients, so that the selection correction introduces $18$ additional control variables.\footnote{We obtain very similar estimates when using higher-order polynomial approximations.}

\medskip

\noindent \textit{\textbf{Continuous-$\boldsymbol{\kappa}_{t}$ approach: multivariate normal unobservables.}} The two approaches above restrict the unobservables driving selection to enter through the scalar index $v_{jt}$. As an alternative, we consider the approach in Online Appendix~\ref{app: cont kappa}, which allows for multi-dimensional common-knowledge unobservables, $\boldsymbol{\kappa}_{t} = \boldsymbol{\xi}_{t}$, under the assumption that these are jointly normally distributed \citep[similar to][]{ciliberto2021market}. As shown in Online Appendix~\ref{app: cont kappa}, the selection-bias function then is:
\begin{equation*}
    \lambda_{j}(\boldsymbol{x}_{t})
    \; = \;
    \sum_{k} \varrho_{jk} \,
    \frac{\partial \log \overline{P}_{j}(\boldsymbol{x}_{t})}{\partial x_{k1t}},
\end{equation*}
where $x_{k1t}$ is a continuous regressor that shifts airline $k$'s latent entry index, and the composite coefficients $\varrho_{jk}$ encode the covariances between the demand unobservables. We implement this correction using the same probit entry model with $L=1$ from Table~\ref{tab_app_entry_model} as in the two approaches above. For each carrier $k$, we use its own combined hub size at the route's endpoints as the special regressor $x_{k1t}$. Thus, for every airline $j$, the correction includes seven derivatives of $\log \overline{P}_{j}(\boldsymbol{x}_{t})$: one with respect to $j$'s own hub size and six with respect to the competitors' hub sizes. With six airlines included in the entry model (all except Others), this yields $6 \times 7 = 42$ additional control variables.\footnote{In the notation of Online Appendix~\ref{app: cont kappa}, we interpret equation~\eqref{eq:latent_entry_index} so that, among the airline-specific hub measures, each latent index $\zeta_{kt}$ includes only airline $k$'s own combined hub size $x_{k1t}$; rivals' hub sizes do not enter $\zeta_{kt}$. Nevertheless, because $P_j(\boldsymbol{x}_{t},\boldsymbol{\xi}_{t})=G_j(\zeta_{1t},\ldots,\zeta_{Jt})$, the ordinary propensity score $\overline P_j(\boldsymbol{x}_{t})$ generally depends on every airline's $x_{k1t}$. Accordingly, the reduced form index of each airline in the entry model includes the combined hub sizes of all airlines.}

\medskip

\noindent \textit{\textbf{Finite-mixture logit with latent types.}} This is the entry model estimated in Section~\ref{sec:empirical_entry_estim}. For each airline $j$ and latent type $\ell = 1, \ldots, L-1$, the selection-function regressor is as in equation~\eqref{eq:r_hat}. The $L-1$ regressors per airline are stacked into $\widehat{\boldsymbol{r}}_{jt} = (\widehat{r}_{j1t}, \widehat{r}_{j2t}, \dots, \widehat{r}_{j,L-1,t})^{\prime}$, so with six airlines in the entry model the selection correction adds $6(L-1)$ controls: $6$ for $L=2$, $12$ for $L=3$, and $18$ for $L=4$.

\medskip

\noindent For all the two-stage least squares (2SLS) estimators of \eqref{eq:empirical_nested_logit}, we use as instruments, for each of an airline's competitors, airport presence if it is a potential entrant in the market (and zero otherwise), together with the number of competitors that are potential entrants. As detailed in Online Appendix~\ref{app:bootstrap}, standard errors are computed using a linearized common-weight bootstrap with $200$ replications.

Table~\ref{tab_app_demand} reports demand-parameter estimates and Table~\ref{tab_app_elast} their implied median price elasticities and Lerner indexes. Correcting for the endogeneity of prices and within-nest market shares by moving from OLS to 2SLS without selection correction more than doubles the magnitude of the median own-price elasticity ($-1.03$ to $-2.30$) and lowers the median Lerner index from $97.1\%$ to $43.4\%$. Accounting additionally for endogenous product entry yields median own-price elasticities ranging from $-4.75$ to $-2.37$ and median Lerner indexes from $21.1\%$ to $42.3\%$ across the six selection-corrected estimators. Median cross-price elasticities range from $0.81$ to $1.68$, versus $0.91$ under 2SLS, with the largest increases under the continuous-$\boldsymbol{\kappa}_{t}$ and finite-mixture corrections. Despite variation across corrections, accounting for endogenous product entry generally yields more elastic demand estimates, with implications for any subsequent analyses using them as inputs.

The last four rows of Table~\ref{tab_app_demand} report Hausman-type specification tests comparing each selection-corrected estimator with the baseline 2SLS, based on the statistic $\mathcal{H}(\text{2SLS}, \cdot)$ defined in equation~\eqref{eq:hausman_general} of Section~\ref{sec:estimating_L} and computed with the linearized common-weight bootstrap of Online Appendix~\ref{app:bootstrap} using 200 replications. We report the joint test on the demand parameters $(\alpha,\sigma,\beta_{\text{dist}},\beta_{\text{dist}^2},\beta_{\text{pres}})$ ($\chi^2(5)$) and the restricted test on the price and nesting parameters $(\alpha,\sigma)$ ($\chi^2(2)$), which govern the price elasticities in Table~\ref{tab_app_elast}. The $\chi^2(5)$ test rejects equality with the baseline 2SLS at the $5\%$ level or lower for every selection-corrected estimator except $L=4$; the restricted $\chi^2(2)$ test rejects for the three $L=1$ corrections and for $L=3$, but not for $L=2$. The non-rejections in the $L=4$ column reflect the wide bootstrap standard errors induced by the imprecise estimation of the entry model.

{
	\let\savedendtable\endtable
	\renewcommand{\endtable}{%
		\par\vspace{0.5em}
		\centering
		\begin{minipage}{0.90\linewidth}
			\footnotesize
			\textit{Notes:} The table reports estimates of the nested-logit demand parameters across the baseline OLS and 2SLS specifications and across the alternative selection-correction specifications described in the text. Standard errors are computed using the linearized common-weight bootstrap of Online Appendix~\ref{app:bootstrap} with 200 replications. BIC$_{\text{2nd}}$ is the second-step BIC-style criterion defined in equation~\eqref{eq:BIC_2}. The last four rows report the Hausman-type specification test $\mathcal{H}(\text{2SLS}, \cdot)$ from equation~\eqref{eq:hausman_general}, comparing each selection-corrected estimator against the baseline 2SLS: the $\chi^2(5)$ statistic is joint over $(\alpha,\sigma,\beta_{\text{dist}},\beta_{\text{dist}^2},\beta_{\text{pres}})$, while the $\chi^2(2)$ statistic is joint over $(\alpha,\sigma)$.
		\end{minipage}
		\par
		\savedendtable}
	\begin{table}[ht]
\caption{Estimation of Demand Parameters \label{tab_app_demand}}
\centering\resizebox{0.9\textwidth}{!}{
    \begin{tabular}{l|cc|cccccc}
    \hline \hline
    & \multicolumn{2}{c|}{\textit{Not control. for sel.}}
    & \multicolumn{6}{c}{\textit{Controlling for endogenous selection}} \\
    & {OLS} & {2SLS} & {2SLS} & {2SLS} & {2SLS} & {2SLS} & {2SLS} & {2SLS} \\
    & & & {Heckman} & {Semipar.} & {Cont.\ $\kappa$} & {Fin.-Mix.} & {Fin.-Mix.} & {Fin.-Mix.} \\
    & & & $L=1$ & $L=1$ & $L=1$ & $L=2$ & $L=3$ & $L=4$ \\
    \hline
    Price (100\$) ($\alpha$) & $-0.342$ & $-0.837$ & $-1.293$ & $-1.384$ & $-1.786$ & $-0.851$ & $-0.954$ & $-0.993$ \\
     & $(0.015)$ & $(0.157)$ & $(0.222)$ & $(0.224)$ & $(0.311)$ & $(0.157)$ & $(0.158)$ & $(0.201)$ \\
    & & & & & & & & \\
    Within share ($\sigma$) & $0.675$ & $0.592$ & $0.473$ & $0.437$ & $0.556$ & $0.602$ & $0.602$ & $0.621$ \\
     & $(0.009)$ & $(0.034)$ & $(0.041)$ & $(0.043)$ & $(0.061)$ & $(0.034)$ & $(0.035)$ & $(0.047)$ \\
    & & & & & & & & \\
    Distance (1000mi) & $-0.016$ & $0.513$ & $0.844$ & $0.902$ & $1.416$ & $0.421$ & $0.603$ & $0.686$ \\
     & $(0.055)$ & $(0.169)$ & $(0.228)$ & $(0.231)$ & $(0.318)$ & $(0.168)$ & $(0.166)$ & $(0.210)$ \\
    & & & & & & & & \\
    Distance$^2$ & $-0.112$ & $-0.180$ & $-0.168$ & $-0.159$ & $-0.261$ & $-0.139$ & $-0.185$ & $-0.209$ \\
     & $(0.022)$ & $(0.031)$ & $(0.037)$ & $(0.038)$ & $(0.047)$ & $(0.032)$ & $(0.030)$ & $(0.037)$ \\
    & & & & & & & & \\
    Airport presence & $4.593$ & $7.438$ & $7.566$ & $7.646$ & $8.820$ & $7.313$ & $7.801$ & $7.573$ \\
     & $(0.221)$ & $(0.550)$ & $(0.655)$ & $(0.666)$ & $(0.723)$ & $(0.544)$ & $(0.598)$ & $(0.892)$ \\
    \hline
    Airline$\times$Quarter FE & Y & Y & Y & Y & Y & Y & Y & Y \\
    Airport FE & Y & Y & Y & Y & Y & Y & Y & Y \\
    \# control var.\ entry & 0 & 0 & 6 & 18 & 42 & 6 & 12 & 18 \\
    Observations  & $37,184$ & $37,184$ & $37,184$ & $37,184$ & $37,184$ & $37,184$ & $37,184$ & $37,184$ \\
    & & & & & & & & \\
    BIC$_{\text{2nd}}$ & & $-21916.8$ & & & & $-22057.7$ & $-22433.4$ & $-22677.7$ \\
    Hausman $\chi^2(5)$ vs 2SLS & & & $189.80$ & $165.49$ & $83.12$ & $55.71$ & $12.23$ & $3.09$ \\
    \quad p-value & & & $[0.000]$ & $[0.000]$ & $[0.000]$ & $[0.000]$ & $[0.032]$ & $[0.686]$ \\
    Hausman $\chi^2(2)$ vs 2SLS & & & $101.06$ & $97.29$ & $64.81$ & $3.98$ & $6.74$ & $1.94$ \\
    \quad p-value & & & $[0.000]$ & $[0.000]$ & $[0.000]$ & $[0.137]$ & $[0.034]$ & $[0.380]$ \\
    \hline \hline
    \end{tabular}}
\end{table}

}

{
	\let\savedendtable\endtable
	\renewcommand{\endtable}{%
		\par\vspace{0.5em}
		\centering
		\begin{minipage}{0.90\linewidth}
			\footnotesize
			\textit{Notes:} The table reports median own-price elasticities, cross-price elasticities, and Lerner indexes implied by the demand estimates in Table~\ref{tab_app_demand}. For each specification, these quantities are first computed at the airline-market level using the corresponding column's demand parameter estimates; the table then reports their medians, overall and separately by airline. Cross-price elasticities are reported by price-setting airline.
		\end{minipage}
		\par
		\savedendtable}
	\begin{table}[ht]
\caption{Median Price Elasticities and Lerner Indexes
\label{tab_app_elast}}
\centering\resizebox{0.9\textwidth}{!}{
    \begin{tabular}{l|cc|cccccc}
    \hline \hline
    & \multicolumn{2}{c|}{\textit{Not control. for sel.}}
    & \multicolumn{6}{c}{\textit{Controlling for endogenous selection}} \\
    & {OLS} & {2SLS} & {2SLS} & {2SLS} & {2SLS} & {2SLS} & {2SLS} & {2SLS} \\
    & & & {Heckman} & {Semipar.} & {Cont.\ $\kappa$} & {Fin.-Mix.} & {Fin.-Mix.} & {Fin.-Mix.} \\
    & & & $L=1$ & $L=1$ & $L=1$ & $L=2$ & $L=3$ & $L=4$ \\
    \hline
    \multicolumn{1}{r|}{\textit{Own-Price Elasticity}} & $-1.030$ & $-2.304$ & $-3.220$ & $-3.369$ & $-4.750$ & $-2.366$ & $-2.651$ & $-2.812$ \\
    \multicolumn{1}{r|}{\textit{AA}} & $-1.269$ & $-2.653$ & $-3.620$ & $-3.780$ & $-5.430$ & $-2.738$ & $-3.068$ & $-3.291$ \\
    \multicolumn{1}{r|}{\textit{DL}} & $-1.073$ & $-2.486$ & $-3.598$ & $-3.791$ & $-5.189$ & $-2.543$ & $-2.849$ & $-3.002$ \\
    \multicolumn{1}{r|}{\textit{UA}} & $-1.273$ & $-2.817$ & $-3.973$ & $-4.139$ & $-5.841$ & $-2.896$ & $-3.246$ & $-3.438$ \\
    \multicolumn{1}{r|}{\textit{US}} & $-1.078$ & $-2.352$ & $-3.306$ & $-3.475$ & $-4.817$ & $-2.416$ & $-2.706$ & $-2.884$ \\
    \multicolumn{1}{r|}{\textit{WN}} & $-0.771$ & $-1.842$ & $-2.755$ & $-2.913$ & $-3.889$ & $-1.879$ & $-2.105$ & $-2.205$ \\
    \multicolumn{1}{r|}{\textit{LCC}} & $-0.953$ & $-2.017$ & $-2.671$ & $-2.763$ & $-4.079$ & $-2.083$ & $-2.334$ & $-2.504$ \\
    \multicolumn{1}{r|}{\textit{Others}} & $-1.174$ & $-2.357$ & $-2.943$ & $-2.994$ & $-4.673$ & $-2.445$ & $-2.740$ & $-2.980$ \\
    & & & & & & & & \\
    \multicolumn{1}{r|}{\textit{Cross-Price Elasticity}} & $0.532$ & $0.910$ & $0.874$ & $0.810$ & $1.680$ & $0.963$ & $1.079$ & $1.215$ \\
    \multicolumn{1}{r|}{\textit{AA}} & $0.629$ & $1.074$ & $1.029$ & $0.952$ & $1.983$ & $1.138$ & $1.274$ & $1.436$ \\
    \multicolumn{1}{r|}{\textit{DL}} & $0.774$ & $1.323$ & $1.267$ & $1.173$ & $2.441$ & $1.401$ & $1.569$ & $1.768$ \\
    \multicolumn{1}{r|}{\textit{UA}} & $0.555$ & $0.949$ & $0.908$ & $0.842$ & $1.751$ & $1.005$ & $1.125$ & $1.267$ \\
    \multicolumn{1}{r|}{\textit{US}} & $0.593$ & $1.014$ & $0.971$ & $0.899$ & $1.871$ & $1.074$ & $1.202$ & $1.355$ \\
    \multicolumn{1}{r|}{\textit{WN}} & $0.514$ & $0.880$ & $0.845$ & $0.783$ & $1.625$ & $0.932$ & $1.043$ & $1.175$ \\
    \multicolumn{1}{r|}{\textit{LCC}} & $0.428$ & $0.732$ & $0.702$ & $0.650$ & $1.352$ & $0.775$ & $0.868$ & $0.978$ \\
    \multicolumn{1}{r|}{\textit{Others}} & $0.292$ & $0.499$ & $0.478$ & $0.443$ & $0.921$ & $0.529$ & $0.592$ & $0.667$ \\
    & & & & & & & & \\
    \multicolumn{1}{r|}{\textit{Lerner Index}} & $97.1\%$ & $43.4\%$ & $31.1\%$ & $29.7\%$ & $21.1\%$ & $42.3\%$ & $37.7\%$ & $35.6\%$ \\
    \multicolumn{1}{r|}{\textit{AA}} & $78.8\%$ & $37.7\%$ & $27.6\%$ & $26.5\%$ & $18.4\%$ & $36.5\%$ & $32.6\%$ & $30.4\%$ \\
    \multicolumn{1}{r|}{\textit{DL}} & $93.2\%$ & $40.2\%$ & $27.8\%$ & $26.4\%$ & $19.3\%$ & $39.3\%$ & $35.1\%$ & $33.3\%$ \\
    \multicolumn{1}{r|}{\textit{UA}} & $78.6\%$ & $35.5\%$ & $25.2\%$ & $24.2\%$ & $17.1\%$ & $34.5\%$ & $30.8\%$ & $29.1\%$ \\
    \multicolumn{1}{r|}{\textit{US}} & $92.7\%$ & $42.5\%$ & $30.3\%$ & $28.8\%$ & $20.8\%$ & $41.4\%$ & $37.0\%$ & $34.7\%$ \\
    \multicolumn{1}{r|}{\textit{WN}} & $129.7\%$ & $54.3\%$ & $36.3\%$ & $34.3\%$ & $25.7\%$ & $53.2\%$ & $47.5\%$ & $45.4\%$ \\
    \multicolumn{1}{r|}{\textit{LCC}} & $105.0\%$ & $49.6\%$ & $37.4\%$ & $36.2\%$ & $24.5\%$ & $48.0\%$ & $42.8\%$ & $39.9\%$ \\
    \multicolumn{1}{r|}{\textit{Others}} & $85.2\%$ & $42.4\%$ & $34.0\%$ & $33.4\%$ & $21.4\%$ & $40.9\%$ & $36.5\%$ & $33.6\%$ \\
    \hline
    Airline$\times$Quarter FE & Y & Y & Y & Y & Y & Y & Y & Y \\
    Airport FE & Y & Y & Y & Y & Y & Y & Y & Y \\
    \# control var.\ entry & 0 & 0 & 6 & 18 & 42 & 6 & 12 & 18 \\
    Observations  & $37,184$ & $37,184$ & $37,184$ & $37,184$ & $37,184$ & $37,184$ & $37,184$ & $37,184$ \\
    \hline \hline
    \end{tabular}}
\end{table}

}

Following the combined selection strategy of Section~\ref{sec:estimating_L}, we select $L=3$ using BIC$_{\text{1st}}$ in Table~\ref{tab_app_entry_model}, BIC$_{\text{2nd}}$ in Table~\ref{tab_app_demand}, and the pairwise Hausman-type tests in Online Appendix Table~\ref{tab_hausman}. Although BIC$_{\text{2nd}}$ alone favors $L=4$, that specification ranks last by BIC$_{\text{1st}}$, has an imprecisely estimated entry model, and yields demand estimates statistically indistinguishable from those for $L=3$. By contrast, the $L=3$ estimates differ significantly from those for $L=2$ ($\chi^2(5)=27.32$, $p<0.01$), and moving from $L=2$ to $L=3$ improves BIC$_{\text{2nd}}$.

Online Appendix Figures~\ref{fig_distribution_elasticities} and~\ref{fig_distribution_crosselast} report the empirical distributions of estimated own- and cross-price elasticities under 2SLS, the continuous-$\boldsymbol{\kappa}_{t}$ correction, and the finite-mixture correction with $L=3$. Relative to 2SLS, both selection corrections shift the own-price distributions leftward and increase their dispersion for every airline, most markedly under the continuous-$\boldsymbol{\kappa}_{t}$ correction; they also imply stronger substitution across airlines. Thus, ignoring endogenous entry understates the magnitude of both own- and cross-price elasticities.

\FloatBarrier

\subsection{Estimation of costs and profits \label{sec_counterfactuals}}

Although our main focus is demand estimation with endogenous product entry, as discussed in Section~\ref{sec:identification_costs}, our framework also allows us to identify the remaining ingredients required for a wide range of counterfactual experiments: marginal costs, equilibrium variable profits, and fixed costs. In this section, we illustrate these ideas. We first estimate marginal costs and then use these estimates, together with the first-step type-specific propensity scores and the second-step demand parameters, to construct equilibrium variable profits and to recover fixed costs. Finally, we introduce an ANOVA--Sobol variance decomposition of the net entry-profit index that quantifies the contribution of unobserved market-type heterogeneity to endogenous selection, distinguishing the expected-variable-profit channel from the fixed-cost channel.

\subsubsection{Marginal costs}
Following \cite{ciliberto2021market}, we assume that the airlines engage in Bertrand--Nash competition and that each airline has a marginal cost function that does not depend on output. Then, given demand equation \eqref{eq:empirical_nested_logit}, the marginal cost of airline $j$ in demand market $t$ can be estimated from the following pricing equation: 
\begin{equation}\label{eq:mc_1}
    p_{jt} + \frac{1-\sigma}{\alpha(1-\sigma s_{jt|g}-(1-\sigma)s_{jt})} = mc_{jt},
\end{equation}
where $g$ denotes the nest that contains all the airlines, $s_{jt|g} \equiv s_{jt}/(1-s_{0t})$, and $mc_{jt}$ denotes the marginal cost. We then estimate the following marginal cost function:
\begin{equation}\label{eq:mc_2}
    \ln(mc_{jt}) \; = \;
    (\boldsymbol{x}_{jt}^{\mathrm{mc}})^{\prime}\;\boldsymbol{\varphi} + \lambda^{\text{mc}}_{j}(\boldsymbol{x}_{t}) + \widetilde{\omega}_{jt},
\end{equation}
where $\boldsymbol{x}_{jt}^{\mathrm{mc}}$ includes market distance and market distance squared, together with airline $\times$ quarter and airport indicators, $\lambda^{\text{mc}}_{j}(\boldsymbol{x}_{t})$ is a selection-bias function specified as for the demand equation \eqref{eq:empirical_nested_logit}, and $\widetilde{\omega}_{jt}\equiv\omega_{jt}-\lambda^{\text{mc}}_{j}(\boldsymbol{x}_{t})$ is the residual variable-cost unobservable, with conditional mean zero in the selected sample.

Online Appendix Table~\ref{tab_app_mc} reports the average marginal costs obtained from equation~\eqref{eq:mc_1} using the demand estimates of Table~\ref{tab_app_demand} (see Online Appendix Figure~\ref{fig_distribution_mc} for the corresponding empirical distributions), while Online Appendix Table~\ref{tab_app_cost} presents the estimates of $\boldsymbol{\varphi}$ from equation~\eqref{eq:mc_2}. The $\ln(mc_{jt})$ regressand in each column is constructed from the demand estimates of the corresponding column of Table~\ref{tab_app_demand}, while equation~\eqref{eq:mc_2} is always estimated by OLS: once selection is controlled for, its regressors are exogenous. Two facts stand out. First, price and within-nest-share endogeneity matters sizably: the demand estimates that do not correct for it (OLS column) imply average marginal costs close to zero, with roughly half of the observations dropped because the implied marginal cost is not positive. Second, after accounting for demand-side selection when constructing the dependent variable \(\ln(mc_{jt})\), endogenous selection in the marginal-cost regression still matters, although less than in demand: only the \(L=1\) corrections yield estimates significantly different from the baseline 2SLS.

\subsubsection{Contribution of unobservables to endogenous selection}

The preceding estimates show that unobserved market heterogeneity generates selection bias in demand estimation and, to a lesser extent, in marginal-cost estimation. We now assess its role in airlines' entry decisions by quantifying the fraction of cross-market variation in the net entry-profit index driven by the latent market state \(\boldsymbol{\kappa}_{t}\) and, within this latent-type channel, the parts operating through expected variable profits and fixed entry costs. We do this for our preferred specification with $L=3$ using an airline-by-airline ANOVA--Sobol variance decomposition. All derivations and implementation details are in Online Appendix~\ref{app:sobol}.

Following Section~\ref{sec:identification_costs}, let $VP_{j\ell t}\equiv VP^P_{j\ell}(\boldsymbol{x}_t)$ denote expected variable profit at entry, which averages equilibrium variable profits over rivals' possible entry configurations using their type-specific propensity scores. Let $fc_{j\ell t}\equiv fc_{j\ell}(\boldsymbol{x}_{jt}^{\mathrm{fc}})$ denote the component of fixed entry cost excluding the private shock $\sigma_{\eta_j}\eta_{jt}$, and let $\Pi_{j\ell t}\equiv VP_{j\ell t}-fc_{j\ell t}$ denote the net entry-profit index before that shock. As detailed in Online Appendix~\ref{app:sobol}, we estimate $\sigma_{\eta_j}$ by 2SLS from the inverted entry equation and recover $fc_{j\ell t}$. The variance of each $Y\in\{\Pi,VP,fc\}$ across markets and latent types decomposes as $V_j^Y=V_{j,x}^Y+V_{j,\ell}^Y+V_{j,\mathrm{res}}^Y$, where the three components capture variation attributable to observables alone, latent types alone, and their interaction, respectively. We define $V_{j,\mathrm{unobs}}^Y\equiv V_{j,\ell}^Y+V_{j,\mathrm{res}}^Y$ as the variation associated with latent types, either directly or through their interaction with observables. Since $\Pi=VP-fc$,
\begin{equation*}
\frac{V_{j,\mathrm{unobs}}^{\Pi}}{V_j^{\Pi}}
=
\frac{V_{j,\mathrm{unobs}}^{VP}}{V_j^{\Pi}}
+
\frac{V_{j,\mathrm{unobs}}^{fc}}{V_j^{\Pi}}
-
\frac{2\,\operatorname{Cov}^{VP,fc}_{j,\mathrm{unobs}}}{V_j^{\Pi}}.
\end{equation*}
Table~\ref{tab_sobol} reports, for each airline, $\widehat{\sigma}_{\eta_j}$, the corresponding first-stage $F$-statistic, the unobservable share of net-entry-profit variance and its pure-type and interaction components, and the three terms in this identity, each normalized by $V_j^\Pi$.

{
\let\savedendtable\endtable
\renewcommand{\endtable}{%
\par\vspace{0.5em}
\centering
\begin{minipage}{0.9\linewidth}
\footnotesize
\textit{Notes:} The table reports airline-specific measures from the ANOVA--Sobol decomposition of the net entry-profit index, computed, for each airline, over its potential-entry markets. The sample is restricted to route-quarters for which both directional demand markets are observed, since expected variable profits are constructed by aggregating the two directions of each route. All definitions and estimation details are in Online Appendix~\ref{app:sobol}. $\widehat{\sigma}_{\eta_j}$ is the airline-specific scale parameter estimated from the 2SLS regression in equation~\eqref{eq:sobol_sigma_eta}; $F$-stat is the corresponding first-stage statistic, clustered by route-quarter.
\end{minipage}
\par
\savedendtable}
\begin{table}[ht]
\caption{Contribution of Unobservables to Endogenous Selection
\label{tab_sobol}}
\centering\resizebox{0.9\textwidth}{!}{
    \begin{tabular}{l|cc|ccc|ccc}
    \hline \hline
    Airline
    & $\widehat{\sigma}_{\eta}$
    & $F$-stat
    & $\displaystyle \frac{V^{\Pi}_{\mathrm{unobs}}}{V^{\Pi}}$
    & $\displaystyle \frac{V^{\Pi}_{\ell}}{V^{\Pi}}$
    & $\displaystyle \frac{V^{\Pi}_{\mathrm{res}}}{V^{\Pi}}$
    & $\displaystyle \frac{V^{VP}_{\mathrm{unobs}}}{V^{\Pi}}$
    & $\displaystyle \frac{V^{fc}_{\mathrm{unobs}}}{V^{\Pi}}$
    & $\displaystyle \frac{-2 \, \operatorname{Cov}^{VP,fc}_{\mathrm{unobs}}}{V^{\Pi}}$ \\
    \hline
    AA
    & $125.65$ & $492.75$
    & $0.60$ & $0.27$ & $0.34$
    & $0.27$ & $0.89$ & $-0.55$ \\
    DL
    & $47.45$ & $550.33$
    & $0.53$ & $0.06$ & $0.47$
    & $0.37$ & $1.01$ & $-0.85$ \\
    UA
    & $88.27$ & $1235.89$
    & $0.36$ & $0.05$ & $0.31$
    & $0.43$ & $0.86$ & $-0.92$ \\
    US
    & $152.97$ & $387.13$
    & $0.62$ & $0.16$ & $0.46$
    & $0.24$ & $0.95$ & $-0.57$ \\
    WN
    & $348.12$ & $1275.94$
    & $0.40$ & $0.02$ & $0.38$
    & $1.04$ & $1.62$ & $-2.26$ \\
    LCC
    & $364.51$ & $218.36$
    & $0.42$ & $0.05$ & $0.37$
    & $0.32$ & $0.80$ & $-0.70$ \\
    \hline \hline
    \end{tabular}}
\end{table}

}

The estimated private-shock scales differ substantially across airlines, and all six first stages are strong. Latent market types and their interactions with observables account for $36\%$--$62\%$ of the variation in the net entry-profit index. The interaction component exceeds the pure-type component for every airline, with the latter sizable mainly for AA and US. For every airline, the gross fixed-cost contribution exceeds the expected-variable-profit contribution. The unobservable components of expected variable profits and fixed costs also covary positively: latent market states associated with higher expected variable profits tend to entail higher fixed costs, so the two channels partially offset in the net entry-profit index. These results complement our earlier finding that endogenous selection matters more for demand estimation than for marginal-cost estimation: the ANOVA--Sobol decomposition additionally reveals that fixed-cost heterogeneity is the larger gross channel through which latent market states affect airlines' entry decisions.

\FloatBarrier

\section{Conclusions \label{sec:conclusions}}

Firms offer products in markets where they expect to earn profits, using information about market conditions partly unobserved by researchers. When this information correlates with demand unobservables, entry creates endogenous selection bias in demand estimation. Existing approaches either restrict firms' information at entry or jointly estimate a full structural model of demand, pricing, and entry, making demand estimates heavily dependent on supply-side assumptions.

This paper addresses this form of endogenous selection with a two-step identification approach. When firms' market-entry decisions are correlated through latent market states, we show that the selection-bias function in demand can be represented using latent propensity scores: entry probabilities conditional on observables and latent market states. We provide conditions under which these scores are identified from the joint distribution of entry decisions and can be used to control for endogenous selection. These identification results lead to a computationally convenient two-step estimator of demand that accommodates price endogeneity without either ruling out endogenous selection through restrictions on firms' information or jointly estimating demand, pricing, and entry.

We illustrate the method using data from the airline industry. The results highlight the empirical importance of correcting demand estimates for endogenous product entry allowing for correlation across firms' market-entry decisions. Failing to correct for endogenous entry substantially understates the magnitude of own- and cross-price elasticities and overstates market power.

\singlespacing

\bibliography{references}

\clearpage

\onehalfspacing

\appendix

\renewcommand{\appendixname}{Online Appendix}
\renewcommand{\appendixpagename}{Online Appendix}

\renewcommand{\thetable}{A.\arabic{table}}
\renewcommand{\thefigure}{A.\arabic{figure}}
\setcounter{table}{0}
\setcounter{figure}{0}

\appendixpage

\section{Notation guide\label{app:notation_guide}}

\begin{table}[!ht]
\centering
\caption{Notation Used in the Main Text}
\label{tab:notation_main}
\begingroup
\singlespacing
\fontsize{8}{9}\selectfont
\setlength{\tabcolsep}{2pt}
\renewcommand{\arraystretch}{1.00}
\begin{tabular}{@{}>{\raggedright\arraybackslash}p{0.17\textwidth}
                    >{\raggedright\arraybackslash}p{0.29\textwidth}
                    |
                    >{\raggedright\arraybackslash}p{0.17\textwidth}
                    >{\raggedright\arraybackslash}p{0.29\textwidth}@{}}
\hline \hline
Symbol & Definition & Symbol & Definition \\
\hline
\multicolumn{2}{@{}l|}{\textit{Indices and observed variables}}
& \multicolumn{2}{l@{}}{\textit{Entry and selection}} \\
$h$; $j\in\mathcal J\equiv\{1,\ldots,J\}$;
$t\in\{1,\ldots,T\}$
& Consumer; potential product and firm (in the single-product model); market.
& $\pi_j^P(\boldsymbol x_t,\boldsymbol\kappa_t,
\boldsymbol\eta_{jt})$;
$P_{jt}\equiv P_j(\boldsymbol x_t,\boldsymbol\kappa_t)$
& Entry profit averaged over rivals' BNE choices; BNE entry probability, or
latent propensity score. \\
$\boldsymbol{x}_{t}$;
$\boldsymbol{x}_{jt}^{D},\boldsymbol{x}_{jt}^{\mathrm{mc}},
\boldsymbol{x}_{jt}^{\mathrm{fc}}$
& All exogenous variables observed by the researcher, including every potential
product's characteristics; product-$j$ demand, variable-cost, and fixed-cost
subvectors.
& $\overline P_j(\boldsymbol x_t)$
& Ordinary propensity score,
$\Pr(a_{jt}=1\mid\boldsymbol x_t)=
\int P_j(\boldsymbol x_t,\boldsymbol\kappa)\,dF_{\kappa}(\boldsymbol\kappa)$. \\
$a_{jt}$; $\boldsymbol a_t$;
$\mathcal J_t^{\boldsymbol a}$
& Offer indicator; vector of offer indicators; set of products offered in
market $t$.
& $\mu_j(\boldsymbol\kappa)$
& $\mathbb E(\xi_{jt}\mid\boldsymbol\kappa_t=\boldsymbol\kappa)$, the
conditional mean demand shock. \\
$p_{jt}$; $s_{jt}$; $\boldsymbol s_t^{\boldsymbol a}$
& Price; market share; vector of shares of the offered products.
& $\lambda_j(\boldsymbol x_t)$; $\widetilde\xi_{jt}$
& Selection-bias function
$\mathbb E(\xi_{jt}\mid\boldsymbol x_t,a_{jt}=1)$; residual
$\xi_{jt}-\lambda_j(\boldsymbol x_t)$. \\
\hline
\multicolumn{2}{@{}l|}{\textit{Demand}}
& \multicolumn{2}{l@{}}{\textit{Finite-mixture representation}} \\
$\alpha,\boldsymbol\beta,\boldsymbol\sigma,\xi_{jt}$
& Price coefficient, observed-characteristic coefficients,
random-coefficient parameters, and mean-zero demand unobservable.
& $L$; $\{\mathcal K_\ell\}_{\ell=1}^{L}$
& Generic or candidate number of latent market types; positive-probability
$L$-cell partition of $\mathcal K$. \\
$\delta_{jt}$
& Mean utility,
$\alpha p_{jt}+(\boldsymbol x_{jt}^{D})'\boldsymbol\beta+\xi_{jt}$.
& $\widetilde f_\ell$
& Type probability $F_{\kappa}(\mathcal K_\ell)$. \\
$\boldsymbol\upsilon_{ht},F_{\upsilon}$;
$\boldsymbol\Omega_{\boldsymbol\sigma}$; $\varepsilon_{hjt}$
& Consumer heterogeneity and its distribution; random-coefficient loading
matrix; i.i.d. type-I extreme-value shock.
& $\widetilde P_{j,\ell}(\boldsymbol x_t)$;
$\widetilde\mu_{j,\ell}$
& Type-specific propensity score and mean demand shock: the respective cell
averages of $P_j(\boldsymbol x_t,\boldsymbol\kappa_t)$ and $\xi_{jt}$ over
$\mathcal K_\ell$. \\
$d_{jt}^{-1}(\boldsymbol s_t^{\boldsymbol a},\boldsymbol\sigma)$
& Product-$j$ mean utility obtained by inverting the offered-product demand
system.
& $L^*$
& Population number of latent market types satisfying
Assumption~\ref{assumption_2}; the minimal number reproducing the joint entry
distribution. \\
$\boldsymbol\theta\equiv
(\alpha,\boldsymbol\beta',\boldsymbol\sigma')'$
& Demand-parameter vector. In the nested-logit specialization,
$\boldsymbol\sigma$ reduces to the scalar nesting parameter $\sigma$.
& $r_{j\ell t}$; $\boldsymbol r_{jt}$; $\boldsymbol R_{jt}$
& Selection-function regressor
$\widetilde f_\ell(\widetilde P_{j,\ell}-
\widetilde P_{j,L^*})/\overline P_j$; its $L^*-1$ stack; corresponding
$J(L^*-1)$-vector with only product $j$'s block nonzero. \\
\hline
\multicolumn{2}{@{}l|}{\textit{Costs and information at entry}}
& \multicolumn{2}{l@{}}{\textit{Identification and estimation}} \\
$\omega_{jt}$; $mc_{jt}$; $fc_{jt}$;
$\boldsymbol\xi_t,\boldsymbol\omega_t$
& Mean-zero variable-cost unobservable, marginal cost, and fixed entry cost;
vectors of product-level demand and variable-cost unobservables.
& $\widetilde{\boldsymbol\mu}$;
$\boldsymbol\vartheta\equiv
(\boldsymbol\theta',\widetilde{\boldsymbol\mu}')'$
& Stacked product-specific selection-function coefficients; full second-step
parameter vector. \\
$\boldsymbol\kappa_t$; $\mathcal K$; $F_{\kappa}$
& Latent market state, unobserved by the researcher and common knowledge
among firms; its support and marginal distribution.
& $\boldsymbol z_{jt}$; $\boldsymbol w_{jt}$;
$\boldsymbol m(\boldsymbol\vartheta)$;
$\boldsymbol M(\boldsymbol\vartheta_0)$
& Excluded instruments; vector of excluded instruments and included regressors;
pooled selected-sample moment function; its Jacobian at the true value. \\
$\boldsymbol\eta_{jt}$; $F_{\boldsymbol\eta}$;
$\eta_{jt}$; $F_\eta$; $\sigma_{\eta_j}$
& General private fixed-cost information and distribution (vectors i.i.d.
across firms); scalar recovery specialization, known CDF, and firm-specific
scale.
& $\boldsymbol b_t,N_X$;
$\boldsymbol\gamma_{j,\ell},\boldsymbol\gamma$;
$\Lambda$; $\widetilde{\boldsymbol f}$
& Vector of $N_X$ sieve basis functions; product- and type-specific
coefficients and their stack; logistic CDF; type-probability vector. \\
\hline \hline
\end{tabular}
\endgroup
\end{table}

\clearpage

\begin{table}[!ht]
\centering
\caption{Additional Notation Used in the Online Appendix}
\label{tab:notation_appendix}
\begingroup
\singlespacing
\fontsize{9}{10.5}\selectfont
\setlength{\tabcolsep}{3pt}
\renewcommand{\arraystretch}{1.03}
\begin{tabular}{@{}>{\raggedright\arraybackslash}p{0.30\textwidth}
                    >{\raggedright\arraybackslash}p{0.65\textwidth}@{}}
\hline \hline
Symbol & Definition \\
\hline
\multicolumn{2}{@{}l}{\textit{Multi-product firms}} \\
$f\in\mathcal F$; $\mathcal J_f$
& Firm index and set of products owned by firm $f$; in this extension, $j$
indexes products only. \\
$\boldsymbol a_f,\boldsymbol a_{-f}$;
$\boldsymbol a_{ft},\boldsymbol a_{-ft}$
& Generic firm-$f$ portfolio and rival-product entry profile; their market-$t$
counterparts. \\
$\boldsymbol x_{ft}^{\mathrm{fc}}$;
$\boldsymbol\eta_{ft}\equiv
(\eta_{ft}(\boldsymbol a_f):\boldsymbol a_f\in
\{0,1\}^{|\mathcal J_f|})$
& Portfolio fixed-cost observables and vector of portfolio-specific private
cost shocks. \\
\hline
\multicolumn{2}{@{}l}{\textit{Dynamic game of product entry and exit}} \\
$t$; $\boldsymbol x_t^{ex}$;
$\boldsymbol x_t\equiv(\boldsymbol x_t^{ex},
\boldsymbol a_{t-1})$
& Time within a market (whose index is suppressed); exogenous state; dynamic
observed state, including lagged availability. \\
$fc(a_{jt},a_{j,t-1},\boldsymbol x_{jt}^{\mathrm{fc}},
\boldsymbol\kappa_t,\boldsymbol\eta_{jt})$
& Dynamic fixed and transition cost. \\
$P=(P_i:i\in\mathcal J)$; $v_j^P$
& Candidate profile of conditional choice probabilities (CCPs), which are
latent propensity scores at an MPE; difference between firm $j$'s
choice-specific value functions for $a_{jt}=1$ and $a_{jt}=0$ under $P$. \\
$F_{\kappa\mid x}(\cdot\mid\boldsymbol x_t)$
& Conditional distribution of the latent market state in the dynamic model. \\
\hline
\multicolumn{2}{@{}l}{\textit{Identification under normal heterogeneity}} \\
$\boldsymbol\tau_j$;
$\zeta_{jt}\equiv\boldsymbol x_t'\boldsymbol\tau_j+\xi_{jt}$;
$\boldsymbol\zeta_t$; $G_j,G_{jk}$
& Latent-index coefficients, product-$j$ index, and vector of indexes;
$P_j(\boldsymbol x_t,\boldsymbol\xi_t)=G_j(\boldsymbol\zeta_t)$ and
$G_{jk}=\partial G_j/\partial\zeta_k$. \\
$x_{k1t}$; $\tau_{k1}$;
$\Sigma_{jk}\equiv\operatorname{Cov}(\xi_{jt},\xi_{kt})$;
$\varrho_{jk}\equiv\Sigma_{jk}/\tau_{k1}$
& Product-$k$ special regressor and its index coefficient; demand-shock
covariance; composite coefficient on the corresponding derivative regressor. \\
\hline
\multicolumn{2}{@{}l}{\textit{Linearized common-weight bootstrap}} \\
$c_\ell$; $\boldsymbol\psi$
& Log ratio $\log(\widetilde f_\ell/\widetilde f_1)$, with $c_1=0$;
unconstrained first-step parameter vector combining the log ratios and
$\boldsymbol\gamma$. \\
$\ell_t^{\mathrm{1st}}$; $\dot\ell_t^{\mathrm{1st}}$;
$\widehat{\boldsymbol m}_t$; $\widehat{\boldsymbol H}$
& Market first-step log-likelihood contribution and score; market
second-step moment contribution; information-matrix estimator (distinct from
$H_t$). \\
$w_t^{\ast(b)}$; $b=1,\ldots,B_{\mathrm{boot}}$;
$\widehat{\boldsymbol\psi}^{\ast(b)}$;
$\widehat{\boldsymbol g}^{\ast(b)}$;
$\widehat{\boldsymbol W}^{\ast(b)}$
& Common market weight and replication index; updated first-step parameters,
weighted second-step moments, and GMM weighting matrix in replication $b$. \\
$\widehat{\boldsymbol\Delta}_{AB}^{\ast(b)}$;
$\widehat{\boldsymbol V}_{AB}$
& Bootstrap difference
$\widehat{\boldsymbol\theta}_q^{B,\ast(b)}-
\widehat{\boldsymbol\theta}_q^{A,\ast(b)}$ for the common parameter subvector;
its estimated variance-covariance matrix. \\
\hline
\multicolumn{2}{@{}l}{\textit{ANOVA--Sobol decomposition}} \\
$d\in\{1,2\}$; $T_j$
& Direction index within non-directional entry market $t$ (distinct from the
demand map $d_{jt}$); number of retained potential-entry markets for airline
$j$. \\
$VP_{j\ell t}\equiv VP_{j\ell}^{P}(\boldsymbol x_t)$;
$fc_{j\ell t}$; $\Pi_{j\ell t}\equiv VP_{j\ell t}-fc_{j\ell t}$;
$Y_{j\ell t}\in\{\Pi_{j\ell t},VP_{j\ell t},fc_{j\ell t}\}$
& Type-specific expected variable profit; fixed-cost component excluding the
private shock; net entry-profit index before the private shock (distinct from
$\Pi_{jt}$); generic object being decomposed. \\
$\overline Y_j$; $m_j^Y(t)$; $n_j^Y(\ell)$;
$u_{j\ell t}^Y$
& Overall, market-specific, and type-specific means; market--type interaction
residual (the $m_j^Y$ notation is unrelated to GMM moments). \\
$V_j^Y$; $V_{j,x}^Y,V_{j,\ell}^Y,V_{j,\mathrm{res}}^Y$;
$V_{j,\mathrm{unobs}}^Y$;
$\operatorname{Cov}_{j,\ell}^{VP,fc},
\operatorname{Cov}_{j,\mathrm{res}}^{VP,fc},
\operatorname{Cov}_{j,\mathrm{unobs}}^{VP,fc}$
& Total variance; observable, latent-type, interaction, and total
unobservable components; corresponding covariance components of expected
variable profit and fixed cost. \\
\hline \hline
\end{tabular}
\endgroup
\end{table}

\clearpage

\section{Multi-product firms \label{sec_multiproduct}}

This section sketches how the
argument of Proposition~\ref{proposition_3} can be extended to
multi-product firms. Products remain indexed by $j\in\mathcal{J}$, and firms
are indexed by $f\in\mathcal{F}$. Let
$\{\mathcal{J}_{f}:f\in\mathcal{F}\}$ be a partition of $\mathcal{J}$, where
$\mathcal{J}_{f}$ is the set of products owned by firm $f$. Firm $f$'s
portfolio choice in market $t$ is represented by
$\boldsymbol{a}_{ft}\equiv(a_{jt}:j\in\mathcal{J}_{f})
\in\{0,1\}^{|\mathcal{J}_{f}|}$, while
$\boldsymbol{a}_{-ft}\equiv
(a_{jt}:j\in\mathcal{J}\setminus\mathcal{J}_{f})$
collects the entry indicators for products owned by other firms.

Proposition~\ref{proposition_1} continues to hold because its inversion argument depends only on the demand system, not on product ownership.
Consequently, for any product $j$ and values of $\boldsymbol{x}_{t}$ such that
$\overline{P}_{j}(\boldsymbol{x}_{t})>0$, the product-level selection-bias
function remains
$\mathbb{E}(\xi_{jt}\mid\boldsymbol{x}_{t},a_{jt}=1)$.

At the entry stage, a firm's private information must be indexed by the
portfolios it may choose. Let
$\boldsymbol{a}_{f}\equiv(a_j:j\in\mathcal{J}_{f})
\in\{0,1\}^{|\mathcal{J}_{f}|}$ denote a generic portfolio
and define
$\boldsymbol{\eta}_{ft}\equiv
(\eta_{ft}(\boldsymbol{a}_{f}):
\boldsymbol{a}_{f}\in\{0,1\}^{|\mathcal{J}_{f}|})$.
Let $\boldsymbol{x}_{ft}^{\mathrm{fc}}$ collect the components of
$\boldsymbol{x}_{t}$ that enter firm $f$'s portfolio fixed cost, including the
relevant market-level variables and characteristics of the products in
$\mathcal{J}_{f}$. We maintain the analogue of
Assumption~\ref{assumption_1}[c]: the vectors $\boldsymbol{\eta}_{ft}$ are
independent across firms and independent of
$(\boldsymbol{\xi}_{t},\boldsymbol{\omega}_{t},
\boldsymbol{\kappa}_{t},\boldsymbol{x}_{t})$, while their components may be
dependent within a firm. The analogue of Assumption~\ref{assumption_1}[d] is
also maintained, with
$\{\boldsymbol{\eta}_{ft}:f\in\mathcal{F}\}$ replacing
$\{\boldsymbol{\eta}_{jt}:j\in\mathcal{J}\}$. For example, for a two-product firm,
$\boldsymbol{\eta}_{ft}\equiv
(\eta_{ft}(0,0),\eta_{ft}(0,1),\eta_{ft}(1,0),\eta_{ft}(1,1))$, where
$\eta_{ft}(1,0)$ is the private component of fixed cost when the firm offers
the first product but not the second. We normalize the entire fixed cost of the
empty portfolio to zero, including $\eta_{ft}(0,0)=0$ in this example.

Under this information structure, and using $\boldsymbol{\xi}_{t}$ in the
following expected-profit expression as shorthand for both demand and
variable-cost unobservables, the analogue of
equation~\eqref{eq_expected_profit} for a profile of rival product
choices
$\boldsymbol{a}_{-f}\in\{0,1\}^{J-|\mathcal{J}_{f}|}$ is
\begin{equation*}
\begin{aligned}
    \pi_{f}\left(
        \boldsymbol{a}_{f},
        \boldsymbol{a}_{-f},
        \boldsymbol{x}_{t},
        \boldsymbol{\kappa}_{t},
        \boldsymbol{\eta}_{ft}
    \right)
    &=
    \int
    VP_{f}\left(
        \boldsymbol{a}_{f},
        \boldsymbol{a}_{-f},
        \boldsymbol{x}_{t},
        \boldsymbol{\xi}_{t}
    \right)
    \,dF_{f,\xi}\left(
        \boldsymbol{\xi}_{t}
        \mid
        \boldsymbol{\kappa}_{t}
    \right) \\
    &\quad-
    fc\left(
        \boldsymbol{a}_{f},
        \boldsymbol{x}_{ft}^{\mathrm{fc}},
        \boldsymbol{\kappa}_{t},
        \boldsymbol{\eta}_{ft}
    \right).
\end{aligned}
\end{equation*}
Here $F_{f,\xi}(\cdot\mid\boldsymbol{\kappa}_{t})$ denotes firm $f$'s
conditional beliefs about $\boldsymbol{\xi}_{t}$. Maintaining
product-separable variable costs,
$VP_{f}(\boldsymbol{a}_{f},\boldsymbol{a}_{-f},
\boldsymbol{x}_{t},\boldsymbol{\xi}_{t})$
is firm $f$'s indirect variable profit: the sum of the equilibrium variable
profits of the products in $\mathcal{J}_{f}$ offered under
$\boldsymbol{a}_{f}$, evaluated at the Bertrand--Nash equilibrium selected
under the convention in the main text, in which every multi-product firm
jointly sets the prices of its offered products. The
fixed-cost function depends on $\boldsymbol{\eta}_{ft}$ through the component
$\eta_{ft}(\boldsymbol{a}_{f})$ associated with the chosen portfolio.

Given these payoffs, a BNE is characterized, conditional on
$(\boldsymbol{x}_{t},\boldsymbol{\kappa}_{t})$, by one probability distribution
over mutually exclusive product portfolios for each firm, induced by each
firm's maximization of expected profit after averaging over rival firms'
portfolio choices. For any product $j\in\mathcal{J}_{f}$, firm $f$'s conditional
portfolio-choice probabilities induce the latent propensity score for product
$j$:
\[
P_j(\boldsymbol{x}_{t},\boldsymbol{\kappa}_{t})
\equiv
\Pr(a_{jt}=1\mid\boldsymbol{x}_{t},\boldsymbol{\kappa}_{t})
=
\sum_{\boldsymbol{a}_{f}:\,a_j=1}
\Pr(\boldsymbol{a}_{ft}=\boldsymbol{a}_{f}
\mid\boldsymbol{x}_{t},\boldsymbol{\kappa}_{t}).
\]
Because firm $f$'s equilibrium portfolio choice depends on the unobservables
only through $\boldsymbol{\eta}_{ft}$ conditional on
$(\boldsymbol{x}_{t},\boldsymbol{\kappa}_{t})$, the maintained independence of
$\boldsymbol{\eta}_{ft}$ from
$(\boldsymbol{\xi}_{t},\boldsymbol{\omega}_{t},
\boldsymbol{\kappa}_{t},\boldsymbol{x}_{t})$
implies that $a_{jt}$ and $\xi_{jt}$ are conditionally independent given
$(\boldsymbol{x}_{t},\boldsymbol{\kappa}_{t})$. Together with
the maintained analogue of Assumption~\ref{assumption_1}[d],
Proposition~\ref{proposition_3} therefore
continues to hold: for
$\overline{P}_{j}(\boldsymbol{x}_{t})>0$, the selection-bias function
$\lambda_j(\boldsymbol{x}_{t})$ admits the mixture representation in
equation~\eqref{eq_prop_3_mixture}, with the latent propensity score defined
above. This argument requires no conditional independence among the entry
indicators of products owned by the same firm. Multi-product ownership
therefore changes how the latent propensity score for each product is obtained
from firm-level portfolio choices, but not the general selection-bias
representation.

\section{Dynamic game of product entry and exit \label{sec_dynamic_entry_exit}}

This section sketches how our framework extends when forward-looking firms'
product-availability decisions arise from a Markov Perfect Equilibrium (MPE) of
a dynamic game of product entry and exit. We write firm $j$'s fixed and
transition cost as
$fc(a_{jt},a_{j,t-1},\boldsymbol{x}_{jt}^{\mathrm{fc}},
\boldsymbol{\kappa}_{t},\boldsymbol{\eta}_{jt})$. The configurations
$(a_{jt},a_{j,t-1})=(1,0),(0,1),(1,1)$, and $(0,0)$ correspond, respectively,
to entry, exit, continued operation, and continued inactivity; we normalize
$fc(0,0,\boldsymbol{x}_{jt}^{\mathrm{fc}},\boldsymbol{\kappa}_{t},
\boldsymbol{\eta}_{jt})=0$.

\bigskip

\phantomsection\label{assumption_1_dyn}
\noindent \textbf{ASSUMPTION 1-Dyn.} \textit{Suppose that $t$ indexes time
within a market; we suppress the market index. Conditions [a]--[c] of
Assumption~\ref{assumption_1} continue to hold, with
$\boldsymbol{\eta}_{jt}$ representing private information about current
fixed and transition costs. We do not impose condition [d], because the
observed state now contains lagged availability decisions and may therefore be
correlated with the latent market state. The following additional conditions
hold.
\begin{itemize}
    \item [e.] Let
    $\boldsymbol{x}_{t}\equiv
    (\boldsymbol{x}^{ex}_{t},\boldsymbol{a}_{t-1})$
    denote the observed state at period $t$, where
    $\boldsymbol{x}^{ex}_{t}$ collects the exogenous market- and product-level
    observables and
    $\boldsymbol{a}_{t-1}\equiv
    (a_{j,t-1}:j\in\mathcal{J})$
    collects lagged availability decisions.
    \item [f.] The joint process
    $\{(\boldsymbol{x}^{ex}_{t},\boldsymbol{\kappa}_{t})\}_{t}$
    is first-order Markov, with time-invariant components allowed.
    \item [g.] The private-information vectors
    $\boldsymbol{\eta}_{jt}$ are i.i.d. across firms and over time.
    \item [h.] Conditional on the latent market state
    $\boldsymbol{\kappa}_{t}$, the current demand and variable-cost
    unobservables
    $(\boldsymbol{\xi}_{t},\boldsymbol{\omega}_{t})$
    are independent of $\boldsymbol{x}^{ex}_{t}$ and the past market history.
    Consequently,
    $\mathbb{E}(\xi_{jt}\mid\boldsymbol{x}_{t},
    \boldsymbol{\kappa}_{t})
    =\mu_{j}(\boldsymbol{\kappa}_{t})$.
    \qquad $\blacksquare$
\end{itemize}
}

\noindent The conditions in \assumptiondyn{} are standard in the literature on
empirical dynamic games of oligopoly competition
\citep[see][]{aguirregabiria_collard_2021}. Given a candidate profile $P$, let
$P_{it}\equiv P_i(\boldsymbol{x}_{t},\boldsymbol{\kappa}_{t})$ denote firm $i$'s
conditional choice probability (CCP) of choosing $a_{it}=1$ at state
$(\boldsymbol{x}_{t},\boldsymbol{\kappa}_{t})$, and let
$v^{P}_{j}(\boldsymbol{x}_{t},\boldsymbol{\kappa}_{t},
\boldsymbol{\eta}_{jt})$ denote the difference between firm $j$'s
choice-specific value functions associated with $a_{jt}=1$ and $a_{jt}=0$.
This difference equals the difference in current profits plus the discounted
difference in expected continuation values.

\bigskip

\begin{define}
\label{definition_mpe}
    \textbf{Markov Perfect Equilibrium.} Under \assumptiondyn{}, an MPE can be
    represented as a $J$-tuple of CCPs
    $\{P_{jt}\equiv P_j(\boldsymbol{x}_{t},\boldsymbol{\kappa}_{t}):
    j\in\mathcal{J}\}$ that solves the following system of $J$ fixed-point
    equations at every state:
    \begin{equation*}
        P_{jt}
        \; = \;
        \displaystyle \int
        \mathbb{1} \left\{
            v^{P}_{j}(\boldsymbol{x}_{t}, \boldsymbol{\kappa}_{t}, \boldsymbol{\eta}_{jt})
            \geq 0
        \right\}
        \;
        dF_{\boldsymbol{\eta}} \left( \boldsymbol{\eta}_{jt} \right).
        \qquad \blacksquare
    \end{equation*}
\end{define}

\noindent At an MPE, $P_{jt}$ is also the latent propensity score for firm
$j$'s product-availability decision. Conditional on
$(\boldsymbol{x}_{t},\boldsymbol{\kappa}_{t})$, this decision is a function of
$\boldsymbol{\eta}_{jt}$; hence, Assumption~\ref{assumption_1}[c] implies that
$a_{jt}$ and $\xi_{jt}$ are conditionally independent given
$(\boldsymbol{x}_{t},\boldsymbol{\kappa}_{t})$. Together with
\assumptiondyn{}[h], the conditioning argument underlying
Proposition~\ref{proposition_3} yields
\begin{align*}
\overline{P}_j(\boldsymbol{x}_{t})
&\equiv \Pr(a_{jt}=1\mid\boldsymbol{x}_{t})
=\int P_j(\boldsymbol{x}_{t},\boldsymbol{\kappa})
 \,dF_{\kappa\mid x}(\boldsymbol{\kappa}\mid\boldsymbol{x}_{t}),\\
\lambda_j(\boldsymbol{x}_{t})
&=\int
\left[
\frac{P_j(\boldsymbol{x}_{t},\boldsymbol{\kappa})}
{\overline{P}_j(\boldsymbol{x}_{t})}
\right]
\mu_j(\boldsymbol{\kappa})
 \,dF_{\kappa\mid x}(\boldsymbol{\kappa}\mid\boldsymbol{x}_{t}),
\qquad \overline{P}_j(\boldsymbol{x}_{t})>0,
\end{align*}
where $F_{\kappa\mid x}(\cdot\mid\boldsymbol{x}_{t})$ is the
conditional distribution of the latent market state given the observed state. The
conditional rather than marginal distribution is required because
$\boldsymbol{x}_{t}$ contains lagged availability decisions and may therefore
be correlated with $\boldsymbol{\kappa}_{t}$.

\section{Proofs of Propositions \label{app:proofs}}

\subsection{Proof of Proposition \ref{proposition_1} \label{app:proof_proposition_1}}

Fix a market $t$, its prices and observed demand characteristics, a configuration
$\boldsymbol{a}_{t}$, and a value of $\boldsymbol{\sigma}$. Let
$n\equiv|\mathcal{J}_{t}^{\boldsymbol{a}}|$, and write
$\boldsymbol{\delta}\equiv\boldsymbol{\delta}_{t}^{\boldsymbol{a}}$ and
$\boldsymbol{d}(\boldsymbol{\delta})\equiv
\boldsymbol{d}_{t}^{\boldsymbol{a}}(\boldsymbol{\delta},\boldsymbol{\sigma})$,
suppressing the fixed arguments. If $n=0$, both $\mathbb{R}^{0}$ and
$\mathcal{S}^{\boldsymbol{a}}$ contain only the empty vector, so the result is
immediate. Hence, suppose that $n\geq1$.

For $j\in\{0\}\cup\mathcal{J}_{t}^{\boldsymbol{a}}$, let
$\widetilde{d}_{j}(\boldsymbol{\delta},\boldsymbol{\upsilon})$ denote product
$j$'s conditional demand share implied by equation~\eqref{eq_market_shares},
that is, its logit choice probability given $\boldsymbol{\upsilon}$. Define
\[
    d_j(\boldsymbol{\delta})
    \equiv
    \int
    \widetilde{d}_{j}(\boldsymbol{\delta},\boldsymbol{\upsilon})
    \,dF_{\upsilon}(\boldsymbol{\upsilon}),
    \qquad
    j\in\{0\}\cup\mathcal{J}_{t}^{\boldsymbol{a}}.
\]
For $j\in\mathcal{J}_{t}^{\boldsymbol{a}}$, this is the $j$-th component of
$\boldsymbol{d}(\boldsymbol{\delta})$; $d_0(\boldsymbol{\delta})$ is the
outside share. Pointwise, every conditional share is strictly positive and
\[
    \widetilde{d}_{0}(\boldsymbol{\delta},\boldsymbol{\upsilon})
    +
    \sum_{j\in\mathcal{J}_{t}^{\boldsymbol{a}}}
    \widetilde{d}_{j}(\boldsymbol{\delta},\boldsymbol{\upsilon})
    =1.
\]
The derivatives of the conditional logit shares with respect to
$\boldsymbol{\delta}$ are continuous and uniformly bounded, so differentiation
under the integral is valid and $\boldsymbol{d}$ is continuously differentiable.
Its $n\times n$ Jacobian $\boldsymbol{D}(\boldsymbol{\delta})$ has entries
\[
    D_{ji}(\boldsymbol{\delta})
    =
    \int
    \widetilde{d}_{j}(\boldsymbol{\delta},\boldsymbol{\upsilon})
    \left[
        \mathbb{1}\{i=j\}
        -
        \widetilde{d}_{i}(\boldsymbol{\delta},\boldsymbol{\upsilon})
    \right]
    \,dF_{\upsilon}(\boldsymbol{\upsilon}),
    \qquad
    i,j\in\mathcal{J}_{t}^{\boldsymbol{a}}.
\]
For every nonzero
$\boldsymbol{q}=(q_j:j\in\mathcal{J}_{t}^{\boldsymbol{a}})\in\mathbb{R}^{n}$,
\[
\begin{aligned}
    \boldsymbol{q}^{\prime}\boldsymbol{D}(\boldsymbol{\delta})\boldsymbol{q}
    =
    \int
    \Bigg\{
        \sum_{j\in\mathcal{J}_{t}^{\boldsymbol{a}}}
        \widetilde{d}_{j}(\boldsymbol{\delta},\boldsymbol{\upsilon})q_j^2
        -
        \left[
            \sum_{j\in\mathcal{J}_{t}^{\boldsymbol{a}}}
            \widetilde{d}_{j}(\boldsymbol{\delta},\boldsymbol{\upsilon})q_j
        \right]^2
    \Bigg\}
    \,dF_{\upsilon}(\boldsymbol{\upsilon})
    >0.
\end{aligned}
\]
Indeed, the Cauchy--Schwarz inequality and the fact that the conditional inside
shares sum to $1-\widetilde{d}_0$ imply that the expression in braces is at
least
$\widetilde{d}_0\sum_{j\in\mathcal{J}_{t}^{\boldsymbol{a}}}
\widetilde{d}_jq_j^2>0$, where the arguments of the conditional shares are
suppressed. Thus, $\boldsymbol{D}(\boldsymbol{\delta})$ is symmetric positive
definite and therefore nonsingular everywhere.

For any $\boldsymbol{\delta}^{(0)}\neq\boldsymbol{\delta}^{(1)}$, let
$\boldsymbol{q}\equiv\boldsymbol{\delta}^{(1)}-\boldsymbol{\delta}^{(0)}$.
The fundamental theorem of calculus gives
\[
    \boldsymbol{q}^{\prime}
    \left[
        \boldsymbol{d}(\boldsymbol{\delta}^{(1)})
        -
        \boldsymbol{d}(\boldsymbol{\delta}^{(0)})
    \right]
    =
    \int_0^1
    \boldsymbol{q}^{\prime}
    \boldsymbol{D}(\boldsymbol{\delta}^{(0)}+\tau\boldsymbol{q})
    \boldsymbol{q}
    \,d\tau
    >0.
\]
Hence, distinct mean-utility vectors generate distinct market-share vectors, so
$\boldsymbol{d}$ is injective. Moreover, $\boldsymbol{d}(\mathbb{R}^{n})\subseteq
\mathcal{S}^{\boldsymbol{a}}$, because every inside share is positive and their
sum equals $1-d_0(\boldsymbol{\delta})<1$. Let
$\mathcal{I}\equiv\boldsymbol{d}(\mathbb{R}^{n})$. Since the Jacobian is
nonsingular everywhere, the inverse function theorem implies that $\mathcal{I}$
is open in $\mathbb{R}^{n}$ and therefore relatively open in
$\mathcal{S}^{\boldsymbol{a}}$.

To show that $\mathcal{I}$ is also relatively closed, consider any sequence
\[
    \boldsymbol{d}(\boldsymbol{\delta}^{(m)})
    \longrightarrow
    \boldsymbol{s}
    \in
    \mathcal{S}^{\boldsymbol{a}}.
\]
If $\{\boldsymbol{\delta}^{(m)}\}$ were unbounded, then there would be a fixed
$j\in\mathcal{J}_{t}^{\boldsymbol{a}}$ and a subsequence, not relabeled, along
which either $\delta_j^{(m)}\to-\infty$ or
$\delta_j^{(m)}\to+\infty$. In the first case,
$\widetilde{d}_{j}(\boldsymbol{\delta}^{(m)},\boldsymbol{\upsilon})\to0$
pointwise, so dominated convergence implies
$d_j(\boldsymbol{\delta}^{(m)})\to0$, contradicting $s_j>0$. In the second
case, $\widetilde{d}_{0}(\boldsymbol{\delta}^{(m)},\boldsymbol{\upsilon})\to0$
pointwise, so dominated convergence implies
$d_0(\boldsymbol{\delta}^{(m)})\to0$. But convergence of the inside shares also
implies
\[
    d_0(\boldsymbol{\delta}^{(m)})
    \longrightarrow
    1-
    \sum_{j\in\mathcal{J}_{t}^{\boldsymbol{a}}}s_j
    >0,
\]
a contradiction. Therefore, $\{\boldsymbol{\delta}^{(m)}\}$ is bounded. It has
a convergent subsequence, and continuity of $\boldsymbol{d}$ implies that
$\boldsymbol{s}\in\mathcal{I}$. Thus, $\mathcal{I}$ is relatively closed in
$\mathcal{S}^{\boldsymbol{a}}$. The set $\mathcal{S}^{\boldsymbol{a}}$ is convex and hence connected. Since
$\mathcal{I}$ is nonempty and both relatively open and relatively closed in
$\mathcal{S}^{\boldsymbol{a}}$, it follows that
$\mathcal{I}=\mathcal{S}^{\boldsymbol{a}}$. Therefore,
$\boldsymbol{d}_{t}^{\boldsymbol{a}}(\cdot,\boldsymbol{\sigma})$ is a
bijection from $\mathbb{R}^{n}$ onto $\mathcal{S}^{\boldsymbol{a}}$, and
$\left(\boldsymbol{d}_{t}^{\boldsymbol{a}}\right)^{-1}
(\cdot,\boldsymbol{\sigma})$ is well defined on
$\mathcal{S}^{\boldsymbol{a}}$. \qquad $\blacksquare$

\subsection{Proof of Proposition \ref{proposition_3} \label{app:proof_proposition_3}}

Fix $\boldsymbol{x}_{t}$ such that $\overline{P}_{j}(\boldsymbol{x}_{t})>0$,
as in the proposition. By definition,
\begin{equation}
    \lambda_{j}(\boldsymbol{x}_{t})
    \; = \;
    \mathbb{E} \left( \xi_{jt} \mid
    \boldsymbol{x}_{t}, a_{jt}=1 \right)
    \; = \;
    \frac{
        \mathbb{E}\left(
            \xi_{jt} \, a_{jt}
            \mid
            \boldsymbol{x}_{t}
        \right)
    }{
        \Pr\left(
            a_{jt}=1
            \mid
            \boldsymbol{x}_{t}
        \right)
    }.
\label{proof_prop_3_eq_1}
\end{equation} 
Under the equilibrium entry rule, $a_{jt}$ is a function of
$(\boldsymbol{x}_{t},\boldsymbol{\kappa}_{t},\boldsymbol{\eta}_{jt})$.
Assumption~\ref{assumption_1}[c] makes $\boldsymbol{\eta}_{jt}$ independent of
$\xi_{jt}$ conditional on
$(\boldsymbol{x}_{t},\boldsymbol{\kappa}_{t})$; hence, $a_{jt}$ and $\xi_{jt}$
are conditionally independent given
$(\boldsymbol{x}_{t},\boldsymbol{\kappa}_{t})$. It follows that
\begin{equation}
    \begin{array}[c]{rcl}
    \mathbb{E}\left(
        \xi_{jt} \, a_{jt}
        \mid
        \boldsymbol{x}_{t}
    \right)
    & = &
    \mathbb{E}\left(
        \mathbb{E}\left(
            \xi_{jt} \, a_{jt}
            \mid
            \boldsymbol{x}_{t}, \boldsymbol{\kappa}_{t}
        \right)
        \mid
        \boldsymbol{x}_{t}
    \right) \\ \\
    & = &
    \mathbb{E}\left(
        \mathbb{E}\left(
            \xi_{jt}
            \mid
            \boldsymbol{x}_{t}, \boldsymbol{\kappa}_{t}
        \right)
        \mathbb{E}\left(
            a_{jt}
            \mid
            \boldsymbol{x}_{t}, \boldsymbol{\kappa}_{t}
        \right)
        \mid
        \boldsymbol{x}_{t}
    \right) \\ \\
    & = &
    \mathbb{E}\left(
        \mu_{j}(\boldsymbol{\kappa}_{t}) \,
        P_{j}(\boldsymbol{x}_{t}, \boldsymbol{\kappa}_{t})
        \mid
        \boldsymbol{x}_{t}
    \right).
    \end{array}
\label{proof_prop_3_eq_2}
\end{equation}
The first equality is the law of iterated expectations, and the second uses the
conditional independence just established. The third uses the definition of
the latent propensity score,
\[
\mathbb{E}\left(
    a_{jt}
    \mid
    \boldsymbol{x}_{t}, \boldsymbol{\kappa}_{t}
\right)
= P_{j}(\boldsymbol{x}_{t}, \boldsymbol{\kappa}_{t}),
\]
and Assumption~\ref{assumption_1}[d], which implies
\[
\mathbb{E}\left(
    \xi_{jt}
    \mid
    \boldsymbol{x}_{t}, \boldsymbol{\kappa}_{t}
\right)
=
\mathbb{E}\left(
    \xi_{jt}
    \mid
    \boldsymbol{\kappa}_{t}
\right)
= \mu_{j}(\boldsymbol{\kappa}_{t}).
\]
Substituting \eqref{proof_prop_3_eq_2} into \eqref{proof_prop_3_eq_1} and using
$\Pr(a_{jt}=1\mid\boldsymbol{x}_{t})=\overline{P}_{j}(\boldsymbol{x}_{t})$
gives
\begin{equation*}
\lambda_{j}(\boldsymbol{x}_{t})
\; = \;
\frac{
    \mathbb{E}\left(
        \mu_{j}(\boldsymbol{\kappa}_{t}) \,
        P_{j}(\boldsymbol{x}_{t}, \boldsymbol{\kappa}_{t})
        \mid
        \boldsymbol{x}_{t}
    \right)
}{
    \overline{P}_{j}(\boldsymbol{x}_{t})
}.
\end{equation*}
Assumption~\ref{assumption_1}[d] also implies that the conditional distribution of
$\boldsymbol{\kappa}_{t}$ given $\boldsymbol{x}_{t}$ equals its marginal
distribution $F_{\kappa}$. Hence,
\begin{equation*}
    \lambda_{j}(\boldsymbol{x}_{t})
    \; = \;
    \displaystyle
    \int 
    \left[
        \frac{P_{j}(\boldsymbol{x}_{t}, \boldsymbol{\kappa})}
        {\overline{P}_{j}(\boldsymbol{x}_{t})}
    \right]
    \, \mu_{j}(\boldsymbol{\kappa}) \,
    dF_{\kappa}(\boldsymbol{\kappa}).
\end{equation*} 
\noindent This proves Proposition \ref{proposition_3}. \qquad $\blacksquare$

\section{Identification under normal heterogeneity\label{app: cont kappa}}

This section develops the first of the two identification routes introduced in
Section~\ref{sec:identification}. Here, the latent market state coincides with
the jointly normal vector of demand unobservables,
$\boldsymbol{\kappa}_{t}=\boldsymbol{\xi}_{t}$. Under the index, exclusion, and
smoothness restrictions stated below, a multivariate version of Stein's lemma
expresses the selection-bias function $\lambda_j(\boldsymbol{x}_{t})$ as a
linear combination of derivatives of the logarithm of the ordinary propensity
score.

\subsection{Setup}

We specialize the general framework of Section~\ref{sec:identification} by setting
$\boldsymbol{\kappa}_{t}=\boldsymbol{\xi}_{t}\equiv
(\xi_{1t},\ldots,\xi_{Jt})^{\prime}$, so that the demand unobservables exhaust
the latent market state and are common knowledge at entry. The conditional-mean
function in Proposition~\ref{proposition_3} then satisfies
$\mu_j(\boldsymbol{\kappa})=
\mathbb{E}(\xi_{jt}\mid\boldsymbol{\kappa}_{t}=\boldsymbol{\kappa})=\kappa_j$,
where $\kappa_j$ is the $j$-th component of $\boldsymbol{\kappa}$. Hence,
equation~\eqref{eq_prop_3_mixture} becomes
\begin{equation}
    \lambda_j(\boldsymbol{x}_{t})
    \;=\;
    \frac{1}{\overline{P}_{j}(\boldsymbol{x}_{t})}
    \mathbb{E}\!\left[
    \xi_{jt}P_j(\boldsymbol{x}_{t},\boldsymbol{\xi}_{t})
    \right].
\end{equation}
We assume that the latent propensity score $P_j(\boldsymbol{x}_{t}, \boldsymbol{\xi}_{t})$
depends on $(\boldsymbol{x}_{t}, \boldsymbol{\xi}_{t})$ only through the vector of product-level
latent indexes $\boldsymbol{\zeta}_{t} = (\zeta_{1t}, \ldots, \zeta_{Jt})^{\prime}$, where
\begin{equation}
    \zeta_{jt} \;\equiv\; \boldsymbol{x}_{t}^{\prime}\boldsymbol{\tau}_{j} + \xi_{jt}.
\label{eq:latent_entry_index}
\end{equation}
That is, $P_j(\boldsymbol{x}_{t}, \boldsymbol{\xi}_{t}) = G_j(\zeta_{1t}, \ldots, \zeta_{Jt})$ for some
function $G_j : \mathbb{R}^J \to [0,1]$. Write $G_{jk}(\boldsymbol{\zeta}) \equiv \partial
G_j(\boldsymbol{\zeta}) / \partial \zeta_k$ for the partial derivative with respect to the $k$-th
argument. The additional assumptions required for this approach are collected below.

\bigskip

\begin{assumption}
\label{assumption_normal}
\begin{itemize}
    \item[\normalfont(i)] \textit{Normality.} The demand unobservables are jointly normal:
        $\boldsymbol{\xi}_{t}
        \sim \mathcal{N}(\boldsymbol{0}, \boldsymbol{\Sigma})$, where
        $\boldsymbol{\Sigma}$ is a positive definite $J \times J$ covariance matrix with
        $(j,k)$ entry $\Sigma_{jk}$.
    \item[\normalfont(ii)] \textit{Smoothness.} For every $j \in \mathcal{J}$, the
        function $G_j(\boldsymbol{\zeta})$ is continuously differentiable in
        $\boldsymbol{\zeta}$. The regularity conditions required to apply Stein's lemma
        and differentiate under the integral sign hold, and
        $\mathbb{E}\,|G_{jk}(\boldsymbol{\zeta}_{t})| < \infty$ for all
        $j,k \in \mathcal{J}$.
    \item[\normalfont(iii)] \textit{Index relevance and exclusion.} For every product
        $k \in \mathcal{J}$, let $x_{k1t}$ denote a continuously distributed component
        of $\boldsymbol{x}_{t}$ associated with product $k$. It enters $\zeta_{kt}$ with
        coefficient $\tau_{k1} \neq 0$, is excluded from $\zeta_{it}$ for every
        $i \neq k$, and has support, conditional on the remaining regressors, containing
        an open interval. Thus, $x_{k1t}$ affects every latent propensity score only
        through $\zeta_{kt}$. \qquad $\blacksquare$
\end{itemize}
\end{assumption}

\subsection{Multivariate Stein's lemma and selection-bias representation}

We use the following standard result. If $\boldsymbol{\xi}_{t} \sim \mathcal{N}(\boldsymbol{0},
\boldsymbol{\Sigma})$ and $m : \mathbb{R}^J \to \mathbb{R}$ is continuously differentiable
with the required integrability conditions, then
\begin{equation}
    \mathbb{E}\!\left(\boldsymbol{\xi}_{t} \; m(\boldsymbol{\xi}_{t})\right)
    \;=\;
    \boldsymbol{\Sigma} \; \mathbb{E}\!\left(\nabla_{\!\boldsymbol{\xi}}\, m(\boldsymbol{\xi}_{t})
    \right).
\end{equation}
Componentwise, for every $j \in \mathcal{J}$,
\begin{equation}
    \mathbb{E}\!\left(\xi_{jt} \; m(\boldsymbol{\xi}_{t})\right)
    \;=\;
    \sum_{k=1}^{J} \Sigma_{jk} \;
    \mathbb{E}\!\left(\frac{\partial m(\boldsymbol{\xi}_{t})}{\partial \xi_{kt}}\right).
\label{eq:stein_component}
\end{equation}

Applying this lemma to the latent propensity score yields the following
representation of the selection-bias function.

\begin{prop}
\label{proposition_normal}
Under the index structure above and Assumptions~\ref{assumption_1} and~\ref{assumption_normal}, for every $j \in \mathcal{J}$ and every
$\boldsymbol{x}_{t}$ such that $\overline{P}_{j}(\boldsymbol{x}_{t}) > 0$,
\begin{equation}
    \lambda_j(\boldsymbol{x}_{t})
    \;=\;
    \mathbb{E}\!\left(\xi_{jt} \mid \boldsymbol{x}_{t},\; a_{jt} = 1\right)
    \;=\;
    \sum_{k=1}^{J}
    \frac{\Sigma_{jk}}{\tau_{k1}}
    \;\frac{\partial \log \overline{P}_{j}(\boldsymbol{x}_{t})}{\partial x_{k1t}}.
    \qquad \blacksquare
\end{equation}
\end{prop}

\bigskip

\begin{proof}
\textit{Step 1: Selection representation.}
By definition,
$\lambda_j(\boldsymbol{x}_{t})=
\mathbb{E}(\xi_{jt}a_{jt}\mid\boldsymbol{x}_{t})/
\overline{P}_{j}(\boldsymbol{x}_{t})$.
The law of iterated expectations and the definition of the latent propensity score give
$\mathbb{E}(\xi_{jt}a_{jt}\mid\boldsymbol{x}_{t})=
\mathbb{E}[\xi_{jt}P_j(\boldsymbol{x}_{t},\boldsymbol{\xi}_{t})
\mid\boldsymbol{x}_{t}]$. Assumption~\ref{assumption_1}[d] makes the
distribution of $\boldsymbol{\xi}_{t}$ independent of $\boldsymbol{x}_{t}$.
Therefore,
\begin{equation}
    \lambda_j(\boldsymbol{x}_{t})
    \;=\;
    \frac{
        \mathbb{E}\!\left(\xi_{jt}\; P_j(\boldsymbol{x}_{t}, \boldsymbol{\xi}_{t})\right)
    }{
        \overline{P}_{j}(\boldsymbol{x}_{t})
    }.
\label{eq:proof_bayes}
\end{equation}
Here and below, expectations integrate over the marginal distribution of
$\boldsymbol{\xi}_{t}$, with $\boldsymbol{x}_{t}$ held fixed.

\textit{Step 2: Stein's lemma.}
Apply equation~\eqref{eq:stein_component} with $m(\boldsymbol{\xi}_{t}) = P_j(\boldsymbol{x}_{t},
\boldsymbol{\xi}_{t}) = G_j(\zeta_{1t}, \ldots, \zeta_{Jt})$. Because
$\xi_{kt}$ enters only $\zeta_{kt}$, with unit coefficient, the chain rule gives
\begin{equation}
    \frac{\partial P_j(\boldsymbol{x}_{t}, \boldsymbol{\xi}_{t})}{\partial \xi_{kt}}
    \;=\; G_{jk}(\zeta_{1t}, \ldots, \zeta_{Jt}).
\end{equation}
Stein's lemma then gives
\begin{equation}
    \mathbb{E}\!\left(\xi_{jt}\; P_j(\boldsymbol{x}_{t}, \boldsymbol{\xi}_{t})\right)
    \;=\;
    \sum_{k=1}^{J} \Sigma_{jk}\;
    \mathbb{E}\!\left(G_{jk}(\zeta_{1t}, \ldots, \zeta_{Jt})\right).
\label{eq:proof_stein_applied}
\end{equation}

\textit{Step 3: Connecting $\boldsymbol{\xi}_{t}$-derivatives to $\boldsymbol{x}_{t}$-derivatives.}
By Assumption~\ref{assumption_normal}(iii), $x_{k1t}$ affects
$P_j(\boldsymbol{x}_{t},\boldsymbol{\xi}_{t})$ only through $\zeta_{kt}$, and
$\partial\zeta_{kt}/\partial x_{k1t}=\tau_{k1}$. The chain rule therefore gives
\begin{equation}
    \frac{\partial P_j(\boldsymbol{x}_{t}, \boldsymbol{\xi}_{t})}{\partial x_{k1t}}
    \;=\;
    \tau_{k1} \; G_{jk}(\zeta_{1t}, \ldots, \zeta_{Jt}).
\end{equation}
Moreover, Assumption~\ref{assumption_1}[d] implies
$\overline{P}_{j}(\boldsymbol{x}_{t})=
\mathbb{E}[P_j(\boldsymbol{x}_{t},\boldsymbol{\xi}_{t})]$, where the
distribution of $\boldsymbol{\xi}_{t}$ does not depend on
$\boldsymbol{x}_{t}$. Assumption~\ref{assumption_normal}(ii) permits
differentiation under this expectation. Taking expectations of the preceding
identity therefore gives
\begin{equation}
    \mathbb{E}\!\left(G_{jk}(\zeta_{1t}, \ldots, \zeta_{Jt})\right)
    \;=\;
    \frac{1}{\tau_{k1}}
    \;\frac{\partial \overline{P}_{j}(\boldsymbol{x}_{t})}{\partial x_{k1t}}.
\label{eq:proof_link}
\end{equation}

\textit{Step 4: Final expression.}
Substituting \eqref{eq:proof_stein_applied} and \eqref{eq:proof_link} into
\eqref{eq:proof_bayes} yields
\begin{equation}
    \lambda_j(\boldsymbol{x}_{t})
    \;=\;
    \frac{1}{\overline{P}_{j}(\boldsymbol{x}_{t})}
    \sum_{k=1}^{J}
    \frac{\Sigma_{jk}}{\tau_{k1}}
    \;\frac{\partial \overline{P}_{j}(\boldsymbol{x}_{t})}{\partial x_{k1t}}
    \;=\;
    \sum_{k=1}^{J}
    \frac{\Sigma_{jk}}{\tau_{k1}}
    \;\frac{\partial \log \overline{P}_{j}(\boldsymbol{x}_{t})}{\partial x_{k1t}},
\end{equation}
where the second equality uses
$\partial\log\overline{P}_{j}/\partial x=
(1/\overline{P}_{j})\,\partial\overline{P}_{j}/\partial x$, which is valid
because $\overline{P}_{j}(\boldsymbol{x}_{t})>0$.
\end{proof}

\subsection{Regression representation}

Define the composite parameters $\varrho_{jk} \equiv \Sigma_{jk} / \tau_{k1}$. Proposition~%
\ref{proposition_normal} implies that, for observations with $a_{jt} = 1$, the demand
equation takes the form
\begin{equation}
    d_{jt}^{-1}\!\left(\boldsymbol{s}_{t}^{\boldsymbol{a}}, \boldsymbol{\sigma}\right)
    \;=\;
    \alpha \, p_{jt}
    + \boldsymbol{x}_{jt}^{D \prime}\boldsymbol{\beta}
    + \sum_{k=1}^{J} \varrho_{jk} \;
      \frac{\partial \log \overline{P}_{j}(\boldsymbol{x}_{t})}{\partial x_{k1t}}
    + \widetilde{\xi}_{jt},
\label{eq:demand_normal_corrected}
\end{equation}
where $\mathbb{E}(\widetilde{\xi}_{jt} \mid \boldsymbol{x}_{t}, a_{jt} = 1) = 0$. The correction therefore uses derivatives of the logarithm of the ordinary propensity score
$\overline{P}_{j}(\boldsymbol{x}_{t})$ with respect to the continuous regressors $x_{k1t}$
that shift the corresponding latent indexes $\zeta_{kt}$. The coefficients
$\varrho_{jk}=\Sigma_{jk}/\tau_{k1}$ combine the covariance between $\xi_{jt}$ and
$\xi_{kt}$ with the inverse index loading $1/\tau_{k1}$.

\section{Linearized common-weight bootstrap\label{app:bootstrap}}

The estimator in Section~\ref{sec:estimation} has two steps. The first estimates the type probabilities and type-specific propensity scores by sieve MLE. The second uses these estimates to construct the selection-function regressors and applies GMM to estimate the demand parameters $\boldsymbol{\theta}$ jointly with the stacked product-specific selection-function coefficient vector $\widetilde{\boldsymbol{\mu}}$ defined in Section~\ref{sec:identification_finite_mixture}. Under standard regularity conditions for semiparametric two-step estimators \citep{das2003,newey_2009}, the resulting estimator is asymptotically linear. We describe a linearized common-weight bootstrap that accounts for first-step estimation error and its covariance with the second-step moments without re-estimating the first-step MLE in every replication. General bootstrap results for semiparametric estimators with preliminary nonparametric components are provided by \cite{chen_linton_van_keilegom_2003}; \cite{armstrong_2014} and \cite{goncalves_2023} study related computationally efficient bootstrap procedures for multi-step estimators.

Fix $L$ and $N_X$, and treat the resulting finite-dimensional first-step specification as correctly specified. We present the likelihood-based update for the finite-mixture estimator. For another correctly specified finite-dimensional likelihood-based first step, its log-likelihood and smooth mapping into the selection-function regressors can be substituted below; a non-likelihood first step instead requires its corresponding first-order influence-function update. Let $\boldsymbol{\psi}$ combine $\boldsymbol{\gamma}$ with an unconstrained reparameterization of the type probabilities. In particular, set $c_1=0$ and define $c_{\ell}\equiv\log(\widetilde{f}_{\ell}/\widetilde{f}_{1})$ for $\ell=2,\ldots,L$. The corresponding softmax map is
\[
\widetilde{f}_{1}=\frac{1}{1+\sum_{k=2}^{L}\exp(c_k)},
\qquad
\widetilde{f}_{\ell}=\frac{\exp(c_{\ell})}{1+\sum_{k=2}^{L}\exp(c_k)},
\qquad \ell=2,\ldots,L.
\]
Thus, $\boldsymbol{\psi}\equiv(c_2,\ldots,c_L,\boldsymbol{\gamma}^{\prime})^{\prime}$ can be updated in Euclidean coordinates, while the softmax map preserves the simplex constraint on $\widetilde{\boldsymbol{f}}$. Because the type probabilities are pairwise distinct, their full-sample labeling is locally stable. We retain those labels in every bootstrap replication; the probability that a first-order update crosses a labeling boundary converges to zero as $T\to\infty$.

Let $\ell_t^{\mathrm{1st}}(\boldsymbol{\psi})$ denote market $t$'s contribution to the first-step log-likelihood in equation~\eqref{eq:finite_mixture_logit}, and let $\dot\ell_t^{\mathrm{1st}}(\boldsymbol{\psi})\equiv\partial\ell_t^{\mathrm{1st}}(\boldsymbol{\psi})/\partial\boldsymbol{\psi}$ be its score. Also let $\widehat{\boldsymbol{m}}_t(\boldsymbol{\theta},\widetilde{\boldsymbol{\mu}};\boldsymbol{\psi})$ denote the stacked second-step GMM moment contribution of market $t$, aggregating over the selected products in that market. Define the outer-product-of-gradients estimator of the information matrix by
\begin{equation*}
\widehat{\boldsymbol{H}}
\equiv
\frac{1}{T}\sum_{t=1}^{T}
\dot\ell_t^{\mathrm{1st}}(\widehat{\boldsymbol{\psi}})
\dot\ell_t^{\mathrm{1st}}(\widehat{\boldsymbol{\psi}})^{\prime}.
\end{equation*}
Under correct specification and a nonsingular Fisher information matrix, $\widehat{\boldsymbol{H}}$ consistently estimates both the per-market Fisher information and the negative expected Hessian of a market-level log-likelihood contribution, or equivalently of the average first-step log-likelihood.

\begin{enumerate}
    \item \textbf{Step 1: Draw common bootstrap weights.} For replication $b$, draw multinomial weights $\{w_t^{\ast(b)}:t=1,\ldots,T\}$ satisfying $\sum_{t=1}^{T}w_t^{\ast(b)}=T$, equivalently by sampling markets with replacement. Use the same weights in both steps.

    \item \textbf{Step 2: Update the first step.} Construct the one-step Fisher-scoring update
    \begin{equation*}
    \widehat{\boldsymbol{\psi}}^{\ast(b)}
    =
    \widehat{\boldsymbol{\psi}}
    +
    \widehat{\boldsymbol{H}}^{-1}
    \left[
    \frac{1}{T}\sum_{t=1}^{T}
    \big(w_t^{\ast(b)}-1\big)
    \dot\ell_t^{\mathrm{1st}}(\widehat{\boldsymbol{\psi}})
    \right].
    \end{equation*}
    This is the first-order approximation to re-estimating the first-step likelihood on the weighted bootstrap sample. Recover $\widehat{\widetilde{\boldsymbol{f}}}^{\ast(b)}$ from the updated log ratios using the softmax map; the remaining components give $\widehat{\boldsymbol{\gamma}}^{\ast(b)}$.

    \item \textbf{Step 3: Reconstruct the selection-function regressors.} First compute the bootstrap type-specific and ordinary propensity scores:
    \begin{align*}
    \widehat{\widetilde{P}}_{j,\ell}^{\ast(b)}(\boldsymbol{x}_t)
    &=
    \Lambda\!\left(\boldsymbol{b}_t^{\prime}\widehat{\boldsymbol{\gamma}}_{j,\ell}^{\ast(b)}\right),\\
    \widehat{\overline{P}}_j^{\ast(b)}(\boldsymbol{x}_t)
    &\equiv
    \sum_{\ell=1}^{L}\widehat{\widetilde{P}}_{j,\ell}^{\ast(b)}(\boldsymbol{x}_t)\widehat{\widetilde{f}}_{\ell}^{\ast(b)}.
    \end{align*}
    Then construct
    \begin{equation*}
    \widehat r_{j\ell t}^{\ast(b)}
    =
    \frac{
    \widehat{\widetilde{P}}_{j,\ell}^{\ast(b)}(\boldsymbol{x}_t)
    -
    \widehat{\widetilde{P}}_{j,L}^{\ast(b)}(\boldsymbol{x}_t)
    }
    {\widehat{\overline{P}}_j^{\ast(b)}(\boldsymbol{x}_t)}
    \widehat{\widetilde{f}}_{\ell}^{\ast(b)},
    \qquad \ell=1,\ldots,L-1.
    \end{equation*}
    \item \textbf{Step 4: Re-estimate the second step.} Define the weighted bootstrap moment vector
    \begin{equation*}
    \widehat{\boldsymbol{g}}^{\ast(b)}
    (\boldsymbol{\theta},\widetilde{\boldsymbol{\mu}})
    =
    \frac{1}{T}\sum_{t=1}^{T}
    w_t^{\ast(b)}
    \widehat{\boldsymbol{m}}_t
    (\boldsymbol{\theta},\widetilde{\boldsymbol{\mu}};
    \widehat{\boldsymbol{\psi}}^{\ast(b)}).
    \end{equation*}
    Obtain $(\widehat{\boldsymbol{\theta}}^{\ast(b)},\widehat{\widetilde{\boldsymbol{\mu}}}^{\ast(b)})$ by minimizing
    \[
    \widehat{\boldsymbol{g}}^{\ast(b)}
    (\boldsymbol{\theta},\widetilde{\boldsymbol{\mu}})^{\prime}
    \widehat{\boldsymbol{W}}^{\ast(b)}
    \widehat{\boldsymbol{g}}^{\ast(b)}
    (\boldsymbol{\theta},\widetilde{\boldsymbol{\mu}}),
    \]
    where $\widehat{\boldsymbol{W}}^{\ast(b)}$ is constructed by the same rule as the weighting matrix in the original estimator. For a specification without an estimated first step, Steps~2--3 are omitted and Step~4 uses the common bootstrap weights directly.
\end{enumerate}

\noindent Repeating these steps for $b=1,\ldots,B_{\mathrm{boot}}$ yields draws $\{\widehat{\boldsymbol{\theta}}^{\ast(b)},\widehat{\widetilde{\boldsymbol{\mu}}}^{\ast(b)}\}_{b=1}^{B_{\mathrm{boot}}}$. We use their sample variance-covariance matrix to estimate that of the second-step estimator. For the Hausman-type comparison of specifications $A$ and $B$ in Section~\ref{sec:estimating_L}, let $\widehat{\boldsymbol{\theta}}_q^A$ and $\widehat{\boldsymbol{\theta}}_q^B$ denote estimates of the same $q$-dimensional subvector of $\boldsymbol{\theta}$, with $q=\dim(\boldsymbol{\theta})$ for the full-vector test. Apply the respective procedures to the same sample of markets, using the same market-level weights in every replication, and define
\begin{align*}
\widehat{\boldsymbol{\Delta}}_{AB}^{\ast(b)}
&\equiv
\widehat{\boldsymbol{\theta}}_q^{B,\ast(b)}
-\widehat{\boldsymbol{\theta}}_q^{A,\ast(b)},
&
\overline{\boldsymbol{\Delta}}_{AB}^{\ast}
&\equiv
\frac{1}{B_{\mathrm{boot}}}
\sum_{b=1}^{B_{\mathrm{boot}}}
\widehat{\boldsymbol{\Delta}}_{AB}^{\ast(b)},\\
\widehat{\boldsymbol{V}}_{AB}
&\equiv
\frac{1}{B_{\mathrm{boot}}-1}
\sum_{b=1}^{B_{\mathrm{boot}}}
\big(\widehat{\boldsymbol{\Delta}}_{AB}^{\ast(b)}-\overline{\boldsymbol{\Delta}}_{AB}^{\ast}\big)
\big(\widehat{\boldsymbol{\Delta}}_{AB}^{\ast(b)}-\overline{\boldsymbol{\Delta}}_{AB}^{\ast}\big)^{\prime}.
\end{align*}
The common weights preserve the covariance between the two estimators, so $\widehat{\boldsymbol{V}}_{AB}$ consistently estimates the variance-covariance matrix of $\widehat{\boldsymbol{\theta}}_q^B-\widehat{\boldsymbol{\theta}}_q^A$ and is used in equation~\eqref{eq:hausman_general}.

First-order validity requires the usual smoothness, moment, and rank conditions for the two-step estimator. In particular, conditional on fixed $L$ and $N_X$, the finite-dimensional first-step specification must be correctly specified, its score must be sufficiently smooth, and its Fisher information matrix must be nonsingular. The type probabilities must be pairwise distinct, and the mapping from the first-step parameters into the selection-function regressors must be smooth. For the finite-mixture estimator, the ordinary propensity scores $\overline{P}_j(\boldsymbol{x}_t)$ must also be bounded away from zero uniformly over the support of $\boldsymbol{x}_t$. Under these conditions, the one-step update is first-order equivalent to re-estimating the first step on each weighted sample. Writing $\widehat{\boldsymbol{\vartheta}}\equiv(\widehat{\boldsymbol{\theta}}^{\prime},\widehat{\widetilde{\boldsymbol{\mu}}}^{\prime})^{\prime}$, consistency of the simulated variance additionally requires conditional uniform integrability of $T\lVert\widehat{\boldsymbol{\vartheta}}^{\ast}-\widehat{\boldsymbol{\vartheta}}\rVert^2$ and $B_{\mathrm{boot}}\to\infty$. For a Hausman-type comparison, these conditions must hold jointly for the fixed pair of specifications. The procedure conditions on each specification's $L$ and $N_X$ and therefore does not incorporate uncertainty from selecting either.

\section{ANOVA--Sobol decomposition of entry payoffs \label{app:sobol}}

We construct airline-specific ANOVA--Sobol decompositions of the net entry-profit index before the private shock and of its expected-variable-profit and fixed-cost components. Each decomposition separates variation attributable to observable market characteristics, latent market types, and their interaction. Throughout this section, ``unobservables'' refers to latent-type heterogeneity and its interaction with observables, not to the private entry shock, which is excluded from the index. We do not pool across airlines.

\subsection{Setup and entry-payoff objects}

Under the scalar private-entry-shock specification introduced in Section~\ref{sec:identification_costs}, the type-specific propensity scores satisfy $\widetilde{P}_{j,\ell}(\boldsymbol{x}_{t}) = F_{\eta}\!\left(\left[VP^{P}_{j\ell}(\boldsymbol{x}_{t}) - fc_{j\ell}(\boldsymbol{x}_{jt}^{\mathrm{fc}})\right]/\sigma_{\eta_j}\right)$, where $VP^{P}_{j\ell}(\boldsymbol{x}_{t})$ is airline $j$'s expected variable profit at latent market type $\ell$ and market characteristics $\boldsymbol{x}_{t}$, $fc_{j\ell}(\boldsymbol{x}_{jt}^{\mathrm{fc}})$ is the component of fixed entry cost excluding $\sigma_{\eta_j}\eta_{jt}$, and $\sigma_{\eta_j}$ is the airline-specific scale of the private entry shock. Consistent with the logistic first-step specification in Section~\ref{sec:estimation}, we take $F_{\eta}=\Lambda$ in the empirical fixed-cost recovery and work with fitted propensity scores bounded away from zero and one. We suppress hats on plug-in objects except where they clarify the fixed-cost recovery. Inverting gives
\begin{equation}\label{eq:entry_inversion}
    VP^{P}_{j\ell}(\boldsymbol{x}_{t}) - fc_{j\ell}(\boldsymbol{x}_{jt}^{\mathrm{fc}})
    = \sigma_{\eta_j} \, \Lambda^{-1}\!\left(\widetilde{P}_{j,\ell}(\boldsymbol{x}_{t})\right).
\end{equation}
For each airline $j$, latent type $\ell=1,\ldots,L$, and non-directional entry market $t$, define
\[
VP_{j\ell t} \equiv VP^{P}_{j\ell}(\boldsymbol{x}_{t}), \qquad fc_{j\ell t} \equiv fc_{j\ell}(\boldsymbol{x}_{jt}^{\mathrm{fc}}), \qquad \Pi_{j\ell t} \equiv VP_{j\ell t}-fc_{j\ell t}.
\]
Thus, $\Pi_{j\ell t}$ is the type-specific net entry-profit index before the private shock, rather than the realized active-firm profit $\Pi_{jt}$ defined in Section~\ref{sec:model}. Each object is a deterministic function of $\boldsymbol{x}_{t}$ and $\ell$ through the type-specific propensity scores, demand and marginal-cost parameters, and type-specific conditional means, and is defined for every market in which airline $j$ is a potential entrant---not only for markets in which it is observed to enter.

\subsection{Computing expected variable profits}

We compute $VP^{P}_{j\ell}(\boldsymbol{x}_{t})$ as in Section~\ref{sec:identification_costs}, averaging airline $j$'s equilibrium variable profit over its rivals' entry configurations $\boldsymbol{a}_{-j}\in\{0,1\}^{J-1}$:
\begin{equation}\label{eq:sobol_vpp}
    VP^{P}_{j\ell}(\boldsymbol{x}_{t})
    = \sum_{\boldsymbol{a}_{-j} \in \{0,1\}^{J-1}}
    VP_{j\ell}(a_j{=}1, \boldsymbol{a}_{-j}, \boldsymbol{x}_{t})
    \prod_{i \neq j}
    \widetilde{P}_{i,\ell}(\boldsymbol{x}_{t})^{a_i}
    \left[1 - \widetilde{P}_{i,\ell}(\boldsymbol{x}_{t})\right]^{1-a_i}.
\end{equation}
Within this subsection, let $d\in\{1,2\}$ index the two directional demand markets associated with the non-directional entry market $t$. Under the type-level restriction used for fixed-cost recovery, for each configuration $\boldsymbol{a}$ and latent type $\ell$ we set $\xi_{itd}=\widetilde{\mu}_{i,\ell}$ and $\omega_{itd}=\widetilde{\mu}^{\text{mc}}_{i,\ell}$ for every airline $i$. For each direction $d$, we then solve the Bertrand--Nash equilibrium of the nested logit demand system in equation~\eqref{eq:empirical_nested_logit}. Let $\boldsymbol{x}_{itd}^{D}$ and $\boldsymbol{x}_{itd}^{\mathrm{mc}}$ denote the demand and marginal-cost covariate vectors specified in equations~\eqref{eq:empirical_nested_logit} and~\eqref{eq:mc_2}, respectively. Airline $i$'s mean utility is $\delta_{i\ell td}=\alpha p_{i\ell td}+(\boldsymbol{x}_{itd}^{D})^{\prime}\boldsymbol{\beta}+\widetilde{\mu}_{i,\ell}$, and its marginal cost is $mc_{i\ell td}=\exp((\boldsymbol{x}_{itd}^{\mathrm{mc}})^{\prime}\boldsymbol{\varphi}+\widetilde{\mu}^{\text{mc}}_{i,\ell})$. Suppressing the dependence of equilibrium prices and shares on $\boldsymbol{a}$, equilibrium prices solve
\begin{equation*}
    p_{i\ell td} + \frac{1-\sigma}{\alpha\big(1-\sigma s_{i\ell td|g}-(1-\sigma)s_{i\ell td}\big)} = mc_{i\ell td},
    \qquad \forall\, i \text{ active in } \boldsymbol{a},\quad d\in\{1,2\},
\end{equation*}
which is equation~\eqref{eq:mc_1} evaluated at the type-$\ell$ primitives. For each active airline $j$, configuration-level variable profit is
\begin{equation*}
VP_{j\ell}(\boldsymbol{a},\boldsymbol{x}_{t})
\equiv
\sum_{d=1}^{2}H_{td}s_{j\ell td}
\big(p_{j\ell td}-mc_{j\ell td}\big),
\end{equation*}
where $H_{td}$ is the size of directional demand market $d$. As an application-specific approximation, the ``Others'' group enters each direction's nest inclusive value with its demand index evaluated at a zero demand unobservable and at its observed price, which is held fixed rather than re-optimized. Because entry is non-directional, we restrict the decomposition to route-quarters for which both directional demand markets are observed.

With $J=6$ modeled airlines, equation~\eqref{eq:sobol_vpp} sums over $2^5=32$ rival configurations per focal airline. Across all six focal airlines, these are generated by the 63 nonempty configurations for each type and route direction. For a given market, we restrict the sum to profiles of its potential entrants; equivalently, every profile in which a nonpotential airline enters receives zero weight.

\subsection{Estimation of \texorpdfstring{$\sigma_{\eta_j}$}{sigma\_eta\_j} and recovery of fixed costs}

Given $VP^{P}_{j\ell}(\boldsymbol{x}_{t})$ from the previous step, we estimate the airline-specific scale parameter $\sigma_{\eta_j}$ by exploiting equation~\eqref{eq:entry_inversion}. Since $\Lambda^{-1}(\widetilde{P}_{j,\ell}(\boldsymbol{x}_{t}))=\boldsymbol{b}_{t}^{\prime}\boldsymbol{\gamma}_{j,\ell}$ is the fitted first-step entry index, equation~\eqref{eq:entry_inversion} can be written as
\begin{equation}\label{eq:sobol_sigma_eta}
    VP^{P}_{j\ell}(\boldsymbol{x}_{t})
    = \sigma_{\eta_j}\,(\boldsymbol{b}_{t}^{\prime}\boldsymbol{\gamma}_{j,\ell})
    + fc_{j\ell}(\boldsymbol{x}_{jt}^{\mathrm{fc}}).
\end{equation}
We treat equation~\eqref{eq:sobol_sigma_eta} as a regression of $VP^{P}_{j\ell}(\boldsymbol{x}_{t})$ on the first-step entry index $\boldsymbol{b}_{t}^{\prime}\boldsymbol{\gamma}_{j,\ell}$, with the fixed-cost component $fc_{j\ell}(\boldsymbol{x}_{jt}^{\mathrm{fc}})$ as the unobserved regression component. The entry index is generally endogenous in this regression because it is correlated with $fc_{j\ell}(\boldsymbol{x}_{jt}^{\mathrm{fc}})$. However, under the exclusion restriction that fixed costs depend only on $j$'s own characteristics $\boldsymbol{x}_{jt}^{\mathrm{fc}}$, functions of the components of $\boldsymbol{x}_{t}$ that are excluded from $\boldsymbol{x}_{jt}^{\mathrm{fc}}$ serve as instruments for $\boldsymbol{b}_{t}^{\prime}\boldsymbol{\gamma}_{j,\ell}$, provided they shift expected variable profits while remaining excluded from fixed costs. 

In the application, we use as instruments, for each of airline $j$'s six competitors $m$ (the five other modeled airlines and the Others group), competitor $m$'s share of non-stop activity at the market's endpoints: the sum, over the origin and destination airports, of $m$'s hub size at the airport divided by the combined hub size of all carriers at that airport, set to zero unless $m$ is a potential entrant in the market. Of these six competitors, the five other modeled airlines enter expected variable profits through their type-specific propensity scores, whereas the Others group is not assigned a type-specific propensity score and enters through the inclusive-value approximation described above. The associated rival-presence variables remain excluded from airline $j$'s fixed costs. For each airline $j$, we then estimate $\sigma_{\eta_j}$ by a separate 2SLS, pooling across latent types $\ell=1,\ldots,L$ over the route-quarters in which airline $j$ is a potential entrant and both directional markets are observed, with latent-type-by-quarter fixed effects.

Once the scale parameters are estimated, we recover fixed costs as the residual from equation~\eqref{eq:entry_inversion}:
\begin{equation*}
    \widehat{fc}_{j\ell t}
    = \widehat{VP}^{P}_{j\ell}(\boldsymbol{x}_{t})
    - \widehat{\sigma}_{\eta_j}\,\Lambda^{-1}\!\left(\widehat{\widetilde{P}}_{j,\ell}(\boldsymbol{x}_{t})\right).
\end{equation*}

\subsection{Variance decomposition}

For airline $j$, reindex its $T_j$ retained potential-entry markets as $t=1,\ldots,T_j$ and assign probability $1/T_j$ to each market. Independently, latent type $\ell$ has probabilities $\widetilde{f}_1,\ldots,\widetilde{f}_L$. The resulting product measure is consistent with the independence of $\boldsymbol{x}_{t}$ and the latent market state in Assumption~\ref{assumption_1}[d]. Each object $Y_{j\ell t}\in\{\Pi_{j\ell t},VP_{j\ell t},fc_{j\ell t}\}$ is a deterministic function of these two inputs.

Define the unconditional, market-specific, and type-specific means by
\[
\overline{Y}_j = T_j^{-1}\sum_{t=1}^{T_j}\sum_{\ell=1}^{L}\widetilde{f}_\ell Y_{j\ell t}, \qquad m_j^Y(t) = \sum_{\ell=1}^{L}\widetilde{f}_\ell Y_{j\ell t}, \qquad n_j^Y(\ell) = T_j^{-1}\sum_{t=1}^{T_j}Y_{j\ell t}.
\]
The interaction residual is
\begin{equation*}
u_{j\ell t}^{Y}
\equiv
Y_{j\ell t}-m_j^Y(t)-n_j^Y(\ell)+\overline{Y}_j.
\end{equation*}
The variance components are
\[
\begin{alignedat}{2}
V_j^Y &= T_j^{-1}\sum_{t=1}^{T_j}\sum_{\ell=1}^{L}\widetilde{f}_\ell\big(Y_{j\ell t}-\overline{Y}_j\big)^2,
&\qquad V_{j,x}^Y &= T_j^{-1}\sum_{t=1}^{T_j}\big(m_j^Y(t)-\overline{Y}_j\big)^2, \\[2pt]
V_{j,\ell}^Y &= \sum_{\ell=1}^{L}\widetilde{f}_\ell\big(n_j^Y(\ell)-\overline{Y}_j\big)^2,
&\qquad V_{j,\mathrm{res}}^Y &= T_j^{-1}\sum_{t=1}^{T_j}\sum_{\ell=1}^{L}\widetilde{f}_\ell\big(u_{j\ell t}^Y\big)^2.
\end{alignedat}
\]
Orthogonality of the three ANOVA components under the product measure gives
\begin{equation*}
V_j^Y=V_{j,x}^Y+V_{j,\ell}^Y+V_{j,\mathrm{res}}^Y.
\end{equation*}
The components capture variation due to observables alone, latent types alone, and their interaction, respectively.

\subsection{Measures of the contribution of unobservables to endogenous selection}

For $Y\in\{\Pi,VP,fc\}$ with $V_j^Y>0$, define
\begin{equation*}
V_{j,\mathrm{unobs}}^Y
\equiv
V_{j,\ell}^Y+V_{j,\mathrm{res}}^Y.
\end{equation*}
The ratio $V_{j,\mathrm{unobs}}^Y/V_j^Y$ is the total-effect share of latent market types: the share of variance associated with them either directly or through their interaction with observable market characteristics. Its complement is $V_{j,x}^Y/V_j^Y$.

To distinguish the channels through which latent types affect selection, we decompose $V_{j,\mathrm{unobs}}^{\Pi}$ into the corresponding expected-variable-profit and fixed-cost components. Write $\overline{VP}_j$ and $\overline{fc}_j$ for $\overline{Y}_j$ evaluated at $Y=VP$ and $Y=fc$. The pure-type and interaction covariance components are
\begin{align*}
\operatorname{Cov}^{VP,fc}_{j,\ell}
&\equiv
\sum_{\ell=1}^{L}\widetilde{f}_\ell
\big(n_j^{VP}(\ell)-\overline{VP}_j\big)
\big(n_j^{fc}(\ell)-\overline{fc}_j\big),\\[4pt]
\operatorname{Cov}^{VP,fc}_{j,\mathrm{res}}
&\equiv
\frac{1}{T_j}\sum_{t=1}^{T_j}\sum_{\ell=1}^{L}
\widetilde{f}_\ell\,u_{j\ell t}^{VP}u_{j\ell t}^{fc},\\[4pt]
\operatorname{Cov}^{VP,fc}_{j,\mathrm{unobs}}
&\equiv
\operatorname{Cov}^{VP,fc}_{j,\ell}
+\operatorname{Cov}^{VP,fc}_{j,\mathrm{res}}.
\end{align*}
Because $\Pi_{j\ell t}=VP_{j\ell t}-fc_{j\ell t}$,
\begin{equation}\label{eq:sobol_var_unobs_split}
    V^{\Pi}_{j,\mathrm{unobs}} \;=\;
    V^{VP}_{j,\mathrm{unobs}} \;+\;
    V^{fc}_{j,\mathrm{unobs}} \;-\;
	    2 \, \operatorname{Cov}^{VP,fc}_{j,\mathrm{unobs}}.
\end{equation}
We report the three normalized terms
\begin{equation*}
\frac{V^{VP}_{j,\mathrm{unobs}}}{V_j^{\Pi}},
\qquad
\frac{V^{fc}_{j,\mathrm{unobs}}}{V_j^{\Pi}},
\qquad
\frac{-2\,\operatorname{Cov}^{VP,fc}_{j,\mathrm{unobs}}}{V_j^{\Pi}}.
\end{equation*}
They are, respectively, the gross expected-variable-profit contribution, the gross fixed-cost contribution, and the covariance adjustment. The first two are gross contributions and therefore need not lie between zero and one. By equation~\eqref{eq:sobol_var_unobs_split}, the three terms sum to $V_{j,\mathrm{unobs}}^{\Pi}/V_j^{\Pi}$.

\section{Additional results \label{app:additional_results}}

{
\let\savedendtable\endtable
\renewcommand{\endtable}{%
\par\vspace{0.5em}
\centering
\begin{minipage}{0.9\linewidth}
\footnotesize
\textit{Notes:} The table reports the pairwise Hausman-type test $\mathcal{H}(A,B)$ from equation~\eqref{eq:hausman_general} applied, within each panel, to every pair among the baseline 2SLS and the six selection-corrected estimators of Tables~\ref{tab_app_demand} and~\ref{tab_app_cost}; the OLS estimator is not included in these tests. Each cell gives the $\chi^2$ statistic for the null that the row and column estimators recover the same parameter vector, with $\widehat{\boldsymbol{V}}_{AB}$ computed using the linearized common-weight bootstrap of Online Appendix~\ref{app:bootstrap} with 200 replications. Panel A reports the joint tests on the demand parameters $(\alpha,\sigma,\beta_{\text{dist}},\beta_{\text{dist}^2},\beta_{\text{pres}})$, Panel B restricts the tests to the price and nesting parameters $(\alpha,\sigma)$, and Panel C reports the joint tests on the marginal-cost parameters $(\varphi_{\text{dist}},\varphi_{\text{dist}^2})$. Stars denote rejection of the null of equality: $^{*}$ at the 10\% level, $^{**}$ at the 5\% level, and $^{***}$ at the 1\% level.
\end{minipage}
\par
\savedendtable}
\setlength{\tabcolsep}{4pt}
\begin{table}[ht]
\caption{Pairwise Hausman Specification Tests \label{tab_hausman}}
\centering\resizebox{0.9\textwidth}{!}{
\begin{tabular}{l|cccccc}
    \hline \hline
     & 2SLS & Heckman & Semipar. & {Cont.\ $\kappa$} & $\text{F.M., }L=2$ & $\text{F.M., }L=3$ \\
    \hline
    & \multicolumn{6}{l}{\textit{Panel A: All demand parameters, $\chi^2(5)$}} \\
    Heckman & $189.80^{***}$ & & & & & \\
    Semipar. & $165.49^{***}$ & $48.57^{***}$ & & & & \\
    Cont.\ $\kappa$ & $83.12^{***}$ & $17.64^{***}$ & $33.41^{***}$ & & & \\
    $\text{F.M., }L=2$ & $55.71^{***}$ & $166.48^{***}$ & $149.24^{***}$ & $86.01^{***}$ & & \\
    $\text{F.M., }L=3$ & $12.23^{**}$ & $194.87^{***}$ & $170.68^{***}$ & $81.73^{***}$ & $27.32^{***}$ & \\
    $\text{F.M., }L=4$ & $3.09$ & $139.95^{***}$ & $143.29^{***}$ & $82.92^{***}$ & $16.17^{***}$ & $2.73$ \\
    \hline
    & \multicolumn{6}{l}{\textit{Panel B: Price and nesting parameters, $\chi^2(2)$}} \\
    Heckman & $101.06^{***}$ & & & & & \\
    Semipar. & $97.29^{***}$ & $23.26^{***}$ & & & & \\
    Cont.\ $\kappa$ & $64.81^{***}$ & $6.22^{**}$ & $12.90^{***}$ & & & \\
    $\text{F.M., }L=2$ & $3.98$ & $98.98^{***}$ & $97.13^{***}$ & $66.36^{***}$ & & \\
    $\text{F.M., }L=3$ & $6.74^{**}$ & $86.93^{***}$ & $88.50^{***}$ & $54.66^{***}$ & $4.80^{*}$ & \\
    $\text{F.M., }L=4$ & $1.94$ & $44.47^{***}$ & $59.21^{***}$ & $44.13^{***}$ & $1.32$ & $0.51$ \\
    \hline
    & \multicolumn{6}{l}{\textit{Panel C: Marginal cost parameters, $\chi^2(2)$}} \\
    Heckman & $7.56^{**}$ & & & & & \\
    Semipar. & $9.24^{***}$ & $1.81$ & & & & \\
    Cont.\ $\kappa$ & $13.72^{***}$ & $7.62^{**}$ & $10.64^{***}$ & & & \\
    $\text{F.M., }L=2$ & $1.17$ & $5.74^{*}$ & $6.99^{**}$ & $11.25^{***}$ & & \\
    $\text{F.M., }L=3$ & $5.36^{*}$ & $2.72$ & $3.67$ & $5.36^{*}$ & $3.00$ & \\
    $\text{F.M., }L=4$ & $2.86$ & $19.59^{***}$ & $22.34^{***}$ & $10.96^{***}$ & $1.82$ & $0.07$ \\
    \hline \hline
\end{tabular}}
\end{table}

}

{
\let\savedendtable\endtable
\renewcommand{\endtable}{%
\par\vspace{0.5em}
\centering
\begin{minipage}{0.9\linewidth}
\footnotesize
\textit{Notes:} The table reports average marginal costs implied by the pricing equation~\eqref{eq:mc_1}, evaluated at the observed prices and market shares using the corresponding column's demand parameter estimates from Table~\ref{tab_app_demand}. Marginal costs are first computed at the airline-market level and are then averaged overall and separately by airline.
\end{minipage}
\par
\savedendtable}
\begin{table}[ht]
\caption{Average Marginal Costs \label{tab_app_mc}}
\centering\resizebox{0.9\textwidth}{!}{
    \begin{tabular}{l|cc|cccccc}
    \hline \hline
    & \multicolumn{2}{c|}{\textit{Not control. for sel.}}
    & \multicolumn{6}{c}{\textit{Controlling for endogenous selection}} \\
    & {OLS} & {2SLS} & {2SLS} & {2SLS} & {2SLS} & {2SLS} & {2SLS} & {2SLS} \\
    & & & {Heckman} & {Semipar.} & {Cont.\ $\kappa$} & {Fin.-Mix.} & {Fin.-Mix.} & {Fin.-Mix.} \\
    & & & $L=1$ & $L=1$ & $L=1$ & $L=2$ & $L=3$ & $L=4$ \\
    \hline
    \multicolumn{1}{r|}{\textit{Marginal Cost (100\$)}} & $-0.002$ & $1.180$ & $1.455$ & $1.485$ & $1.653$ & $1.201$ & $1.297$ & $1.338$ \\
    \multicolumn{1}{r|}{\textit{AA}} & $0.265$ & $1.348$ & $1.596$ & $1.622$ & $1.791$ & $1.369$ & $1.459$ & $1.501$ \\
    \multicolumn{1}{r|}{\textit{DL}} & $0.207$ & $1.477$ & $1.776$ & $1.809$ & $1.978$ & $1.498$ & $1.599$ & $1.641$ \\
    \multicolumn{1}{r|}{\textit{UA}} & $0.448$ & $1.571$ & $1.830$ & $1.857$ & $2.025$ & $1.592$ & $1.684$ & $1.725$ \\
    \multicolumn{1}{r|}{\textit{US}} & $0.081$ & $1.254$ & $1.526$ & $1.556$ & $1.723$ & $1.275$ & $1.370$ & $1.411$ \\
    \multicolumn{1}{r|}{\textit{WN}} & $-0.459$ & $0.877$ & $1.193$ & $1.228$ & $1.395$ & $0.897$ & $1.002$ & $1.043$ \\
    \multicolumn{1}{r|}{\textit{LCC}} & $-0.258$ & $0.801$ & $1.043$ & $1.068$ & $1.237$ & $0.823$ & $0.911$ & $0.952$ \\
    \multicolumn{1}{r|}{\textit{Others}} & $-0.004$ & $0.840$ & $1.024$ & $1.040$ & $1.209$ & $0.861$ & $0.937$ & $0.977$ \\
    \hline
    Airline$\times$Quarter FE & Y & Y & Y & Y & Y & Y & Y & Y \\
    Airport FE & Y & Y & Y & Y & Y & Y & Y & Y \\
    \# control var.\ entry & 0 & 0 & 6 & 18 & 42 & 6 & 12 & 18 \\
    Observations  & $37,184$ & $37,184$ & $37,184$ & $37,184$ & $37,184$ & $37,184$ & $37,184$ & $37,184$ \\
    \hline \hline
    \end{tabular}}
\end{table}

}

{
\let\savedendtable\endtable
\renewcommand{\endtable}{%
\par\vspace{0.5em}
\centering
\begin{minipage}{0.9\linewidth}
\footnotesize
\textit{Notes:} The table reports estimates of the marginal-cost parameters in equation~\eqref{eq:mc_2}. The dependent variable $\ln(mc_{jt})$ is constructed from the implied marginal costs obtained from equation~\eqref{eq:mc_1} using the corresponding column's demand parameter estimates from Table~\ref{tab_app_demand}. All columns are estimated by OLS. Standard errors are computed using the linearized common-weight bootstrap of Online Appendix~\ref{app:bootstrap} with 200 replications. The BIC$_{\text{2nd}}$ row reports the analogous second-step BIC-style criterion for the log-marginal-cost regression, in the same spirit as equation~\eqref{eq:BIC_2}. The last two rows report the Hausman-type specification test $\mathcal{H}(\text{2SLS}, \cdot)$ from equation~\eqref{eq:hausman_general}, applied here to the two marginal-cost parameters $(\varphi_{\text{dist}},\varphi_{\text{dist}^2})$ ($\chi^2(2)$). The number of observations varies across columns because observations with a non-positive implied marginal cost from equation~\eqref{eq:mc_1} are excluded.
\end{minipage}
\par
\savedendtable}
\begin{table}[ht]
\caption{Estimation of Marginal Cost Parameters \label{tab_app_cost}}
\centering\resizebox{0.9\textwidth}{!}{
    \begin{tabular}{l|cc|cccccc}
    \hline \hline
    & \multicolumn{2}{c|}{\textit{Not control. for sel.}}
    & \multicolumn{6}{c}{\textit{Controlling for endogenous selection}} \\
    & {OLS} & {2SLS} & {2SLS} & {2SLS} & {2SLS} & {2SLS} & {2SLS} & {2SLS} \\
    & & & {Heckman} & {Semipar.} & {Cont.\ $\kappa$} & {Fin.-Mix.} & {Fin.-Mix.} & {Fin.-Mix.} \\
    & & & $L=1$ & $L=1$ & $L=1$ & $L=2$ & $L=3$ & $L=4$ \\
    \hline
    Distance (1000mi) & $1.368$ & $1.390$ & $0.904$ & $0.864$ & $0.764$ & $1.325$ & $1.143$ & $1.103$ \\
     & $(0.093)$ & $(0.202)$ & $(0.161)$ & $(0.130)$ & $(0.091)$ & $(0.203)$ & $(0.205)$ & $(0.149)$ \\
    & & & & & & & & \\
    Distance$^2$ & $-0.281$ & $-0.318$ & $-0.164$ & $-0.150$ & $-0.134$ & $-0.296$ & $-0.242$ & $-0.231$ \\
     & $(0.030)$ & $(0.058)$ & $(0.042)$ & $(0.033)$ & $(0.023)$ & $(0.059)$ & $(0.058)$ & $(0.039)$ \\
    & & & & & & & & \\
    \hline
    Airline$\times$Quarter FE & Y & Y & Y & Y & Y & Y & Y & Y \\
    Airport FE & Y & Y & Y & Y & Y & Y & Y & Y \\
    \# control var.\ entry & 0 & 0 & 6 & 18 & 42 & 6 & 12 & 18 \\
    Observations  & $19,441$ & $36,390$ & $37,176$ & $37,177$ & $37,184$ & $36,438$ & $36,731$ & $36,809$ \\
    & & & & & & & & \\
    BIC$_{\text{2nd}}$ & & $-61819.4$ & & & & $-63535.7$ & $-75253.9$ & $-78389.5$ \\
    Hausman $\chi^2(2)$ vs 2SLS & & & $7.56$ & $9.24$ & $13.72$ & $1.17$ & $5.36$ & $2.86$ \\
    \quad p-value & & & $[0.023]$ & $[0.010]$ & $[0.001]$ & $[0.558]$ & $[0.069]$ & $[0.240]$ \\
    \hline \hline
    \end{tabular}}
\end{table}

}

{
\let\savedendfigure\endfigure
\renewcommand{\endfigure}{%
\par\vspace{0.5em}
\centering
\begin{minipage}{0.9\linewidth}
\footnotesize
\textit{Notes:} The figure reports the empirical distributions of estimated own-price elasticities at the airline-market level. Each row corresponds to the price-setting airline, while the three columns correspond respectively to the 2SLS estimator without selection correction, the continuous-$\boldsymbol{\kappa}_{t}$ correction with $L=1$, and the finite-mixture correction with $L=3$. Within each column, elasticities are evaluated using the corresponding demand parameter estimates $\left(\widehat{\alpha},\widehat{\sigma}\right)$ at the observed prices and market shares.
\end{minipage}
\par
\savedendfigure}
\begin{figure}[!t]
\caption{Distribution of Estimated Own-Price Elasticities \label{fig_distribution_elasticities}}
    \centering
    \includegraphics[width=10.5cm]{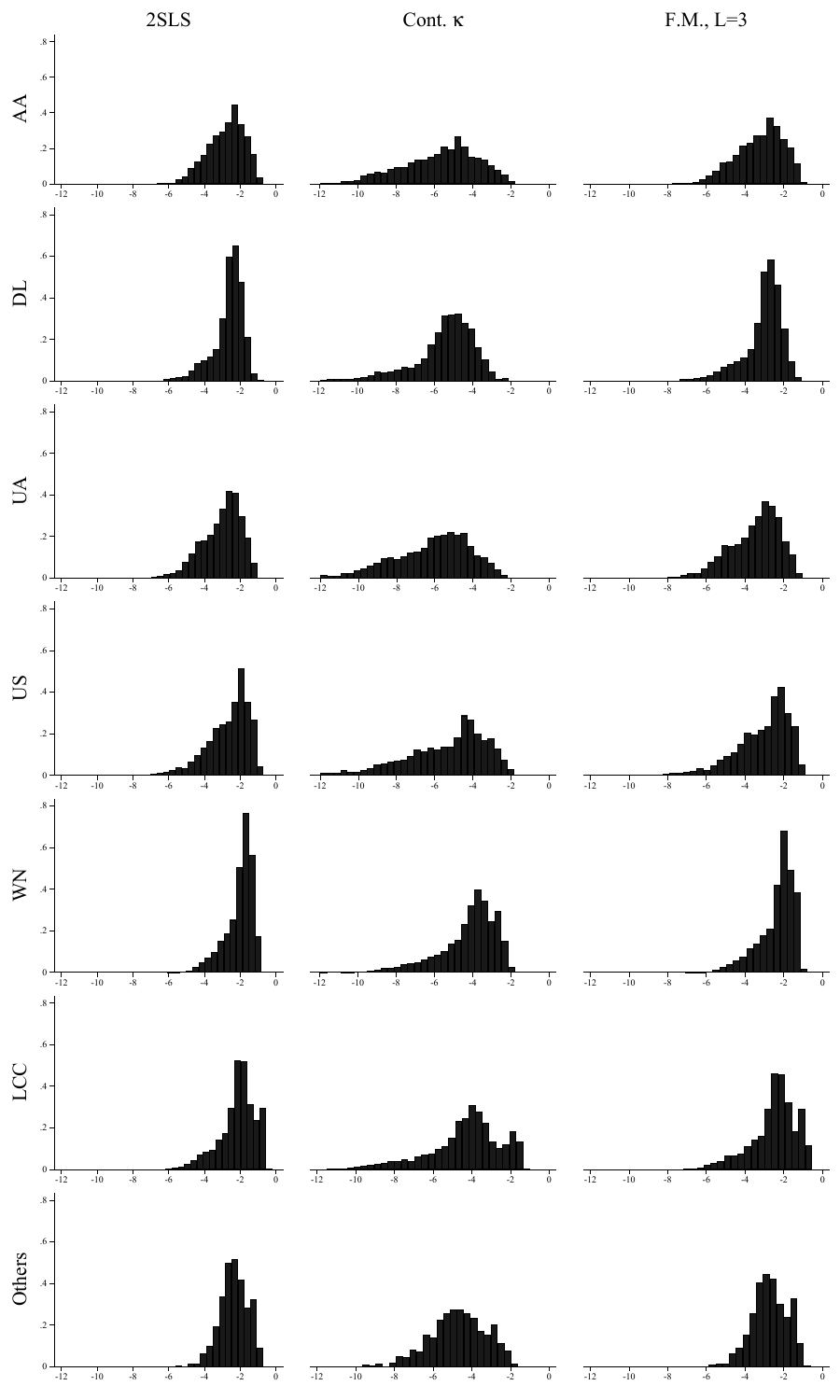}
\end{figure}
}

{
\let\savedendfigure\endfigure
\renewcommand{\endfigure}{%
\par\vspace{0.5em}
\centering
\begin{minipage}{0.9\linewidth}
\footnotesize
\textit{Notes:} The figure reports the empirical distributions of estimated cross-price elasticities at the airline-market level. Each row corresponds to the price-setting airline, while the three columns correspond respectively to the 2SLS estimator without selection correction, the continuous-$\boldsymbol{\kappa}_{t}$ correction with $L=1$, and the finite-mixture correction with $L=3$. Within each column, elasticities are evaluated using the corresponding demand parameter estimates $\left(\widehat{\alpha},\widehat{\sigma}\right)$ at the observed prices and market shares.
\end{minipage}
\par
\savedendfigure}
\begin{figure}[!t]
\caption{Distribution of Estimated Cross-Price Elasticities \label{fig_distribution_crosselast}}
    \centering
    \includegraphics[width=10.5cm]{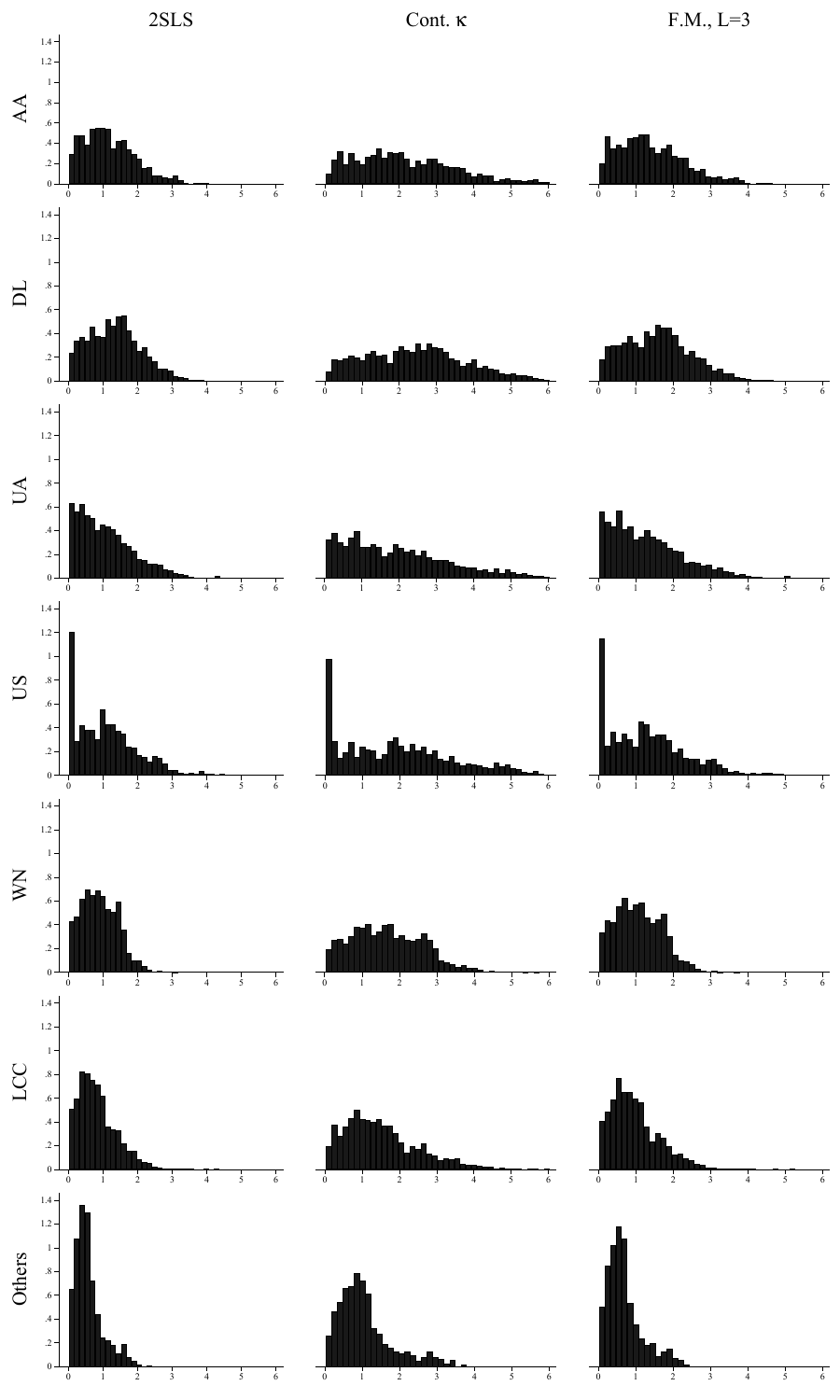}
\end{figure}
}

{
\let\savedendfigure\endfigure
\renewcommand{\endfigure}{%
\par\vspace{0.5em}
\centering
\begin{minipage}{0.9\linewidth}
\footnotesize
\textit{Notes:} The figure reports the empirical distributions of estimated marginal costs at the airline-market level. Each row corresponds to an airline, while the three columns correspond respectively to the 2SLS estimator without selection correction, the continuous-$\boldsymbol{\kappa}_{t}$ correction with $L=1$, and the finite-mixture correction with $L=3$. Within each column, marginal costs are computed from equation~\eqref{eq:mc_1} using the corresponding demand parameter estimates and the observed prices and market shares.
\end{minipage}
\par
\savedendfigure}
\begin{figure}[!t]
\caption{Distribution of Estimated Marginal Costs \label{fig_distribution_mc}}
    \centering
    \includegraphics[width=10.5cm]{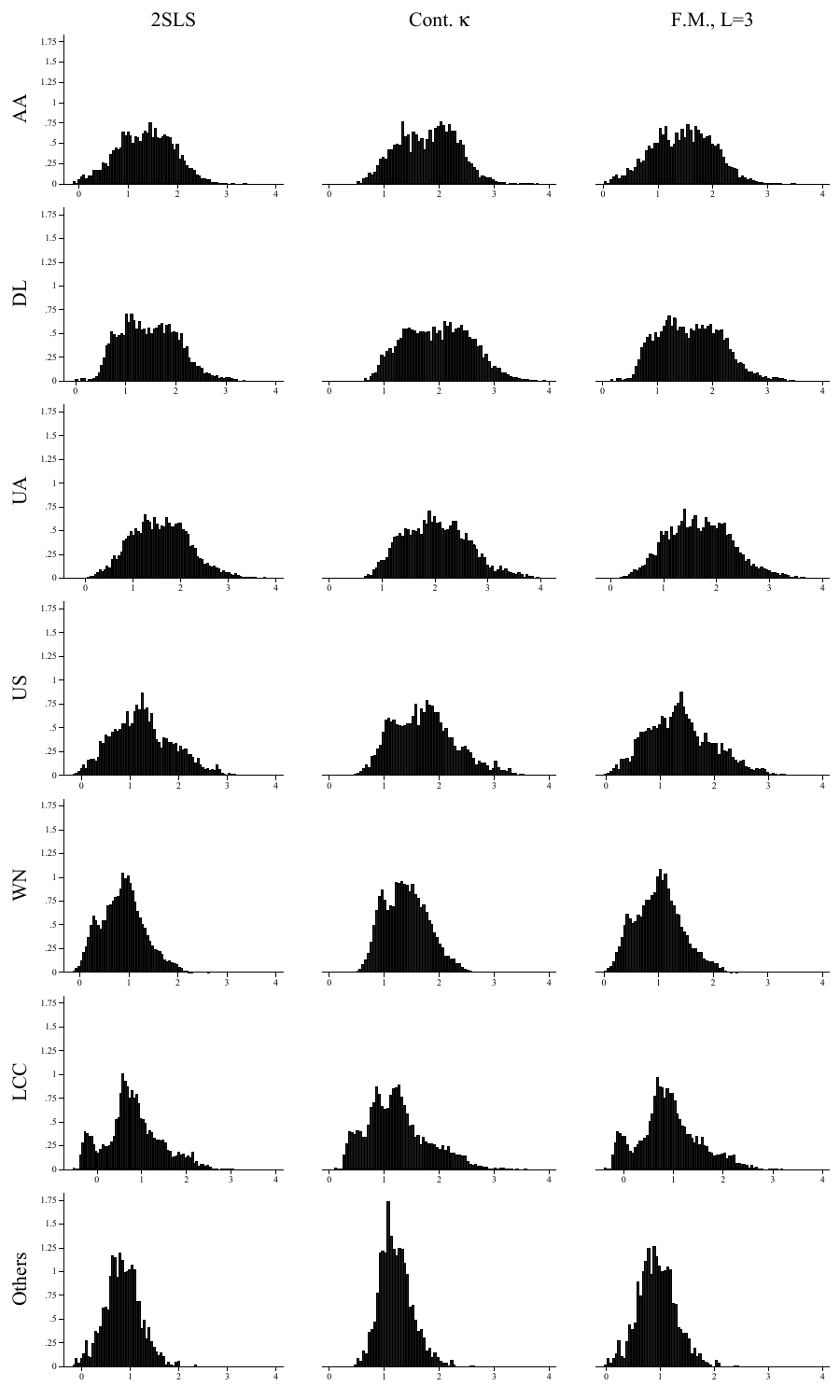}
\end{figure}
}

\end{document}